\documentstyle[aps,prl,epsfig]{revtex}
\begin{document}
\draft
\widetext

\title{Spatially heterogeneous ages in glassy dynamics}
\author{
Horacio~E.~Castillo$^{1,2}$, Claudio~Chamon$^1$,
Leticia~F.~Cugliandolo$^{3,4,5}$, Jos\'e~Luis~Iguain$^{4,6}$, 
Malcolm~P.~Kennett$^{7,8}$
}
\address{ 
$^1$ Physics Department, Boston University, Boston, MA 02215, USA \\
$^2$ Department of Physics and Astronomy, Ohio University,
Athens, OH 45701, USA\\
$^3$ Laboratoire de Physique Th{\'e}orique de l'{\'E}cole Normale
Sup{\'e}rieure, Paris, France \\ 
$^4$ Laboratoire de Physique
Th{\'e}orique et Hautes Energies, Jussieu, Paris, France \\ 
$^5$ Lyman Laboratory of Physics, Harvard University, MA 02138, USA\\
$^6$ D\'epartement de Physique et Groupe de Recherche en Physique et
Technologie des Couches Minces (GMC)\\
Universit\'e de Montr\'eal, Case Postale 6128
Montr\'eal, Qu\'ebec H3H 3J7, Canada.\\ $^7$ Department of Physics,
Princeton University, Princeton, New Jersey 08544, USA\\ $^8$
TCM Group, Cavendish Laboratories, Cambridge University, Madingley Rd, Cambridge
CB3 0HE, UK} \date{\today}

\twocolumn[\hsize\textwidth\columnwidth\hsize\csname@twocolumnfalse\endcsname
\maketitle

\begin{abstract}
We construct a framework for the study of fluctuations in the
nonequilibrium relaxation of glassy systems with and without quenched
disorder. We study two types of two-time local correlators with the
aim of characterizing the heterogeneous evolution in these systems: in
one case we average the local correlators over histories of the
thermal noise, in the other case we simply coarse-grain the local
correlators obtained for a given noise realization. We explain why the
noise-averaged correlators describe the fingerprint of quenched
disorder when it exists, while the coarse-grained correlators are
linked to noise-induced mesoscopic fluctuations.  We predict
constraints on the distribution of the fluctuations of the
coarse-grained quantities. In particular, we show that locally defined
correlations and responses are connected by a generalized local
out-of-equilibrium fluctuation-dissipation relation. We argue that
large-size heterogeneities in the age of the system survive in the
long-time limit. A symmetry of the underlying theory, namely
invariance under reparametrizations of the time coordinates, underlies
these results.  We establish a connection between the
probabilities of spatial distributions of local coarse-grained
quantities and the theory of dynamic random manifolds.  We define, and
discuss the behavior of, a two-time dependent correlation length from
the spatial decay of the fluctuations in the two-time local functions.
We characterize the
fluctuations in the system in terms of their fractal properties.
For concreteness, we present numerical tests performed on disordered
spin models in finite and infinite dimensions.  
Finally, we explain how these ideas can be applied to the analysis of
the dynamics of other glassy systems that can be either spin models
without disorder or atomic and molecular glassy systems.
\end{abstract}
]

\narrowtext


\section{Introduction}

A mean-field theory of glasses has been developed during the last two
decades~\cite{yo,chicos,giorgio}. This approach is based on the study of
fully-connected spin models with disorder.  A fully-connected spin system
with pairwise interactions, the Sherrington-Kirkpatrick ({\sc sk}) model, is
used to model spin-glasses~\cite{Mepavi,Dotsenko,Sompo,Cuku2}. Extensions in
which the spins interact via all possible $p$-uplets, with $p\geq 3$, are
used to describe structural glasses of fragile type~\cite{Kithwo,Cuku}.
These are the so-called $p$-spin models.  Even though structural glasses
consist of molecules moving in a finite dimensional volume, rather than spins
interacting via random exchanges on a complete (hyper) graph, the disordered
$p$-spin models yield a gross description of many important features of the
structural glass phenomenology. For instance, they have dynamic and static
transitions occurring at different values of the external temperature,
mimicking the dynamic slowing down at the freezing temperature, $T_g$, and
the entropy crisis at the Kauzmann temperature, $T_s$~\cite{Kithwo}.  More
strikingly, these models capture the slow non-equilibrium dynamics
characterized by macroscopic observables showing aging effects below
$T_g$~\cite{Cuku2,Cuku}.

Whilst they are successful in many respects, these models lack a geometric
structure and hence cannot inform us about the spatial evolution of the glass
former.  In the context of spin-glasses, there is consensus about there being
some kind of growing order below $T_g$. However, there has been a
long-lasting debate about the characteristics of this
order~\cite{Parisi,Fihu,Fihu2,older,Villain,OMartin,Krzakala,Payo,Lamarcq,Moore,Yoshi}.
In other words, the question as to whether there are only two or many
competing ground states, not related by symmetry, that grow during the
nonequilibrium
evolution~\cite{Fihu2,Yoshi,Eric,Nordblad,BerthierBouchaud,BerthierBouchaud2}
has not been answered yet. In comparison, it is not even clear if there is a
phase transition in structural glasses~\cite{glasses}.

Very recently, a number of experiments have shown the appearance of
mesoscopic regions in supercooled liquids and glasses that have distinctively
different dynamics from the bulk of the
system~\cite{Ediger,heterogeneities,confocal,confocal2,dyn-heterogeneities}.
The position and identity of these ``particles'' changes in time. In general,
these regions are referred to as dynamic heterogeneities and have also
been identified in numerical simulations~\cite{heterogeneities-num,Heuer}.

Developing a theoretical description of the real-space dynamics of glassy
systems is now a major challenge to theoreticians. The purpose of this paper
is to expand on the theoretical framework we presented in
Refs.~\cite{paper1,paper2} that allowed us to predict several properties of
local dynamic fluctuations in spin-glasses. We test the predictions of our
framework against numerical simulations on a spin model with disorder defined
on a finite dimensional lattice, the Edwards-Anderson ({\sc ea})
model~\cite{Mepavi,Dotsenko}.  Following the philosophy described in the
first paragraph, we claim that the main results arising from our analysis
will carry on rather simply to the particle models which are more relevant
for real glasses. In the Conclusions and Perspectives sections we explain how
one should translate our results and predictions to this case.  We also list
a number of glassy models on which our ideas could be tested.

\subsection{Background}

We first briefly summarize previous studies (mostly experimental and numerical)
of dynamic heterogeneities in 
particle and spin systems that set the scene for our analysis.

There is great interest in the experimental observation of dynamic
heterogeneous regions in super-cooled liquids and glasses.  Many experimental
techniques have been used to signal the existence of dynamically
heterogeneous regions in samples of a variety of 
glasses~\cite{Ediger,heterogeneities,confocal,confocal2,dyn-heterogeneities}, 
and to try to characterize their properties.  The confocal microscopy
technique~\cite{confocal,confocal2} is particularly useful for this, as it
allows one to reconstruct the particle trajectories in $3d$ space and have
the complete configuration of the system at chosen times.

In the context of theoretical studies of structural and polymeric glasses,
Bennemann {\it et al}. and Kob {\it et al}.  identified fast moving particles
embedded in a bulk of slow moving ones at temperatures above $T_g$ in the
super-cooled liquid phase of several models using molecular
dynamics~\cite{heterogeneities-num}. Other numerical studies of similar
features appeared in~\cite{Heuer}.

In these experimental systems, and the models used to study them
numerically, there is no quenched disorder. Moreover, the particles
are identical and move in continuous space, so they cannot be
identified by their position on a lattice, as in typical spin
models. However, two possible ways of studying the heterogeneous
dynamics of the system are as follows.  (i) One can tag each particle,
follow their evolution and detect which are the fast and slow moving
particles during a previously chosen time-window around some time
after preparation.  Particles are labeled by an index $i$.  This is
the route followed in Refs.~\cite{heterogeneities-num,Heuer}. (ii) One
can divide the space into boxes of a chosen size and study the
behavior of all particles within each box. The locality is then given
by the position of the box which is labeled by $i$. At the end of this
article we explain why we believe that the second approach will be
very useful to characterize some spatial features of the
nonequilibrium dynamics of glassy systems. An analysis of data
obtained with the confocal microscopy technique of real systems and
molecular dynamics of simple models along the lines described in this
paper will yield valuable information for the future development of a
complete analytic theory for glasses.

In the context of disordered spin models, fast and slow spins that
decorrelate on totally different time-scales were identified in numerical
simulations in Refs.~\cite{Sharon-spins} and~\cite{Baze} for the $3d$ {\sc ea}
model above and below $T_c$, respectively. Ricci-Tersenghi and
Zecchina~\cite{Rize} and Montanari and Ricci-Tersenghi~\cite{Mori} found a
similar separation in the low temperature phase of spin models defined on
random graphs.  Even finite-size samples of mean-field models show important
spin-to-spin fluctuations in the characteristic time-scale for individual
decorrelation, as shown by Brangian and Kob for the disordered Potts model
above its dynamic critical temperature~\cite{Brangian}. A similar behavior,
superposed on aging phenomena, is observed in the Sherrington-Kirkpatrick
model at low temperatures, as shown in Sec.~\ref{finite-size-effects-sk}.  In
all these studies the noise-averaged correlations for fixed disorder were the
focus of the studies.  In this type of analysis the identity of a spin is
determined by its position on the lattice or by the random exchanges.

In Refs.~\cite{paper1} and \cite{paper2} we concentrated instead on the 
$3d$ {\sc ea} model and its two-time coarse-grained (but not noise-averaged) local
quantities.  In Ref.~\cite{paper1} we showed that the action for the slow
part of the local relaxation becomes reparametrization invariant
asymptotically. This (approximate) symmetry allowed us to propose several
properties of the dynamic behavior of the coarse-grained local correlations
and responses that we tested numerically in Ref.~\cite{paper2}. These
quantities are relevant both for spin models (with and without quenched
disorder) and for continuous systems of interacting particles.

\subsection{Plan of the paper}

In this paper we complete the analysis that we started in Refs.~\cite{paper1}
and \cite{paper2}. We study several aspects of the {\it local} dynamics of
the $2d$ and $3d$ {\sc ea} spin-glass,
\begin{equation}
{\cal H}_J = \sum_{\langle i,j \rangle} J_{ij} s_i s_j  
\label{dea}
\; . 
\end{equation}
The sum runs over nearest neighbor sites $i,j$ on a $d$ dimensional cubic
lattice.  The couplings $J_{ij}$ take values $\pm J/\sqrt{2z}$ with
probability $1/2$.  $z$ is the coordination of the lattice, $z=2d$ in the
square/cubic case.  The spins are Ising variables, $s_i=\pm 1$.

We also analyze the dynamic fluctuations in finite dimensional and
fully-connected models with finite size. To test the latter we use the {\sc
sk} mean-field spin-glass model defined by
\begin{equation}
{\cal H}_J = \sum_{ i\neq j } J_{ij} s_i s_j  
\; ,
\label{sk}
\end{equation}
with $J_{ij}$ taken from a bimodal probability distribution with zero mean
and variance $[J^2_{ij}]=J^2/(2N)$. (We expect to find similar results using
a Gaussian distribution of exchanges.) Here and in what follows we use square
brackets to denote the average over disorder.

We fix the value of $J$ in such a way that the critical temperatures are at
$T_c\sim 1.1$ for the $3d$ {\sc ea} model and at $T_c=1$ for the 
{\sc sk} model. The
$2d$ {\sc ea} model has a zero-temperature phase transition. We set the
Boltzmann constant, $k_B$, to one.

We focus on two types of locally defined correlations and susceptibilities:

\noindent
(i) coarse-grained local quantities
$C^{cg}_i(t,t_w)$ and $\chi_i^{cg}(t,t_w)$.

\noindent 
(ii) noise-averaged local
quantities $C_i^{na}(t,t_w)$ and $\chi_i^{na}(t,t_w)$. 

\noindent
The two time dependence reflects the out-of-equilibrium dynamics of these
systems after the quench at time $t=0$. $t_w$ denotes the waiting-time
elapsed after preparation and $t$ a longer time, $t\geq t_w$.  We present a
detailed comparison of the behavior of these local non-equilibrium
correlations that have been averaged differently.

The numerical simulations were performed as follows.  To study the finite 
$d$ {\sc ea} model we evolved a cubic (square) system with side-length $L$ and
periodic boundary conditions from a random initial condition using 
Metropolis
dynamics at temperature $T$.  The random initial configuration represents the
result of an infinitely rapid quench from infinite temperature performed at
$t=0$.  In the $3d$ case we considered $L=32$ and $L=64$ and for $d=2$ we
considered $L=128$. To compute spatially coarse-grained two-time quantities
we used a coarse-graining volume which is a cubic (square) box of linear size
$2M+1$. A coarse-graining time that serves to make the spin variable smooth
varies from study to study and is noted in each plot.  We considered several
values of the external temperature that lie above and below $T_c$ and these
are indicated as necessary. The noise-averaged data we present were obtained
using shorter values of the waiting-time to allow for an average over many
noise-realizations, typically $10^3$ samples.

When studying the {\sc sk} model we evolved systems with $N=128$ and $N=512$
spins with Monte Carlo dynamics at $T=0.4$ also starting from a random
configuration of spins. The noise-averaged data were obtained using,
approximately, $10^3$ samples.

In short, the results in this paper are organized as follows:

First, we establish the dynamic scaling forms of both coarse-grained and
noise-averaged local quantities numerically. In particular, we test scaling
forms that we propose in Sec.~\ref{sec:definitions}.

Second, we study the local relations between
noise-averaged correlation and integrated response and between the same
quantities when coarse-grained. 

Third, we show that the global quantities in
finite size systems in finite dimensions and those 
defined on the complete graph show similar fluctuations as 
the local quantities in finite $d$ models.

Fourth, we propose a relation between the study of the probability 
distribution of local fluctuations and the theory of dynamic
random surfaces. 

Fifth, we define and analyze a dynamic correlation length that depends
on two-times.  
 
Sixth, we present a new way of looking at
geometric properties in spin-glasses that should be relevant to future
experiments with local probes.  We analyze the real-space organization of
local correlations, $C^{cg}_i$ and $C_i^{na}$
by investigating the geometric properties of
the (random) surfaces given by their evaluation on the substrate $d$
dimensional real space. In particular, we study the properties of clusters of
spins with local correlation in the interval $[{\tt C}, {\tt C}+d{\tt C}]$
for which ${\tt C}$ is a parameter taking values between $-1$ and $1$. Again
we compare the behavior of noise-averaged and coarse-grained quantities.  
A similar analysis could be applied to the random surface of local 
susceptibilities, $\chi_i^{cg}$ and $\chi_i^{na}$.

Because real systems do not equilibrate on accessible time-scales,
spatially resolved measurements will not be static; instead, they will
still depend on the age of the system, very much like the bulk or
global measurements. In these measurements one can monitor noise and
response in a {\it mesoscopic} region of the sample. The results of
our analysis will be of relevance to the interpretation and analysis
of these experiments.  They can also be used as a source of
inspiration to analyze dynamic heterogeneities in super-cooled liquids
and glasses~\cite{heterogeneities,heterogeneities-num,Heuer}, as we
discuss in the Conclusions and Perspectives sections.

\subsection{Summary of results}

Before getting into the technical details, let us summarize our results. 

In Ref.~\cite{paper1} we showed that, in the limit of long-times, a
zero mode develops in the dynamics of finite dimensional spin-glasses.
This soft mode is related to the invariance of the effective action
for the slow fields (that are actually two-time functions) under a
global reparametrization of time. Thus, we argued that the least
costly spatial fluctuations should be ones that smoothly change the
local time reparametrization. In Ref.\cite{paper2} we tested these
ideas numerically by evaluating the local coarse-grained correlations
and integrated responses in the $3d$ {\sc ea} model.  We observed that
the fluctuations in these quantities are constrained to follow the
fluctuation-dissipation relation ({\sc fdr}) between the global
quantities as a direct consequence of the existence of the asymptotic
zero mode. In this paper we show further evidence that the
coarse-grained two-time correlators reflect the existence of an
asymptotic zero mode in the underlying theory.  In particular, using
the fact that the dynamics become ``critical'' in a well-chosen long
time limit, we explain why a scaling limit of long-times and large
coarse-graining volumes should exist in which the distributions of
fluctuations approach a stable limit.

In disordered systems such as the $3d$ {\sc ea} model, another set of local
two-time quantities can be defined using a different averaging
procedure. Indeed, one can work with noise-averaged, as opposed to
coarse-grained, two-time functions. Even if these quantities do not
fluctuate in systems without quenched disorder, they do in
spin-glasses and other random systems due to the fingerprint of
disorder.  One can then wonder if these quantities are also coupled
to the asymptotic zero mode and whether their fluctuations are
constrained in the same way as those of the coarse-grained
quantities. We show numerically that this is not the case:
the noise-averaged fluctuations behave in a rather different
way. 

In order to sustain further this claim, we also study the 
mesoscopic fluctuations in disordered models in finite 
and infinite dimensions. We show that 
the fluctuations in the global quantities, that are
due to the finite size of the systems, behave just like
the coarse-grained local quantities in finite dimensional
models. We observed this property in the $3d$ {\sc ea} and the 
{\sc sk} model.

We relate the study of the fluctuations in the local 
correlations (and susceptibilities) to the study of 
the evolution of random surfaces. Indeed, we propose that 
one can derive a ``phenomenological'' effective action 
for the fluctuations in the local quantities from the 
statistical analysis of the surfaces given by the evaluation 
of the two-time quantities on the $d$ dimensional substrate. 
This idea gives us a handle to describe analytically the 
fluctuations in a large variety of systems with slow dynamics. 

We study numerically the random surfaces describing the fluctuations
in space of both the coarse-grained and noise-averaged two-time local
correlators. We show that noise-averaging leads to surfaces that
encode the fingerprint of the disorder realization, and are static if
the two times $t$ and $t_w$ have a fixed ratio
(or $h(t)/h(t_w)$, with a more suitable function $h(t)$, is kept 
fixed).

The coarse-grained surfaces obtained for different pairs of times
$(t,t_w)$, even if the ratio $t/t_w$ [or $h(t)/h(t_w)$] is fixed, 
fluctuate and cross
each other at many points as a function of $t_w$. This result implies,
as we show, that the relative age (as measured using the correlation
value) between two sites $i$ and $j$ in the sample is not static, but
fluctuates as a function of time. These are examples of {\it sorpassi}
that we define in this paper, and show a clear contrast to
noise-averaged local quantities, where the relative age between all
sites in the sample keeps a constant, static, relative rank.

We define a {\it two-time dependent} correlation length $\xi(t,t_w)$
using the spatial correlation of the local two-time correlations. We
study how this correlation length diverges asymptotically in the
glassy phase of the $3d$ {\sc ea} model. We discuss how the ratio
between the coarse-grained volume and the correlation length affects
the probability distributions for the measured quantities. In
particular, we argue that when the coarse-graining length is smaller
than the correlation length, one probes the spatial fluctuations
controlled by the zero mode. When using 
coarse-graining lengths that are larger, but still of the order of the 
correlation length one is measuring mesoscopic
fluctuations of nearly independent finite size systems.
If the coarse-graining length is much larger than the 
correlation length the fluctuations are suppressed.
We show numerically that the qualitative 
features of the local and mesoscopic fluctuations 
are indeed very similar and we conjecture that they may 
have a similar origin. 

Finally, we study the spatial organization of the local correlations
(coarse-grained and noise-averaged).  We propose that the analysis of
the geometric properties of clusters of sites with similar values of
the local two-time correlations can be useful to determine if one is
at or below the lower critical dimension. More precisely, we claim
that the geometric organization of the fluctuations in the $2d$ {\sc
ea} model are different from those in the $3d$ case signaling the fact
that the former does not have a finite $T$ transition.  The
difference is very clear when one looks at the fractal dimension of
clusters, $d_f$.  In $3$ dimensions $d_f \sim 2$, whilst in $2$
dimensions, $d_f \leq 2$.  These values of the fractal dimension are
quite close to each other -- however, the difference between the
fractal dimension and that of the substrate space, $\Delta = d - d_f$,
is very different.  In $3$ dimensions, $\Delta \sim 1$ and in $2$
dimensions $\Delta \ll 1$.  This suggests that the level surfaces in $2$
dimensions lie on a much rougher underlying manifold, as we suggest
from theoretical arguments in Sec.~\ref{sec:randomsurfaceaction}, and
should be linked to the absence of a glass transition in $d=2$.
However, a multifractal analysis gives less clear-cut distinctions,
suggesting that the non-equilibrium aging regime in the $2$ dimensional
case has aspects that are very similar to the $3$ dimensional case.  

In the conclusions we discuss how
to adapt this approach and the picture that emerges 
to super-cooled liquids and structural glasses.

\section{Definitions and discussion}
\label{sec:definitions}

In this section we define the two-time, global and local, correlators that we
study numerically in the rest of the paper. We recall some known properties
of the global correlators. We then discuss possible scaling forms for the
local quantities, as well as the implications of these scaling forms.

\subsection{Two-time dependent global functions}

To date, analytical, numerical, and experimental studies in glassy systems
have mainly focused on the global correlation and integrated response:
\begin{eqnarray}
C(t,t_w) &\equiv& \frac1{N} \sum_{i=1}^N s_i(t) s_i(t_w)
\; , 
\label{globalC}
\\
\chi(t,t_w) &\equiv& \frac1{N} \sum_{i=1}^N \int_{t_w}^t dt' \, 
\left. \frac{\delta s_i(t) \epsilon_i}{\delta \eta_i(t')} \right|_{\eta=0}
\; ,
\label{globalchi}
\end{eqnarray}
where $\eta_i$ is a (site-dependent) magnetic field given by
$\eta_i=\eta\epsilon_i$ with $\epsilon_i=\pm 1$ with probability
$\frac12$ and $\eta$ its magnitude.  The field couples linearly to the
spin, $H_J \to H_J - \sum_i s_i \eta_i$.  The product $s_i(t) \epsilon_i$
is the ``staggered local spin'', {\it i.e.} the projection of $s_i$ on
the direction of the local external field $\eta_i$.  In order to
extract the linear part of the response the variation is evaluated at
zero field, $\eta=0$. The integrated response is usually averaged 
over many realizations of the random field.

For an infinite system
that evolves out of equilibrium these quantities are self-averaging,
and thus averages over the thermal noise and disorder (if existent)
are not required. All the generic analytic arguments we shall develop
assume that the thermodynamic limit, $N\to\infty$, has been taken at
the outset. (We discuss finite size effects in
Sec.~\ref{finite-size-effects}.)

\subsection{Two-time dependent local functions}

Quenched random interactions have a strong effect on the local properties of
spin systems. For instance, Griffiths singularities in the free-energy of
random ferromagnets are due to regions in space with strong ferromagnetic
couplings~\cite{Griffiths}. These lead to dynamic slowing down even in the
disordered phase of the random problem, below the transition temperature of
the pure model. It is natural to expect that heterogeneous
dynamics in spin-glasses arises for similar reasons. In these systems random
exchanges can be very different between one region of the sample and another:
some regions can have purely ferromagnetic interactions, others can have
purely antiferromagnetic ones, others can be frustrated. One can analyze the
fingerprint of the disorder on the local dynamics by choosing not to average
over the random exchanges.

However, heterogeneous dynamics do not arise simply because of quenched
random couplings. Glassy systems with no explicit quenched disorder also
exist in nature~\cite{glasses}. Many models with spin or particle variables
that capture their behavior have been
proposed~\cite{modelswithoutdisorder,Felix,Juanpe}.  Even if there is no
quenched disorder in these systems, one expects to find heterogeneous
dynamics in which some regions evolve differently than
others~\cite{dyn-heterogeneities,Juanpe}.

An extreme example of the latter situation occurs in ferromagnetic domain
growth~\cite{Bray}. At any finite time with respect to the size of the
system, a coarsening system is heterogeneous. Observed on a very short
time-window, spins on interfaces behave very differently from spins in the
bulk of domains. However, there is nothing special about the identity of these
spins. Spins that belong to an interface at one time can later become part of
a domain and even later be part of another wall. Importantly, no local
region (not even of the minimum linear size given by the lattice spacing) can
be considered to be equilibrated while coarsening is taking place.

If one wishes to analyze the local fluctuations in the dynamics of 
spin systems, two natural
functions to monitor are the two-time local correlations and responses. These
can be made continuous (in Ising spin systems) through different averaging
procedures that highlight different properties of the systems. Each
definition has a different theoretical motivation.

First, consider {\it spatially coarse-grained functions}~\cite{paper2}
\begin{eqnarray}
C^{cg}_i(t,t_w) &\equiv& \frac{1}{V}  \sum_{j\in V_i} 
\overline s_j(t) \overline s_j(t_w) 
\; ,
\label{coarse-C}
\\
\chi_i^{cg}(t,t_w) &\equiv& 
\left.
\frac{1}{V}  \sum_{j\in V_i} 
\int_{t_w}^t dt'  \; \frac{\delta \overline s_j(t) \epsilon_j}
{\delta \eta_j(t')} 
\right|_{\eta =0}
\; , 
\label{coarse-chi}
\end{eqnarray}
where $V_i$ is a coarse-graining region centered on site $i$ with volume $V$,
and the overline stands for a coarse-graining over a short 
time-window $\tau_t$ ($\tau_{t_w}$)
around $t$ ($t_w$), $\tau_t \ll t$ ($\tau_{t_w} \ll t_w$). 
(Note that we use the same coarse-graining volume
on all sites.) Only one realization of the thermal noise is used here, which
mimics nature. This definition is natural for the study of finite
dimensional models in which there is a notion of space and neighborhood.
Indeed, a coarse-graining procedure of this type is usually used to derive a
continuum field-theoretical description of a problem originally defined on a
lattice~\cite{book}. Moreover, it is of use if one wants to compare the
behavior of finite dimensional models with and without disorder since it is
non-trivial in both cases. This quantity is also relevant to compare with
experiments in which mesoscopic probes are limited to testing the behavior of
regions with a minimum size that involve a large (though mesoscopic)
number of spins.

Second, one can define {\it single-site noise-averaged
  functions}~\cite{Sharon-spins,Baze,Rize,Mori,Brangian,Pa}:
\begin{eqnarray}
C_i^{na}(t,t_w) &\equiv& 
\langle \overline s_i(t) \overline s_i(t_w) \rangle 
\; ,
\label{noise-averaged-C}
\\
\chi_i^{na}(t,t_w) &\equiv& 
\left.
\int_{t_w}^t dt' \; \frac{\delta \langle 
\overline s_i(t) \epsilon_i\rangle}{\delta \eta_i(t')}
\right|_{\eta=0}
\; .
\label{noise-averaged-chi}
\end{eqnarray}
Here and in what follows the angular brackets represent the average
over thermal histories.
These functions will appear, for example, in a dynamic cavity method applied
to a disordered model~\cite{cavity-statics}. 
This definition is particularly useful for mean-field
(fully-connected and dilute) systems with quenched disorder for which there
is no notion of neighborhood. However, it completely erases the inherent
heterogeneity of the dynamics in a non-disordered system such as
ferromagnetic domain growth. Moreover, a single-spin experimental measurement
is unlikely, and usually a region involving a large number of spins is probed,
implying an effective coarse-graining.

Neither of the two definitions above include, in the case of random
systems, an average over disorder realizations. This may allow us to detect
regions that have special behavior due to the random interactions. We insist
on the fact that the coarse-grained definition still contains noise-induced
fluctuations.

\subsection{Correlation-scales}
\label{correlation-scales}

The relaxation of glassy systems may take place on many different
time scales. A precise definition of ``correlation-scales'' was 
given in~\cite{Cuku2}. Assuming that a chosen two-time correlation, 
$C$, is a monotonic function of both times $t$ and $t_w$,  
in the long waiting-time limit, one can relate
the values it takes at any three times using a time-independent
function. More precisely, $C(t_1,t_3) = f[C(t_1,t_2), C(t_2,t_3)]$
for $t_1\geq t_2\geq t_3$ when all three times are very long. 

The correlation scales are defined as follows.
Within a correlation scale, $f(x,y) =\jmath^{-1} [\jmath(x) \jmath(y)]$
and $C(t_1,t_2)=\jmath^{-1}[h(t_1)/h(t_2)]$ with $h(t)$ a monotonic 
function of time and $\jmath(x)$ another function. 
Between correlation-scales the function $f$ is 
``ultrametric'', $f(x,y)=\min(x,y)$.

To explain this definition with an example, the correlation function 
$C(t,t_w) = (1-q_{\sc ea}) \exp[-(t-t_w)/\tau] + q_{\sc ea} (t_w/t)$ 
decays in two scales that are separated at the value $C=q_{\sc ea}$
that one sees as a plateau in $C$ that develops at 
long $t_w$ in a plot against $t-t_w$ in logarithmic scale.
The first scale is stationary and characterized by $h_{\sc st}(t) 
= \exp(-t/\tau)$,
the second one ages and is characterized by $h_{\sc ag}(t) = t$.

The structure of scales can be different for different 
correlators. 
The local correlations defined in Eq.~(\ref{coarse-C}) [or in 
Eq.~(\ref{noise-averaged-C})] are different observables labelled by $i$. 
Their decay should follow these generic 
rules whenever one can assume that they are monotonic.

\subsection{Behavior of global two-time quantities}
\label{behaviour-global}

In the long waiting-time limit, taken after the thermodynamic limit,
one can prove analytically that a sharp separation of time-scales
characterizes the dynamics of mean-field glassy
models~\cite{Cuku2,Cuku,Kech,Cukule,Frme}.  A similar separation of time-scales
has been observed numerically~\cite{Picco,numerics,KobBarrat,JLBarrat} and
experimentally~\cite{Eric} in a variety of glassy systems.  In short,
one finds:~\cite{clarify}

\noindent
(i) a fast
stationary evolution at short time-differences in which the correlation
approaches a plateau defined as the Edwards-Anderson parameter
\begin{eqnarray}
q_{\sc ea} &\equiv& \lim_{t-t_w\to\infty} C_{\sc st}(t-t_w) \equiv
\lim_{t-t_w\to\infty} \lim_{t_w\to\infty} C(t,t_w) 
\; ,
\label{globalqEA}
\end{eqnarray}
with  the integrated response linked to the correlation
by the fluctuation-dissipation 
theorem ({\sc fdt}) 
\begin{eqnarray}
\chi_{\sc st}(t-t_w) &\equiv&
\lim_{t_w\to\infty} \chi(t,t_w) 
\nonumber\\
&=&
\frac{1}{T} [1-C_{\sc st}(t-t_w)] 
\; ,
\end{eqnarray}
and, in particular, 
\begin{eqnarray}
\lim_{t-t_w\to\infty} \lim_{t_w\to\infty}
\chi(t,t_w)
&=&
\frac{1}{T} (1-q_{\sc ea}) 
\; .
\label{globalchiEA}
\end{eqnarray}
This regime is also called time-translational invariant
({\sc tti}). In general, the decay in this regime is not 
exponential ({\it cfr.} the example given in Sec.~\ref{correlation-scales}).

\noindent
(ii) A slow aging relaxation for longer time-differences
\begin{eqnarray}
C(t,t_w) &\neq& C(t-t_w)
\; ,
\nonumber\\
\chi(t,t_w) &\neq& \chi(t-t_w)
\; ,
\end{eqnarray}
when the value of the global correlation drops below $q_{\sc ea}$ [and the
integrated response goes above $(1-q_{\sc ea})/T$]. 

The number of scales that appear in this second decay depends on the
model considered. For the fully-connected $p$ spin model a single
correlation-scale has been found below $q_{\sc ea}$ in which the
global correlation scales with a power law, $h_{\sc
ag}(t)=t$~\cite{Cuku,KimLatz}.  A {\it sequence} of global-correlation
scales exists in the analytic solution to the {\sc sk}
model~\cite{Cuku2,Kech}.  For a manifold moving in an infinite
dimensional embedding space with a short-ranged random potential one
finds that the Fourier modes of the correlation $C_r(t,t_w) = \int_V
d^dr' \langle \vec \phi(\vec r+\vec r',t) \vec \phi(\vec r',t_w)
\rangle$ decay on two scales that are separated at a $k$-dependent
Edwards-Anderson value, $q_{\sc ea}^k$~\cite{Cukule}. The functions
$h_{\sc st, ag}$ that characterize the two scales are identical for
all wave-vectors.  This has also been found in molecular dynamic
simulations of Lennard-Jones mixtures~\cite{KobBarrat,JLBarrat}.  If
the manifold feels a a long-ranged random potential~\cite{Cukule,Frme}
the $k$-modes decay in a sequence of scales.  If this structure exists
in any non-mean-field problem it is not clear yet. In particular, the
$3d$ {\sc ea} model behaves more like a model with only two
global-correlation scales~\cite{Picco} but this may be due to the
shortness of the simulation times. In more complex problems one could
even find that different correlators decay on totally different time
scales.

In the second time regime the
{\sc fdt} is not satisfied. However, 
in many glassy models evolving out of equilibrium the global correlation
[Eq.~(\ref{globalC})] and the global integrated response [Eq.~(\ref{globalchi})] are
linked in a rather simple manner~\cite{Cuku2}. Indeed, assuming that the
global correlation decays monotonically as a function of $t$, one can invert
this function and write $t=\tilde f_g^{-1}(C(t,t_w),t_w)$ and
\begin{eqnarray}
\chi(t,t_w) &=& \chi(\tilde f_g^{-1}(C(t,t_w),t_w), t_w)  
\; ,
\end{eqnarray}
implying
\begin{eqnarray}
\lim_{t_w\to\infty\; C(t,t_w) = {\tt C}} \chi(t,t_w)
&=& \tilde \chi({\tt C})
\; ,
\label{fdr}
\end{eqnarray} 
where ${\tt C}$ is held fixed as we take the limit which we assume exists.
We call Eq.~(\ref{fdr}) a fluctuation-dissipation relation ({\sc fdr}). In
equilibrium $\tilde \chi({\tt C})=(1-{\tt C})/T$ holds for all values of
${\tt C}$. Out of equilibrium one finds a different relation with a kink: 
when ${\tt C}> q_{\sc ea}$ the equilibrium result holds, $\tilde
\chi({\tt C})=(1-{\tt C})/T$; when ${\tt C}< q_{\sc ea}$ a non-trivial
functional form $\tilde \chi({\tt C})$ is found and its particular form
depends on the model or system considered.  
For disordered spin models of the mean-field type 
two types of curves have been found:

\noindent
(i) For $p$ spin models it is a broken 
line with two different slopes, one being the negative of the inverse 
temperature of the bath for ${\tt C} \in [q_{\sc ea},1]$;
the other has a  different slope and spans the interval
${\tt C} \in [0,q_{\sc ea}]$~\cite{Cuku,KimLatz}. 

\noindent
(ii) For the {\sc sk} model the construction 
also has a breaking point at $q_{\sc ea}$ separating a straight 
line with slope minus the inverse 
temperature of the bath for ${\tt C} \in [q_{\sc ea},1]$, and a 
curved piece for  ${\tt C} \in [0,q_{\sc ea}]$~\cite{Cuku2}.

In the case of the random manifold the form of the modifications of 
{\sc fdt} depend on the range of correlation of the random 
potential~\cite{Cukule}.
One finds a linear relation between the Fourier modes of the 
space-dependent correlator and susceptibility if it is short-ranged 
and a non-linear form if it is long-ranged. Moreover, one finds 
that all modes behave essentially in the same way in both cases.
We shall discuss this issue in Sec.~\ref{local-resp-fdr}.

Several simulations support the fact that the {\sc fdr} of global 
quantities in models for structural 
glasses, such as Lennard-Jones mixtures, behave like a
manifold in a short-range correlated random potential, 
for which the second slope does not 
vanish and is equal for all the 
wave-vectors $k$~\cite{KobBarrat,JLBarrat}. In the 
case of the $3d$ {\sc ea} model the numerical results were interpreted
as supporting the existence of a curve with non-constant slope
below $q_{\sc ea}$~\cite{fdt-3dea}. Similar conclusions were
drawn from the experimental
work presented in~\cite{Herisson}. The numerical and experimental 
data are still rather far from the asymptotic time
limit, and, in our opinion, it is quite difficult to decide from the present 
data if $\tilde\chi(C)$ is indeed curved or a straight line below 
$q_{\sc ea}$.

\subsection{Behavior of local correlations}
\label{behavior-local}

In this section we denote by $C_i$ a generic local correlation that has been
coarse-grained, noise-averaged, or smoothed with any other prescription.  
We first 
discuss the possible time dependences of individual correlations at a
very general level. Later we distinguish between coarse-grained and
noise-averaged values.

\subsubsection{Local Edwards-Anderson parameter}
\label{local-qea}

Similarly to our discussion of the decay of the global correlation,
see Eq.~(\ref{globalqEA}), one can define a
{\it local} Edwards-Anderson parameter as the value of the local correlation
separating fast and slow decays,
\begin{equation}
q^i_{\sc ea} \equiv \lim_{t-t_w\to\infty} \lim_{t_w\to\infty}C_i(t,t_w) 
\; .
\end{equation}
The local correlations should decay quickly towards $q^i_{\sc ea}$ and
then slowly below this value. If the structure of the global
correlation is preserved at the local level, the first step of the
relaxation should be stationary whilst the second one could be
waiting-time dependent.

Site-to-site fluctuations in $q^i_{\sc ea}$ are possible. For some
spins the local Edwards-Anderson parameter may vanish, {\it i.e.} $q^i_{\sc
ea}=0$, and they may only show the first, fast decay. These are the
fast spins identified in Refs.~\cite{Baze,Rize,Mori} in purely noise
averaged local correlations.  However, one expects the fluctuations in
$q^i_{\sc ea}$ to be smoothed by spatial coarse-graining and to
disappear if a sufficiently large coarse-graining volume $V$ is
used~\cite{paper2}.

\subsubsection{Slow local relaxation}
\label{subsubslow}

What about the decay of the local correlations below the plateau at
  $q^i_{\sc ea}$?  Any monotonically decaying two-time correlation
  within a correlation-scale behaves as~\cite{Cuku2}
\begin{equation}
C_i(t,t_w) = f_i\left(\frac{h_i(t)}{h_i(t_w)}\right)
\; .
\label{scaling1}
\end{equation}
There are two special cases that deserve special mention.
On the one
hand, the scaling arguments $h_i(t)$
 could be site-independent, $h_i(t)=h(t)$, and thus
\begin{equation}
C_i(t,t_w) = f_i\left(\frac{h(t)}{h(t_w)}\right)
\; .
\label{localf}
\end{equation}
On the other hand, 
the external function $f_i$ could be independent 
of the site index $i$ and the scaling could take the form 
\begin{equation}
C_i(t,t_w) = f\left(\frac{h_i(t)}{h_i(t_w)}\right)
\; .
\label{eq:localh}
\end{equation}
This allows several combinations. The main ones to be discussed below are:
(i) the scaling in Eq.~(\ref{localf}) holds for all sites in the sample, (ii)
the scaling in Eq.~(\ref{eq:localh}) holds everywhere in the
sample. Evidently, one can also have more complicated behaviors: (iii) parts
of the sample scale as in Eq.~(\ref{localf}), other parts scale as in
Eq.~(\ref{eq:localh}), and still other parts do not satisfy either of the
special forms but are described by the more general form (\ref{scaling1}).
Note that the noise-averaged and coarse-grained local correlations do not
necessarily scale in the same way.

We first explore the consequences of having the behavior in
Eq.~(\ref{localf}) for all sites in the sample.  For simplicity, let
us first assume that the global correlation decays in a single
correlation-scale once its value drops below the global
Edwards-Anderson parameter,
\begin{equation}
C(t,t_w) = f_g\left( \frac{h_g(t)}{h_g(t_w)}\right)
\; , \;\;\; C<q_{\sc ea}
\; .
\label{global}
\end{equation} 
This is the behavior of fully-connected $p$-spin models with $p\geq
3$~\cite{Cuku}. One can expect it to hold in dilute ferromagnetic $p=3$ spin
models~\cite{Mori} and, surprisingly, it is also suggested by numerical
simulations in the $3d$ {\sc ea} model, at least for the waiting and
total times explored~\cite{Picco}. If the hypothesis in Eq.~(\ref{global})
holds, the $h(t)$ in Eq.~(\ref{localf}) must be identical to $h_g(t)$, and
one can write
\begin{equation}
C_i(t,t_w) = f_i\left( f_g^{-1}(C(t,t_w)) \right) = \tilde f_i(C(t,t_w))
\; .
\label{localf2}
\end{equation}
For all pairs of times $(t,t_w)$ leading to 
a fixed value of the global correlation, $C(t,t_w)={\tt C}$,
one has 
\begin{equation}
C_i(t,t_w)=\tilde f_i({\tt C})
\; .
\end{equation}
All local correlations are then site-dependent functions of the global
correlation. Following the line of thought in Ref.~\cite{Cuku2} one can
extend this argument to the case with a global correlation that decays over
many time scales.  If one assumes that the global correlation is monotonic,
one can write $C_i$ as a function of $C(t,t_w)$ and $t_w$ using $t=\tilde
f_g^{-1}(C(t,t_w),t_w)$. If one further assumes that in the limit
$t_w\to\infty$ with $C(t,t_w)={\tt C}$ fixed {\it each} local correlation
approaches a limit, then
\begin{equation}
\lim_{t_w\to\infty, C(t,t_w)={\tt C}}
C_i(t,t_w) 
= \tilde f_i({\tt C})  
\; .
\label{limit1}
\end{equation}
This means that {\it within each} correlation scale {\it all} local
correlations are locked into following the scaling of the global one.

Replacing the local space index $i$ by a wave vector $k$, this is exactly the
behavior of a finite dimensional manifold embedded in an infinite
dimensional space~\cite{Cukule}.  When the random potential is short-range 
correlated the dynamics are such that there are only two
correlation scales in the problem, a {\sc tti} one satisfying {\sc fdt} and 
an aging one in which all $k$-dependent correlators decay below $q^k_{\sc ea}$
as in Eq.~(\ref{localf}). Instead, when the random potential is of 
long-range type, the correlations have a {\sc tti} scale 
that ends at a $k$-dependent Edwards-Anderson plateau,  $q^k_{\sc ea}$,
and a subsequent decay that takes place in a sequence of scales. 
All modes are locked in the sense that for all times
one can write $C_k=f_k(C_0)$ where the argument can be any chosen mode, 
for instance, the $k=0$ one.
The former behavior was also found numerically in the $k$ modes of the 
incoherent scattering function 
of a Lennard-Jones mixture where there is only one correlation scale below
the $k$-dependent $q^k_{\sc ea}$~\cite{KobBarrat}.  Recently it was also
proposed for all site dependent correlators 
in the context of a ferromagnetic model on a random graph in 
Ref.~\cite{Mori} and for the $3d$ {\sc ea} model in Ref.~\cite{Pa}.  In these
cases the correlations and susceptibilities were averaged over the noise with
no spatial coarse graining.

\begin{figure}
\psfig{file=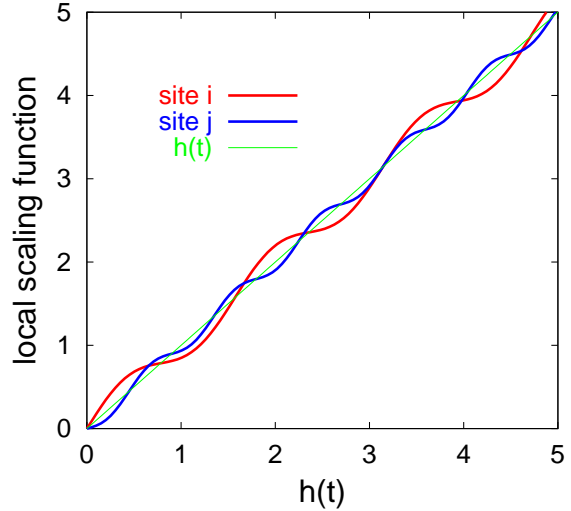,width=8cm}
\vspace{0.2cm}
\caption{Example of scaling functions $h_{i,j}(t)$ {\it vs.} $h(t)$.
Because the functions $h_{i,j}(t)$ cross, there are pairs of times
$(t,t_w)$ for which $C_i>C_j$, but others for which $C_i<C_j$, even
though $h_{i,j}(t)$ only fluctuate slightly around the scaling
function $h(t)$.  }
\label{sketch-exchange}
\end{figure}

Instead, if the hypothesis in Eq.~(\ref{eq:localh}) holds, there are many open
possibilities.  For instance, different sites can evolve on totally
different time-scales. A simple example illustrating this point is
given by two sites $i$ and $j$ that decay below their Edwards-Anderson
values to zero in only one scale but that are different:
$C_i(t,t_w)=q_{\sc ea}^i (\ln t_w)/(\ln t)$ and $C_j(t,t_w)=q_{\sc
ea}^j t_w/t$. In this case, site $i$ decays on a slower scale than
site $j$. There is also the possibility that, while being in the same
correlation scale, the value of the correlations at two
sites $i,j$ may cross each other as a function of time. Let us be more
explicit; consider two sites with local scaling functions
\begin{eqnarray}
h_i(t)&=&h(t)+\delta h_i(t)
\; ,
\\
h_j(t)&=&h(t)+\delta h_j(t)
\; ,
\nonumber
\end{eqnarray}
with $|\delta h_{i,j}(t)|\ll h(t)$, so that the sites evolve within the same
scale. [Note that both $h_i(t)$ and $h_j(t)$ are monotonic functions.]
In this case, one can expand the expression in Eq.~(\ref{eq:localh})
and write:
\begin{eqnarray}
C_i(t,t_w) \approx &&f\left[\frac{h(t)}{h(t_w)}
\left(1+\frac{\delta h_i(t)}{h(t)}
-\frac{\delta h_i(t_w)}{h(t_w)}\right)\right]
\\
\approx && f\left(\frac{h(t)}{h(t_w)}\right)
\nonumber
\\
&&+f'\left(\frac{h(t)}{h(t_w)}\right)\;\frac{h(t)}{h(t_w)}\;
\left(\frac{\delta h_i(t)}{h(t)}
-\frac{\delta h_i(t_w)}{h(t_w)}\right)
\nonumber
\; ,
\end{eqnarray}
and similarly for $C_j(t,t_w)$. At any pair of times $(t,t_w)$, the
difference between the two local correlations is
\begin{equation}
C_i-C_j\propto
\frac{\delta h_i(t)-\delta h_j(t)}{h(t)}
-\frac{\delta h_i(t_w)-\delta h_j(t_w)}{h(t_w)}
\; .
\label{eq:deltah}
\end{equation}
In general, the right-hand-side (r.h.s.) of Eq.~(\ref{eq:deltah})
can change sign as a function of the two times $t$ and $t_w$, allowing for
{\it sorpassi} (passing events) between different sites.  An example
of functions $h_i$ and $h_j$ for which these {\it sorpassi} are
realized is shown in Fig.~\ref{sketch-exchange}.

Finally, one could also find cases in which $\delta h_{i,j}(t)$ are
large [$|\delta h_{i,j}(t)|\ll h(t)$ is not satisfied] and the {\it
sorpassi} can be such that $C_i$ and $C_j$ move across correlation
scales. These variations are quite dramatic. One could expect 
them to be penalized in such a way that they appear less frequently
than the ones in which $|\delta h_{i,j}(t)|\ll h(t)$. We shall return
to this issue in Sec.~\ref{sigma}.

\subsection{Behavior of local responses and fluctuation-dissipation
relation}
\label{local-resp-fdr}

Let us define a generalized {\it local}
fluctuation-dissipation relation ({\sc fdr}) via the limit
\begin{equation}
\lim_{t_w\to\infty\; C_i(t,t_w)={\tt C}_i}
\chi_i(t,t_w) =  \tilde\chi_i({\tt C}_i)
\; .
\label{localchiiCi}
\end{equation}
This limit exists whenever the local correlations are not multivalued, 
{\it i.e.} if they are monotonic functions of $t$ for fixed $t_w$. 
Indeed, in this case one can invert for the time
\begin{equation}
t=\tilde f^{-1}(C_i(t,t_w),t_w) 
\;,
\end{equation}
and write
\begin{equation}
\chi_i(t,t_w) = \chi_i(\tilde f^{-1}(C_i(t,t_w),t_w),t_w)
\; .
\end{equation} 
Taking the limit $t_w\to\infty$ while keeping $C_i(t,t_w)$ fixed
to ${\tt C}_i$ one recovers Eq.~(\ref{localchiiCi}).

The question now arises as to whether the fluctuations 
in $C_i$ and $\chi_i$ are independent or whether 
they are constrained to satisfy certain relations.

Based on thermometric arguments~\cite{Cukupe}, we shall associate 
the variation of $\tilde\chi_i$ with respect to $C_i$ 
to a local  effective-temperature. We would also like to know if the 
values of the local effective-temperatures are constrained or can 
freely fluctuate.

 In the rest of this subsection we discuss different scenarios for the
behavior of the joint probability distribution ({\sc pdf}) $\rho(C_i,\chi_i)$
computed at two fixed times $t$ and $t_w$. We also discuss the
behavior of the fluctuations in the local effective-temperature.  In
order to illustrate different possibilities we present
several plots that sketch the following construction.  Given a pair of
times $t_w\leq t$, we depict the $N$ points $(C_i(t,t_w),T
\chi_i(t,t_w))$ with arrows that represent the velocity of the points
[{\it i.e}, the rate at which the $(C_i,T \chi_i)$ positions change as
one changes $t$], and are located at their position in the $C-T\chi$
plane.  Note that all these points are evaluated {\it at the same}
pair of times $(t,t_w)$.
For the sake of comparison,
in the same plots we also draw the parametric
plot for the global $T \tilde \chi(C)$ constructed as
usual~\cite{Cuku2}: for a fixed $t_w$ we follow the evolution of the
pair $(C,T\chi)$ as time $t$ evolves from $t=t_w$ to $t\to \infty$.

In the figures showing the distributions of $(C_i,\chi_i)$ pairs 
we scale the y-axis by temperature to work with
dimensionless variables. This is important to compare the extent of 
fluctuations in the two directions.

\subsubsection{Similar times}

Let us first discuss the case in which the two times $t$ and $t_w$
 chosen to evaluate the local correlation and integrated linear
 response are close to each other, in such a way that the global
 correlation between them lies above the Edwards-Anderson
 parameter. In this case the global correlator and the global linear
 integrated response are stationary and related by the {\sc fdt}. In
 this regime of times we also expect the local quantities to be linked
 by the {\sc fdt}.

When studying noise-averaged two-time quantities one can
actually use the bound derived in Ref.~\cite{Cukude} to justify this 
claim. For coarse-grained quantities we do not have an 
analytic argument to use but we find this expectation very 
reasonable. Note that the arguments put forward in Sec.~\ref{sigma}
do not apply to these short time-differences since they 
hold for the slowly varying part of the two-time functions only.
We have verified numerically that the fluctuations of the local 
coarse-grained correlations and responses are 
concentrated rather spherically around the global value
$\chi=(1-C)/T$ when $C>q_{\sc ea}$~\cite{paper2}.

\subsubsection{Far away times: no constraint}

We now choose two times $t$ and $t_w$ such that the 
global correlation taken between them is 
less than $q_{\sc ea}$. 

The simplest possibility is that the 
fluctuations in $C_i$ and $\chi_i$ are independent.
If these are not specially constrained the 
pairs $(C_i,T\chi_i)$ can be scattered almost everywhere on the 
$C-T\chi$ plane.

\begin{figure}
\begin{center}
\psfig{file=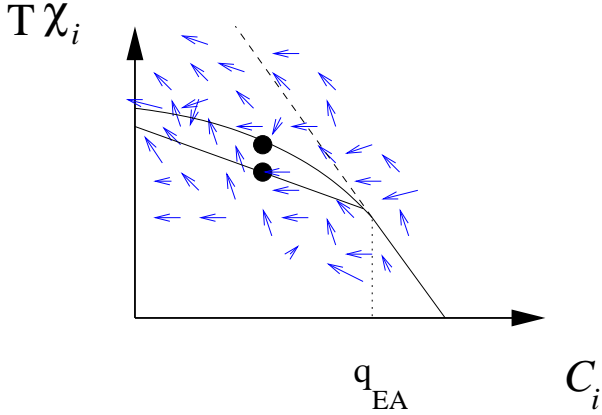,width=8cm}
\end{center}
\caption{Sketch of the local {\sc fdr} between integrated
responses and correlations on all sites, at a fixed pair of times $(t,t_w)$
such that the global correlation goes below $q_{\sc ea}$. In this 
case neither the positions of the pairs ($C_i,T\chi_i)$ nor
the velocities of the points, {\it i.e.} the direction of the arrows
that are associated to the effective temperatures, 
are constrained. This case represents Eqs.~(\ref{scaling1}) 
and~(\ref{localTnoconstraint}). The full line sketches the 
parametric plot $\tilde\chi(C)$ for the global quantities in a model with 
two (straight line below $q_{\sc ea}$)
and a model with a sequence of global correlation-scales
(curve below $q_{\sc ea}$). The dashed line is the continuation of the 
section where the {\sc fdt} holds. The black dots indicate the location of the 
averaged values for the susceptibility and correlation, 
$\chi_{\sc av}=1/N \sum_i \chi_i$ and 
$C_{\sc av}=1/N \sum_i C_i$ at the times $t$ and $t_w$.}
\label{chiCcualq}
\end{figure}

The scaling in Eq.~(\ref{scaling1}) means 
that different sites might evolve on different time-scales.
This could happen, for example, if $h_i(t)=t$ and 
$h_j(t)=\ln(t/t_0)$. 
Hence, one can expect each site to have their own effective temperature, 
\begin{equation}
-\beta^{\sc eff}_i(C_i) \equiv \frac{d\tilde\chi_i(C_i)}{dC_i} 
\; .
\label{localTnoconstraint}
\end{equation}
This situation is sketched in Fig.~\ref{chiCcualq}.

\subsubsection{Far away times: locked scales}

Naturally, one expects that the fluctuations in the local
correlations and integrated linear 
responses at far away times are not completely independent and that
they are constrained to satisfy certain relations. Here we explore the
consequences of the scaling expressed in
Eq.~(\ref{limit1}) for all sites in the sample. Hence,
\begin{equation}
\lim_{t_w\to\infty \;  C(t,t_w) = {\tt C}} \chi_i(t,t_w) 
= \tilde \chi_i(\tilde f_i({\tt C})) =\bar\chi_i({\tt C})
\; ,
\label{limit2}
\end{equation}
with ${\tt C}$ the chosen value of the global correlation.
Note  that there is no constraint 
on the values of $\bar\chi_i({\tt C})$ and $C_i=\tilde f_i({\tt C})$.
Thus, the pairs $(C_i,T\chi_i)$ can be scattered all over the 
plane.

However,
since all sites are locked to evolve in the 
same time-scale as the global correlation for each choice of times
$(t,t_w)$, one expects
\begin{equation}
-\beta^{\sc eff}_i(C_i) \equiv \frac{d\tilde \chi_i(C_i)}{dC_i} 
= -\beta^{\sc eff}({\tt C})
\; ,
\label{localTiindep}
\end{equation}
based on the expectation of having partial equilibration
between observables in the same time-scales~\cite{Cukupe}.
Thus, if the scaling in Eq.~(\ref{limit1}) holds for all sites in the 
sample, there should be  
a {\it site independent} effective temperature.

\begin{figure}
\begin{center}
\psfig{file=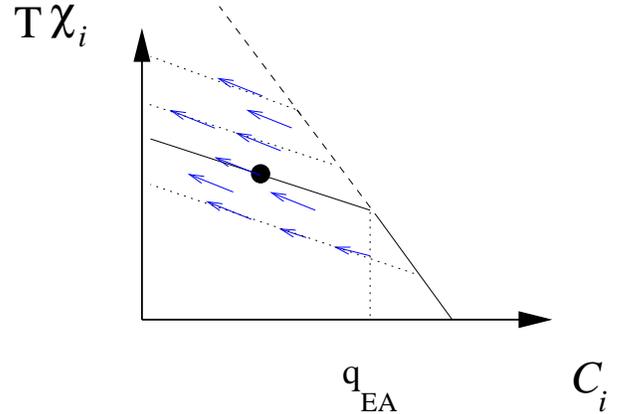,width=8cm}
\end{center}
\caption{Fluctuations in a system with only two correlation
scales defined using the global correlator. The plot 
shows fluctuations in the local Edwards-Anderson parameter, $q_{\sc ea}^i$.
$(t,t_w)$ are such that the global correlation goes below $q_{\sc ea}$
and the pairs $(C_i,T\chi_i)$ are not constrained to lie close to the global 
curve, $\tilde\chi(C)$, indicated with a full broken line.
The effective temperature is the same for all sites,
as shown by the fact that all arrows are parallel to the second
piece of the global curve. Consequently, once a site enters a ``track''
 after leaving $q_{\sc ea}^i$ 
(indicated with straight lines drawn with dots),
it will follow it as time $t$ evolves.
The dashed line is the continuation of the 
part where the {\sc fdt} holds. The black dot indicates the location of the 
values for the susceptibility and correlation averaged over the 
distribution.
}
\label{fig:sketch-chiC1step2}
\end{figure}

\begin{figure}
\begin{center}
\psfig{file=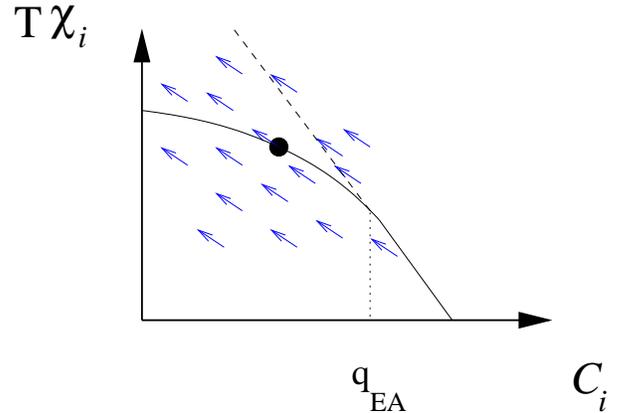,width=8cm}
\end{center}
\caption{Fluctuations in a system with a sequence of correlation
scales defined using the global correlator. 
The points are scattered on the plane with no constraint on their positions.
The local Edwards-Anderson parameter fluctuates.
The velocities with which the
points move are all the same. Their positions are not
constrained but their velocity moduli and directions are 
all identical.  This is  given by the slope at the 
value of the global correlation at the chosen times $(t,t_w)$,
indicated with a black dot on the global line.
}
\label{sketch-chiCthem}
\end{figure}

The content of  Eqs.~(\ref{localf2}) and (\ref{localTiindep}) is 
illustrated in 
Figs.~\ref{fig:sketch-chiC1step2} and 
\ref{sketch-chiCthem} for a system with two correlation-scales
and a system with a continuous sequence of correlation scales,
respectively. 

Note that in this time regime, at longer times $t$ the points will also be 
scattered on the plane but the average of the correlation will
take a smaller value. Since this determines the velocity of the 
points (and their effective temperature) the direction of the 
arrows will be different. We can say that there is a 
nonvanishing acceleration. 

One finds two correlation scales in the noise and disorder
averaged wavevector-dependent two-time functions of 
a $d$ dimensional 
manifold moving in a short-ranged random potential in infinite 
dimensions~\cite{Cukule}.
It has also been found in the noise-averaged 
local dynamics of a dilute ferromagnetic  
model~\cite{Mori} and in molecular dynamic simulations of Lennard-Jones
mixtures~\cite{KobBarrat,JLBarrat} for different wave-vectors.
  
\subsubsection{Far away times: fluctuations in time-scales}

As opposed to the case just discussed, the scaling in
Eq.~(\ref{eq:localh}) means that different sites might evolve on
different time-scales (if $|\delta h_i| \gg h(t)$) and hence have
their own effective temperature.

As we shall argue in Sec.~\ref{sigma}, under certain circumstances the 
local responses and fluctuations are constrained to follow the 
global curve, {\it i.e.} 
\begin{equation}
\tilde\chi_i(C_i) = \tilde\chi(C_i) 
\; ,
\label{globalconstrain}
\end{equation}
with 
$C$ and $\tilde\chi(C)$ defined in Eqs.~(\ref{globalC})
and (\ref{fdr}), respectively.
Then we have
\begin{equation}
-\beta_i^{\sc eff}(C_i) \equiv \frac{d\tilde\chi_i(C_i)}{dC_i} =
\frac{d\tilde\chi(C_i)}{dC_i} =
\left. \frac{d\tilde\chi(C)}{dC}\right|_{C=C_i}
\; .
\label{localTidep}
\end{equation}

If there are only two time-scales for the decay of the global 
correlation and, below the global Edwards-Anderson parameter, 
$\tilde\chi(C)$ is linearly dependent on $C$, this equation yields 
\begin{equation}
\beta_i^{\sc eff} = \beta^{\sc eff}
\; , 
\label{alleq}
\end{equation}
for all sites, and values of $C_i<q_{\sc ea}$. 

If, on the contrary, the global correlation decays in 
a sequence of scales and $\tilde\chi(C)$ is not a linear function of $C$, one
has fluctuations in the local effective temperature due to the 
fluctuations in $C_i$. 

\begin{figure}
\begin{center}
\psfig{file=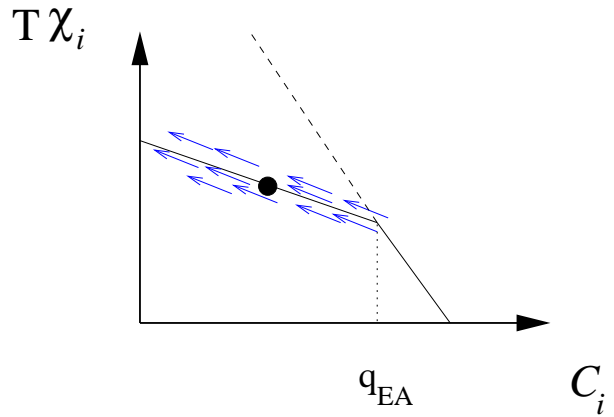,width=8cm}
\end{center}
\caption{Fluctuations in a system with a single correlation scale
below the Edwards-Anderson plateau. The constraint on the location 
of the pairs $(C_i,T\chi_i)$ implies that there are no fluctuations 
of the local Edwards-Anderson parameter and that all pairs are 
concentrated along the global straight line with little dispersion 
perpendicular to it. The arrows have the direction of the 
second slope in the global $\tilde \chi(C)$ curve due to 
the result in Eq.~(\ref{alleq}).
As in previous figures, 
the dashed line continues the {\sc fdt} line and
the black dot indicates the location of the 
values for the susceptibility and correlation averaged over the 
distribution.}
\label{fig:chiCnos1step}
\end{figure}

\begin{figure}
\begin{center}
\psfig{file=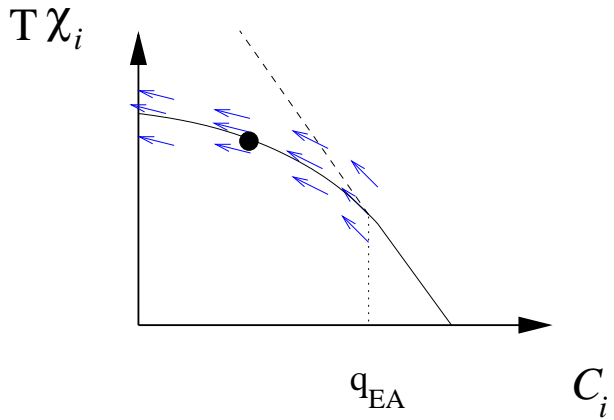,width=8cm}
\end{center}
\caption{This is the possibility expressed in Eqs.~(\ref{globalconstrain}) and
(\ref{localTidep}) for a system with a sequence of correlation scales
below the global $q_{\sc ea}$.  All point positions are constrained to
be near  the global $\tilde\chi(C)$ curve for $C < q_{\sc ea}$
[Eq.~(\ref{globalconstrain})], and their velocities are forced to be
parallel to the global $\tilde\chi(C)$ curve for $C < q_{\sc ea}$
[Eq.~(\ref{localTidep})].  The velocities fluctuate from site to site,
but since they are determined by the global curve, they are identical
for all sites with the same value of the local correlation $C_i$. The
dashed line and the black dot are as in previous figures.}
\label{fig:chiCnos}
\end{figure}

This behavior is sketched in Figs~\ref{fig:chiCnos1step} and 
\ref{fig:chiCnos} for a system with two global correlation-scales
and a system with a sequence of global correlation-scales, respectively.
In~\cite{paper2} we showed that the coarse-grained two-time quantities
in the $3d$ {\sc ea} model behave as in Fig.~\ref{fig:chiCnos}. We expect
that the coarse-grained two-time correlators of structural glasses 
will show a behavior as in Fig.~\ref{fig:chiCnos1step}.


\section{Constraints on the local fluctuations}
\label{sigma}

We recently studied the symmetry properties of the dynamic action for
the aging dynamics of finite, $d$-dimensional
spin-glasses~\cite{paper1}. Using the Schwinger-Keldysh functional
formalism we derived a disorder-averaged dynamic generating
functional that is a path integral over local two-time functions,
$Q^{\alpha\beta}_i(t,t_w)$, with $\alpha,\beta=0,1$ labelling the
Keldysh components (see details below).  This generating functional
becomes the classical one, derived in Ref.~\cite{Sozi}, when $\hbar\to 0$.
This treatment allowed us to discuss the classical and quantum
problems simultaneously. The aim of that work was to show that the
slow part of the action is a fixed point of a group of
time-reparametrizations (R$p$G) and to build upon this result an
argument to constrain how the local fluctuations should behave in time
and space. Here we recall this argument and summarize its main
consequences. Indeed, one of the main results arising from this 
analysis is that the coarse-grained two-time quantities 
should behave as in Fig.~\ref{fig:chiCnos1step} or  Fig.~\ref{fig:chiCnos}
if the global correlations decay on one or a sequence of scales,
respectively.

\subsection{Time-reparametrization transformations}
\label{time-rep}
 
The local two-time fields $Q_i^{\alpha\beta}(t,t_w)$ are related to
the fluctuating local physical two-time functions. For example, the
expectation values $\langle\langle
Q_i^{\alpha\beta}(t,t_w)\rangle\rangle$ (we use
$\langle\langle\cdots\rangle\rangle$ to denote expectation with respect
to the $Q$-action $S[Q]$) give the {\it disorder and thermal averaged} local
correlations and instantaneous responses:
\begin{eqnarray}
&&{\langle\langle Q_i^{00}}(t,t_w)\rangle\rangle 
= {\langle\langle Q_i^{K}}(t,t_w)\rangle\rangle
\nonumber\\
&&= \sum_j K_{ij}
[C_j^{na}(t,t_w)] = \sum_j K_{ij} [ \langle s_j(t) s_j(t_w)\rangle ]
\;, 
\label{localC1} \\
&&{\langle\langle Q_i^{01}}(t,t_w) \rangle\rangle = 
{\langle\langle Q_i^{R}}(t,t_w) \rangle\rangle
\nonumber\\
&&=\sum_j K_{ij} [R_j^{na}(t,t_w)] = \sum_j K_{ij} \left[ 
\left.
\frac{\langle \delta s_j(t)\rangle }{\delta \eta_j(t_w)} \right|_{h=0}
\right]
\; .
\label{localR1}
\end{eqnarray}
The connectivity matrix $K_{ij}$ contains the variance of the random
exchanges $J_{ij}$. For example, for the Edwards-Anderson model with
Gaussian distributed disorder, $K_{ij}=J$ for nearest neighbors
$i,j$, and zero otherwise. Notice that the $\langle\langle
Q_i^{\alpha\beta}\rangle\rangle$ are already an average over a small
region around site $i$ (because of the $K_{ij}$) even before coarse
graining. There are two other components of the $2\times 2$ matrix $Q_i$:
${Q_i^{10}}(t,t_w)={Q_i^{A}}(t,t_w)$ is related to the advanced
response and the remaining two-time function
${Q_i^{11}}(t,t_w)={Q_i^{D}}(t,t_w)$ is related to a correlator whose
average $\langle\langle {Q_i^{D}}(t,t_w)\rangle\rangle$ vanishes for all causal
problems. The two-time fields $Q_i^{\alpha\beta}(t,t_w)$ in the action
are fluctuating quantities.

We defined a time-reparametrization transformation $t\to h(t)$ that acts on
the fluctuating fields $Q_i^{\alpha\beta}(t,t_w)$ as~\cite{paper1,Kech}
\begin{eqnarray}
\hat Q_i^{\alpha\beta}(t,t_w) &\equiv& 
\left(\frac{dh(t)}{dt}\right)^{\Delta^{Q_i}_A} 
\left(\frac{dh(t_w)}{dt_w}\right)^{\Delta^{Q_i}_R}
Q_i^{\alpha\beta}(h(t),h(t_w))
\nonumber\\
\label{RpG}
\end{eqnarray}
where $\Delta_A^{Q_i}$ and $\Delta_R^{Q_i}$ are the advanced and
retarded scaling dimensions of the field $Q_i$ under the rescaling of
the time coordinates~\cite{Kech}. These transformations generalize the
well-known ones for the expected values of the global
$Q$'s~\cite{Dotsenko,Sompo,Cuku2,Frme,Sozi,Ginzburg,Ioffe,SompZ} to
the local and fluctuating case, with $h(t)$ a differentiable
function. This choice of transformation is motivated as follows.

Since the local fields $Q_i$ are related to the local correlation, linear
response and causality breaking correlator, it is natural to associate with
them the same scaling dimensions that correspond to the global counterparts
of these two-point functions. The appearance of an (approximate)
time-reparametrization invariance of the slow part of the relaxation was
noticed long ago in the asymptotic solution of the dynamics of mean-field
models~\cite{Dotsenko,Sompo,Cuku2,Frme,Sozi,Ginzburg,Ioffe,SompZ}. Indeed,
when one studies the effective dynamic equations for the slow decay, one
drops the time-derivatives and approximates the time-integrals.
These approximations can be justified because the neglected terms
are irrelevant in the R$p$G sense~\cite{Kech}. After applying these
approximations the resulting equations become invariant under
time-reparametrizations that transform the global correlator and linear
response according to
\begin{eqnarray}
\label{eq:Crpgtransf}
\hat C(t_1,t_2) &=& C(h(t),h(t_w)) \;,
\\
\label{eq:chirpgtransf}
\hat R(t,t_w) &=& \frac{dh(t_w)}{dt_w} \, R(h(t),h(t_w))\theta(t-t_w) 
\; ,
\end{eqnarray}
for any differentiable and monotonic function $h(t)$~\cite{clarify0}.
Note that once one proposes this transformation of the correlator, the
transformation of the linear response is forced to take the form in
Eq.~(\ref{eq:chirpgtransf}) if one wishes to preserve their link via
the {\sc fdt}.

Extending these definitions to the local and fluctuating
fields $Q_i^{\alpha\beta}$ we propose:
\begin{eqnarray}
(\Delta^{Q_i^K}_A,\Delta^{Q_i^K}_R) &=& (0,0)
\; ,
\\
(\Delta^{Q_i^R}_A,\Delta^{Q_i^R}_R) &=& (0,1)
\; ,
\\
(\Delta^{Q_i^A}_A,\Delta^{Q_i^A}_R) &=&(1,0)
\; ,
 \\
(\Delta^{Q_i^D}_A,\Delta^{Q_i^D}_R) &=& (1,1)
\; .
\end{eqnarray}
This explains the choice of indices $0,1$ for the Schwinger-Keldysh
components, which conveniently label both the matrix components and
scaling dimensions at the same time.

Note that  all sites are transformed in the same way 
under the reparametrization of time just defined.
This is a global transformation that leaves
invariant any local {\sc fdr} of the form 
\begin{equation}
\int_{t_w}^t dt' Q^{01}_i(t,t') 
=f(Q^{00}_i(t,t_w))
\; ,
\end{equation}
{\it cfr.} Eq.~(\ref{RpG}).  This relation is a ``constant of
motion'' with respect to this symmetry. Explicitly:
\begin{eqnarray}
\int_{t_w}^t dt' \hat Q^{01}_i(t,t') 
&=&
\int_{t_w}^t dt'  
\;\left(\frac{dh(t')}{dt'}\right)\;
Q^{01}_i(t,t') 
\nonumber\\
&=&
\int_{h_w}^h dh' Q^{01}_i(h,h') 
=f(Q^{00}_i(h,h_w))
\nonumber\\
&=&
f(\hat Q^{00}_i(t,t_w))
\; .
\end{eqnarray}

These transformations imply that the physical noise and disorder 
averaged saddle-point 
values transform as expected,
\begin{eqnarray}
[C^{na}_i(t,t_w)] &\to& [C^{na}_i(h(t),h(t_w))]
\; , 
\end{eqnarray}
\begin{eqnarray}
[\chi^{na}_i(t,t_w) ]
&\to& 
[\chi^{na}_i(h(t),h(t_w))]
\; ,
\end{eqnarray}
and the {\sc fdr} in Eq.~(\ref{localchiiCi}) is respected.

\subsection{Invariance of the action}

In Ref.~\cite{paper1} we studied the symmetries of the action
$S[Q]$. We showed that the long-time limit of the effective
Landau-Ginzburg aging action for the two-point functions
$Q_i^{\alpha\beta}(t,t_w)$ is a fixed point of the group of time
reparametrizations (R$p$G). By this we mean that, after separating the
field $Q_i^{\alpha\beta}(t,t_w)= {Q_i^{\alpha\beta}}_{\sc
fast}(t,t_w)+ {Q_i^{\alpha\beta}}_{\sc slow}(t,t_w)$, and then
integrating out the fast part of the fields, all terms in the
effective slow action are invariant under a reparametrization of time,
$t\to h(t)$, that transforms the fields as in Eq.~(\ref{RpG}). In
deriving the effective action for the slow contribution
${Q_i^{\alpha\beta}}_{\sc slow}(t,t_w)$, we assumed that there is a
{\em local} separation of time scales. The only other ingredient in
the proof was that the system must be causal.

The approach to the fixed point is asymptotic, and there will be
corrections to scaling at finite times. In particular, the kinetic
contribution to the effective action is irrelevant at long
times. However, irrelevant as it is at long times, this term acts as
a (time-decaying) symmetry breaking field that selects a particular
reparametrization.

The importance of what we have shown is that
it holds for infinite and short-range models alike, and at the level of
the action, not just the equations of motion. Moreover, it suggests an
approach to study spatial fluctuations of aging dynamics, as we
discuss below.

\subsection{Implications of R$p$G invariance -- connection with a sigma model}

In view of this approximate 
(asymptotic) symmetry we constructed an
argument that allowed us to predict how the local fluctuations of the
disordered averaged theory should behave. 
In this section we explain in more detail the arguments sketched in 
\cite{paper1,paper2}.

\subsubsection{Parallel with the $O(N)$ model}

To better explain the argument, it is useful to explore an analogy
with the static, coarse-grained, $O(N)$ theory of magnets in $d$
dimensions:
\begin{equation}
{\cal H} = \int d^d r \left[ (\vec \nabla \cdot \vec m(\vec r))^2 
+ \vec H \cdot \vec m(\vec r) + V(\vec m)
\right] 
\; ,
\label{hamilm}
\end{equation}
where $\vec m(\vec r)$ is a {\it continuous} $N$-dimensional variable,
$\vec m(\vec r) =(m^1(\vec r), m^2(\vec r), \dots, m^N(\vec r))$, that
represents the local magnetization. $V(\vec m)$ is a potential energy
with the form of a Mexican hat. $\vec H$ is an external magnetic
field.  A particular case of this model is the well-known $3d$
Heisenberg ferromagnet obtained when $N=3$ and $d=3$.

The parallel between the two models is as follows.

1. {\it Fields} - 
The two-time fields $Q_i^{\alpha\beta}(t,t_w)$, once coarse-grained
over a volume $V$, play the role of the static magnetization
$\vec m(\vec r)$.

2. {\it Symmetry} - 
When $\vec H=0$ the energy function of the magnetic problem is invariant
under a global rotation of the magnetization, $m^a(\vec r) = R^{ab}
m^b(\vec r)$ ($R^{ab}(\vec r)=R^{ab}$, for all $\vec r$). The
potential $V(\vec m)$ has a zero mode along the bottom of the Mexican
hat potential.

In the dynamic problem, for longer and longer waiting-times, the
symmetry breaking terms become less and less important and the action
progressively acquires a global symmetry (a zero mode develops).

3. {\it Spontaneous symmetry breaking} - 
In the absence of a pinning field, the
ferromagnetic model spontaneously chooses a direction of the vector
$\vec m$ everywhere in real space, $\vec m(\vec r) =\vec m_0$ 
in the broken symmetry phase.

In the R$p$G invariant asymptotic regime, the minima of the dynamic
action satisfy the global reparametrization symmetry. A given
direction in the minima manifold is described by one uniform
time-reparametrization everywhere in space, $h(\vec r,t)=h(t)$,

4. {\it Explicit symmetry breaking} - 
A non-zero magnetic field $\vec H$ breaks the symmetry explicitly by
tilting the Mexican hat potential. It forces the magnetization, $\vec
m$, to point in its direction in the $N$ dimensional space everywhere
in the real $d$-dimensional space.  

R$p$G irrelevant terms, which vanish asymptotically, play the role of
(time-decaying) symmetry breaking fields that select a particular
time-reparametrization. The particular scaling function $h(t)$
chosen by the system is determined by matching the fast and the slow
dynamics. It depends on several details -- the existence of external
forcing, the nature of the microscopic interactions, etc. In other
words, the fast modes which are absent in the slow dynamics act as
symmetry breaking fields for the slow modes.

5. {\it Fluctuations} - 
These correspond to smooth variations in the magnetization as a
function of position, $\vec m(\vec r) =\vec m_0+\delta \vec m(\vec
r)$. There are two types of fluctuations, longitudinal and
transverse. The former change the length of the magnetization vector,
$|\vec m(\vec r)|$; these move off the saddle point manifold of
potential minima, and are thus massive excitations. The latter change
direction only, $\vec m_0 \cdot \delta \vec m(\vec r)=0$; these remain
in the potential minima manifold, and are massless excitations. The
transverse fluctuations, which correspond to smoothly spatially
varying rotations of the magnetization vector, are therefore the most
energetically favored. These are the spin-waves or Goldstone modes in
the $O(N)$ model.

The equivalent of the transverse modes for the dynamic $Q$-theory are
smooth spatially-varying time reparametrizations, $h(\vec
r,t)=h(t)+\delta h(\vec r,t) $. Uniform or global reparametrization is
the symmetry of the model; the smooth spatially fluctuating $\delta
h(\vec r,t)$ are excitations that cost the lowest action, or the
Goldstone modes of the $Q$-action.

What controls the distance scale on which the fluctuations can vary?
The first term in the ferromagnetic model in Eq.~(\ref{hamilm})
restricts the magnitude of the variation of the direction of $\vec m$.
For the second model, one expects that a similar stiffness will be
generated once it is coarse-grained, making sharp variations in
$\delta h(\vec r,t)$ difficult to achieve.

The longitudinal fluctuations in the ferromagnet, those in which the
modulus of the magnetization vector changes, $m = m_0 +\delta m$, cost
more energy due to the Mexican hat potential and hence are less
favorable.  Similarly, in the dynamic problem the fluctuations in
which the ``external form'' [for example, imagine the function $f$ in
Eq.~(\ref{eq:localh}) varying with position] of the coarse-grained
two-time functions changes are less favorable.

\subsubsection{Spatial fluctuations of aging dynamics}

\begin{figure}
\psfig{file=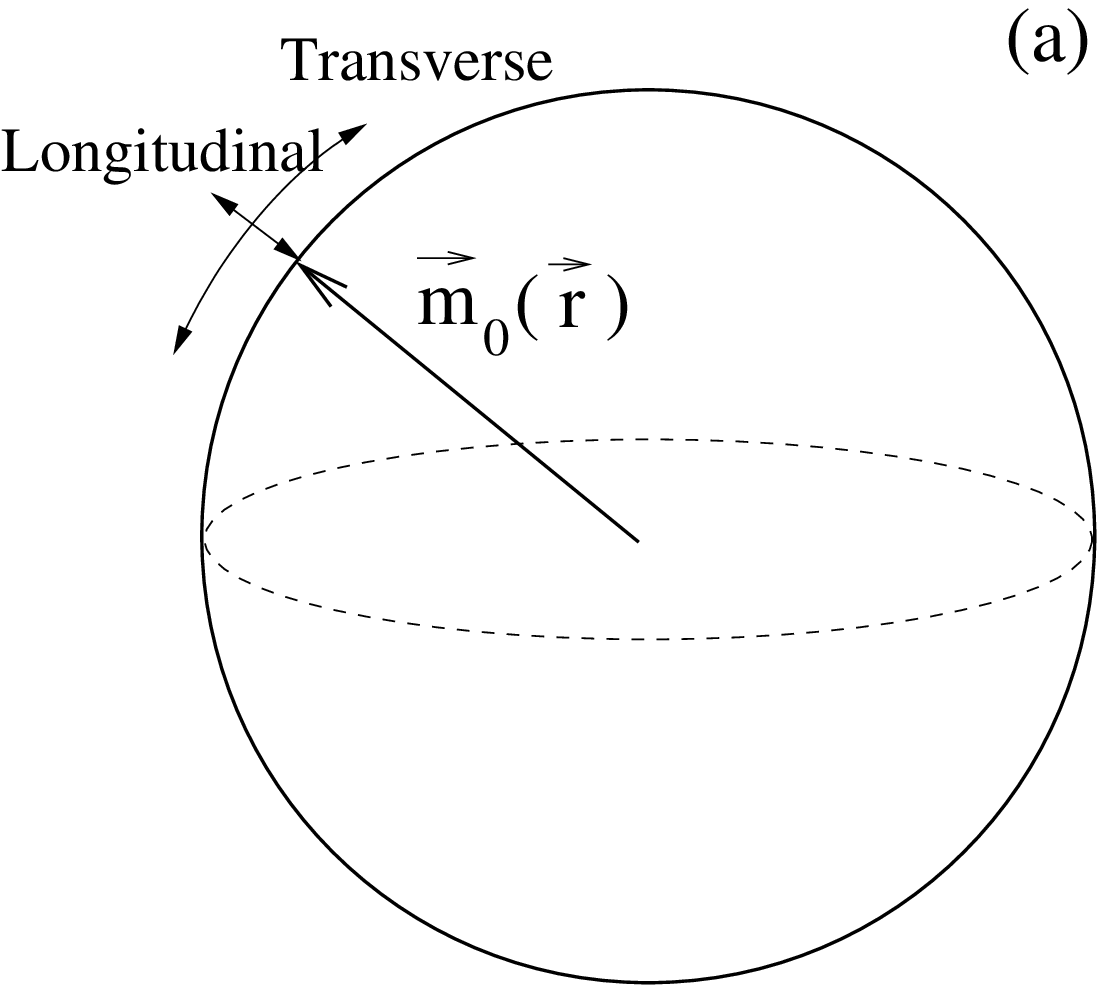,width=6cm}
\psfig{file=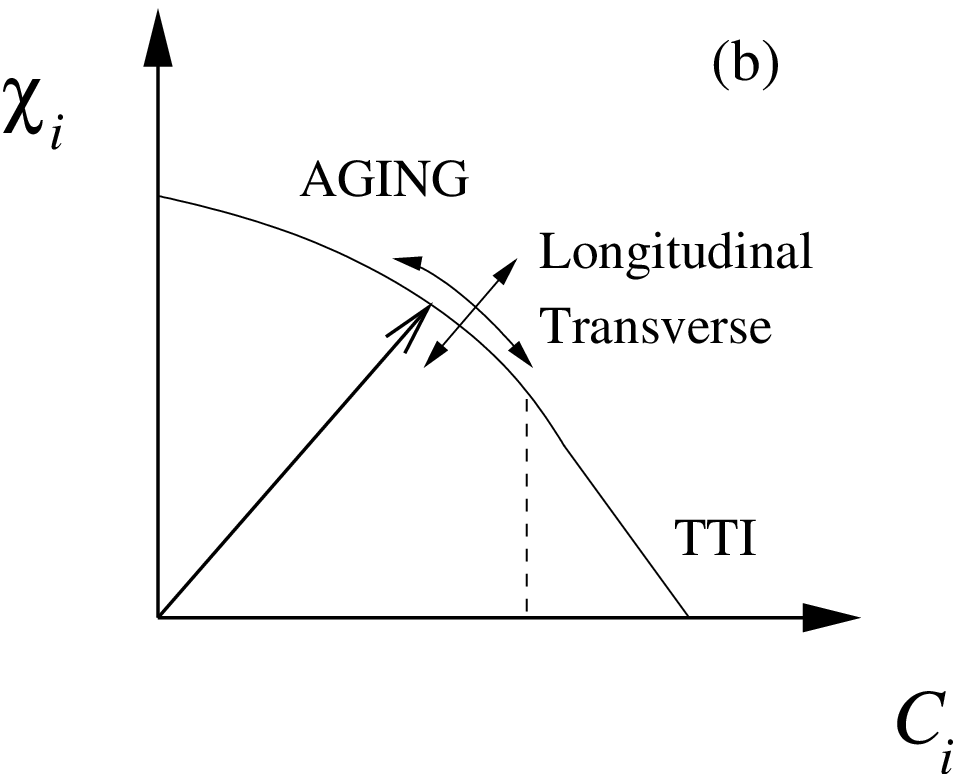,width=7cm}
\vspace{0.5cm}
\caption{Panel a): soft modes in the Heisenberg magnet. 
$N$-dimensional sphere with radius $|\vec m_o(\vec r)|$
centered at the spatial point $\vec r$. The longitudinal and transverse
directions for the fluctuations are indicated with arrows.
Panel b):
soft modes in the spatial fluctuations during aging.
The full line represents the global $\tilde \chi(C)$ curve in a 
model with a sequence of correlation scales below $q_{\sc ea}$. 
The vertical dashed line shows the value of $q_{\sc ea}$ separating the 
slow stationary decay ({\sc tti}) from the slow aging regime. The directions
for longitudinal and transverse fluctuations as explained in the text 
are indicated with arrows.}
\label{fig:modes}
\end{figure}

The penalties for longitudinal fluctuations, and the soft
transverse fluctuations corresponding to local reparametrizations $t\to
h(\vec r,t)$, led us to propose that the {\it coarse-grained local
correlations in the aging regime} scale as in Eq.~(\ref{localf}). In the
$C-T\chi$ plane, these soft modes correspond to displacements along the
global $\tilde\chi(C)$ curve; this is shown in
Fig.~\ref{fig:modes}. Displacements that move points off the $\tilde\chi(C)$
curve are the longitudinal modes, which do not correspond to smoothly
varying time reparametrizations.

A natural consequence of this is the prediction that the fluctuations
in the {\it coarse-grained local} {\sc fdr} should be such that the
distribution of points $\chi_i^{cg}(C^{cg}_i)$ follow the {\it global}
{\sc fdr}, $\tilde \chi(C)$, defined in Eq.~(\ref{fdr}). Thus,
Eq.~(\ref{localTidep}) should hold (see Fig.~\ref{fig:chiCnos}). Given
any pair of times $(t,t_w)$ such that the global correlation equals a
prescribed value, $C(t,t_w)={\tt C}$, the local effective temperature
should be the same for all regions of space having the same value of
the local correlation $C_i(t,t_w)$.

We would like to remark that the above theoretical argument was
developed for the fluctuating fields $Q_i^{\alpha\beta}(t,t_w)$, not
their thermal averages; these correspond, instead, to the expectation
values $\langle\langle Q_i^{\alpha\beta}(t,t_w)\rangle\rangle$. The
average over the disorder is carried from the outset in the
field-theoretical approach, while fluctuations due to noise histories
are kept by writing the $Q$-theory using a path integral
formulation. How can one test the predictions from this disordered
average field theory against numerical simulations? The answer lies on
whether the distributions that one studies are self-averaging or
not. What we find is that, if we study coarse-grained local
correlations and integrated responses~\cite{paper2} through numerical
simulations, their distributions become independent of the disorder
realization for large enough coarse-graining cell sizes. This is an
indication that self-averaging holds for these non-equilibrium distributions, 
and in fact the coarse-graining procedure is somehow averaging
over disorder. This is perhaps another reason to work with
coarse-grained quantities, as opposed to noise averaged quantities.

What can we say about the behavior of noise-averaged (and neither
disorder averaged nor coarse-grained) local two-time functions? We
argue that this procedure corresponds to ``fingerprinting'' the
particular disorder realization of the system. By a ``fingerprint'' of
disorder we mean that noise-averaged correlations and integrated
responses at the different sites purely reflect their disorder
environment, since there is no other source of fluctuations left in
the problem. Of course, further averaging over the disorder would
erase the ``fingerprint'', and simply give the global result. For this
reason, coarse-graining also erases the ``fingerprint'' for a
self-averaging system. The noise-averaging approach should correspond
to the scaling in Eq.~(\ref{localf}), and fluctuations are then not
constrained to follow the global $\tilde\chi(C)$ curve, as in
Fig.~\ref{fig:chiCnos}.

\subsection{Random surface action -- effective action for the 
fluctuations}
\label{sec:randomsurfaceaction}

In the spirit of the usual approach in deriving a coarse-grained
effective action for the relevant fields in a problem~\cite{book}, we search
for the ``minimal'' effective action describing the aging dynamics of the
system in terms of coarse-grained fields $Q^{\alpha\beta}(\vec r; t,t_w)$.
More precisely, in Sec.~\ref{time-rep} we mentioned that the dynamic
generating function is expressed in terms of local two-time fields,
$Q_i^{\alpha\beta}(t,t_w)$, that are coarse-grained over a cube involving
the first neighbors of the site $i$. Since we are interested in postulating
an effective action for the soft fluctuations we shall
consider a further spatial coarse-graining of these fields and work with
$Q^{\alpha\beta}(\vec r,t,t_w)$ that represents a smooth two-time field at the
position $\vec r$ in real space.

In a coarse grained theory, one expects gradient terms to be present
in the effective action. Their effect is 
to penalize rapid variations (in time and space) of the coarse grained fields. 
These terms 
play the role of the term  $\int d^d r \; 
(\vec \nabla \cdot \vec m(\vec r))^2$  in the 
energy Eq.~(\ref{hamilm}) for the Heisenberg ferromagnet. 

As we discussed above, global R$p$G invariance is, in the limit of
long times, a symmetry of the total action and we should preserve
it when constructing the effective action for the fields
$Q_i^{\alpha\beta}(\vec r; t,t_w)$. Thus, this symmetry 
poses strong restrictions on the form of allowed gradient terms. 

In general, the effective action takes the form:
\begin{equation}
S[Q]=S_{\rm grad}[Q]+S_{\rm local}[Q]
\; ,
\end{equation}
with
$S_{\rm local}[Q]$ an R$p$G invariant action comprised solely of local terms
and $S_{\rm grad}[Q]$ including the gradient (non-local)
dependence. Let us propose, for the latter,
\begin{eqnarray}
S_{\rm grad}[Q]&=&
-\int d^d r 
\int_0^\infty \!\!\!\! dt_1 
\int_0^\infty \!\!\!\! dt_2 
\;
\left(\nabla \partial_{t_1} Q^{00}(\vec r; t_1,t_2)\right)
\nonumber\\
&&
\;\;\; \times 
\left(\nabla \partial_{t_2} Q^{00}(\vec r; t_1,t_2)\right)
\; .
\label{eq:rf1}
\end{eqnarray}
 Notice that a global R$p$G transformation $t\to h(t)$, 
$Q^{00}(\vec r; t_1,t_2)\to \hat Q^{00}(\vec r; t_1,t_2)=
Q^{00}(\vec r; h(t_1),h(t_2))$ leaves $S_{\rm
grad}[Q]$ invariant as well, as can be explicitly checked:
\begin{eqnarray}
S_{\rm grad}[\hat Q]&=&
-\int d^d r\;
\int_0^\infty \!\!\!\! dt_1 \int_0^\infty \!\!\!\! dt_2
\;
\left(\nabla \partial_{t_1} \hat Q^{00}(\vec r; t_1,t_2)\right)
\nonumber\\
&&
\;\;\;\;
\times 
\left(\nabla \partial_{t_2} \hat Q^{00}(\vec r; t_1,t_2)\right)
\nonumber\\
&=&
-\int d^d r\;\int_0^\infty \!\!\!\! dh_1 \int_0^\infty \!\!\!\! dh_2
\;
\left(\nabla \partial_{h_1} Q^{00}(\vec r; h_1,h_2)\right)
\nonumber\\
&&
\;\;\;\;
\times 
\left(\nabla \partial_{h_2} Q^{00}(\vec r; h_1,h_2)\right)
= S_{\rm grad}[Q]
\; .
\label{eq:rf2}
\end{eqnarray}

Before proceeding, we would like to remind the reader that $Q^{00}$
and $C^{cg}$ are indeed related, so what we discuss below applies to
the spatially-varying coarse-grained correlation $C^{cg}(\vec r;
t_1,t_2)$. The expectation value $\langle\langle Q^{00}(\vec r;
t_1,t_2)\rangle\rangle$ is related to the noise (and disorder) average
of $C^{cg}(\vec r; t_1,t_2)$ [see Eq.~(\ref{localC1})]. Similarly, one can
show a relation between all $n$-moments $\langle\langle Q^{00}(\vec
r_1; t_1,t_2)\cdots Q^{00}(\vec r_n; t_1,t_2)\rangle\rangle$ and the
noise (and disorder) average of $C$'s at $n$ points. Hence, the
fluctuations of these quantities are akin.

The reader might also note that there are
several other simple R$p$G invariant actions that we could have written down,
involving either $Q^{01}$ and $Q^{10}$, or $Q^{11}$ and $Q^{00}$. However, 
such terms vanish when evaluated on a saddle-point configuration 
that satisfies causality. Since we focus on the causal solution and 
its fluctuations, we do not consider such terms.
The gradient penalizing term in Eq.~(\ref{eq:rf1}) is chosen as one of
the simplest non-trivial terms that respects R$p$G invariance.

In analogy with the study of the 
$O(N)$ model and the derivation of an effective action describing the
spin waves, we start by identifying the uniform (in space) saddle point
configuration and we then consider small fluctuations around it. 

Since the saddle-point solution does not depend on the spatial
position,
$S_{\rm grad}[Q_{\sc sp}]=0$. Thus, $Q_{\sc sp}$ 
is completely determined by
$S_{\rm local}[Q]$ and its precise 
form depends on the details of the model. We
argue that one can still learn a great deal about the spatial
fluctuations by considering simplified, approximate saddle-point
solutions. We know from numerical simulations of the $3d$ {\sc ea}
model~\cite{Picco,BerthierBouchaud} that the global correlation scales
rather well as
\begin{equation}
C(t_1,t_2)\approx q_{\sc ea}\;f
\left(\frac{{\rm min}(h(t_1), h(t_2))}{{\rm max}(h(t_1),h(t_2))}\right)
\; ,
\end{equation}
with $h(t)$ and $f(x)$ two monotonic functions. 
$f(x)$ satisfies $f(1)=1$ and $f(0)=0$.
A very good scaling of the data is obtained using $h(t)= 
\exp(t^{1-\mu}/(1-\mu))$ with $\mu$ taking values that are slightly 
above $1$. The simpler scaling $h(t)=t$ is not perfect but it
provides a rather good approximation. It will be convenient to 
define a new time-dependent function $\phi$ by
\begin{equation}
h(t) = e^{\phi(t)} 
\; .
\end{equation}
The role of spatially varying rotations in the $O(N)$ model
is here played by the spatially varying time
reparametrizations that we write as 
\begin{equation}
h(t) \to h(\vec r,t)=e^{\phi(\vec r,t)}
\; .
\end{equation}
Thus, we express the saddle-point solution as
\begin{equation}
Q_{\sc sp}^{00}(t_1,t_2)\approx
q_{\sc ea}\;f\left(e^{-|\phi(t_1)-\phi(t_2)|}\right)
\; ,
\end{equation} 
and we parametrize the fluctuations around it with
\begin{equation}
Q^{00}(\vec r; t_1,t_2)\approx q_{\sc ea}\;
f\left(e^{-|\phi(\vec r,t_1)-\phi(\vec r,t_2)|}\right)
\; .
\label{eq:rs-q-and-phi}
\end{equation}

Let us now split $\phi(\vec r,t)=\tau[h(t)]+\delta\phi(\vec r,t)$, where
$\tau[h(t)]$ selects the global reparametrization, {\it i.e.} fixes the
``direction'' of the  
saddle-point solution, and $\delta\phi(\vec
r,t)$ controls the small fluctuations around it. Furthermore, it is
best to think of $\tau$ as the proper time variable and work with
$\varphi(\vec r,\tau)\equiv\delta\phi(\vec r,t(\tau))$. Changing
integration variables in $S_{\rm grad}[Q]$ from $t$ to $\tau$, and
expanding around small $\varphi$, we obtain [for the simplest case
$f(x)=x$]:
\begin{eqnarray}
S_{\rm grad}&\approx& q_{\sc ea}^2\int d^d r\;
\int_{-\infty}^\infty d\tau_1 \int_{-\infty}^\infty d\tau_2
\;
\left(\nabla \dot\varphi(\vec r,\tau_1)\right)\;
\nonumber\\
&&
\;\;\; \times 
\left(\nabla \dot\varphi(\vec r,\tau_2)\right)\;
\left[1-|\tau_1-\tau_2|\right]^2\;e^{-2|\tau_1-\tau_2|}
\nonumber\\
&&=\frac{1}{2}\;q_{\sc ea}^2\;\int d^d r\;\int_{-\infty}^\infty d\tau 
\;\left(\nabla \dot\varphi(\vec r,\tau)\right)^2
+
\cdots
\; ,
\label{eq:rf3}
\end{eqnarray}
where $\dot\varphi=\partial\varphi/\partial\tau$. In Eq.~(\ref{eq:rf3}) we
retained only the leading relevant term (we neglected terms with higher
$\tau$ derivatives). It is very simple to show that in the case of a general
function $f(x)$ the resulting action is identical to the above up to a
renormalized stiffness constant, namely:
\begin{equation}
S_{\rm grad}\approx
\frac{1}{2}\;\lambda_{f}\;
q_{\sc ea}^2\;\int d^d r\;\int_{-\infty}^\infty d\tau 
\;\left(\nabla \dot\varphi(\vec r,\tau)\right)^2
+
\cdots
\; ,
\label{eq:rf3a}
\end{equation}
with
\begin{equation}
\lambda_{f}=
4 \int_0^1 dx\;x
\;\left\{ \left[ 1 + \ln x \right] f'(x)+ x \ln x f''(x) \right\}^{2}
\label{eq:rf3b}
\end{equation}
a positive definite constant.  In particular, for a power law
$f(x)=x^\lambda$, the renormalization factor of the stiffness is simply
$\lambda_f=\lambda$.

\begin{figure}
\psfig{file=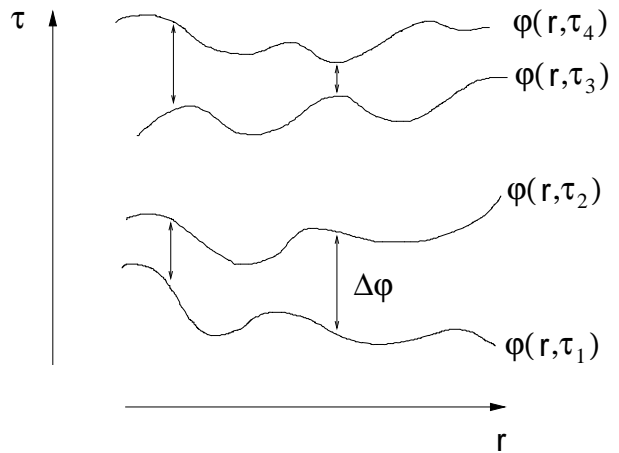,width=8cm}
\vspace{0.2cm}
\caption{Schematics of the height separation between surfaces $\varphi(\vec
r, \tau)$ at different proper times $\tau_1,\cdots,\tau_4$. Notice that height
differences for two different points $\vec r,\vec r'$ can increase or
decrease for different pairs of $\tau$'s. Such fluctuations can explain the
{\it sorpassi} we described previously.}
\label{fig:rand-surf}
\end{figure}

The $\varphi(\vec r,\tau)$ that parametrize the fluctuations can be
thought of as the height of a random surface.
Equation~(\ref{eq:rs-q-and-phi}) relates the fluctuations of the local
correlators with the reparametrization field $\phi$. In the case
$f(x)=x^\lambda$, this relation takes the simple form
\begin{eqnarray}
Q^{00}(\vec r; t_1,t_2) &=& Q_{\sc sp}^{00}(t_1,t_2)\;
e^{- \lambda |\varphi(\vec r,\tau(t_1))-\varphi(\vec r,\tau(t_2))|}
\; ,
\end{eqnarray}
so spatial fluctuations of the correlations $C^{cg}(\vec r; t_1,t_2)$
are related to the fluctuations of the height differences from two
$\varphi$ surfaces evaluated at two proper times $\tau_1$ and
$\tau_2$, as sketched in Fig.~\ref{fig:rand-surf}. It is simple to
understand how these height differences, which fluctuate as a function
of time, can very simply explain the {\it sorpassi} we discussed in
Sec.~\ref{subsubslow}.

What are the statistics of these height differences? Because $\dot\varphi$ is
a Gaussian surface, it is simple to show that
\begin{equation}
\langle \left[\varphi(\vec r,\tau_1)-\varphi(\vec r,\tau_2)\right]^2\rangle
=2G_d(\vec r,\vec r)\;|\tau_1-\tau_2|
\; ,
\label{eq:rf4}
\end{equation}
where $G_d(\vec r,\vec r')$ is the correlation function for a Gaussian random
surface in $d$-dimensions. When $\tau_1$ and $\tau_2$ get close, irrelevant
terms neglected in Eq.~(\ref{eq:rf3}) become important, and $|\tau_1-\tau_2|$
should be replaced by a short-time cut-off $\tau_c$. The details of the
crossover depend on those neglected terms, but one can capture its most
important features by approximately replacing $|\tau_1-\tau_2|\to \tau_c +
|\tau_1-\tau_2|$ in Eq.~(\ref{eq:rf4}).

The simple expression Eq.~(\ref{eq:rf4}) has some powerful
consequences. First, it implies that the statistical distribution of local
correlation functions $C(\vec r; t_1,t_2)$ is controlled by the proper time
difference $\tau_1-\tau_2$. Because $\tau=\ln h(t)$ follows from the
scaling of the global correlations of the {\sc ea} 
model~\cite{Picco,BerthierBouchaud}, 
we can conclude that the
full {\sc pdf} for the local correlations is also a function that scales with
the ratio $h(t)/h(t_w)$; this scaling is confirmed in our numerical
simulations in the $3d$ {\sc ea} model, 
as discussed in Sec.~\ref{sec:distribution_functions}, where 
for simplicity, we use $h(t)\approx t$. 

Second, it suggests that fluctuations of the
height differences are much more pronounced in $2d$ than in $3d$. $G_d(\vec r,
\vec r)$, the correlation function for a Gaussian random surface in
$d$-dimensions, goes to an ultraviolet cut-off dependent constant in $3d$, but
diverges logarithmically with the system size $L$ in $2d$. Hence, this may
provide a simple explanation why in $2d$ fluctuations destroy the global order
in spite of the existence of non-zero local correlations.

Additionally, Eq.~(\ref{eq:rf4}) allows us to obtain more detailed
predictions about the leading order behavior of the fluctuating
reparametrization. Within the present approximation, $\varphi(\vec
r,\tau_1)-\varphi(\vec r,\tau_2)$ is a Gaussian random variable with a
variance given by the square root of the r.h.s. of
Eq.~(\ref{eq:rf4}). Combining this with the physical information that
$ \tau(t) =\ln h(t)$ (where again for simplicity we can take
$h(t)\approx t$ for the scaling of the global functions), and noting
$\phi(\vec r,t)=\tau(t)+\delta\phi(\vec r,t)$, we can write the
following scaling form:
\begin{eqnarray}
\phi(\vec r,t_1) &-& \phi(\vec r,t_2) = \ln t_1 - \ln t_2 
\nonumber \\
\qquad \qquad
&& 
\!\!\!\!\!\!\!\!\!\!\!\!\!\!
+
\left[ G_d(\vec r,
\vec r) ( \tau_c + |\ln t_1 - \ln t_2 | ) \right]^{1/2} X_r(t_1,t_2)
\nonumber \\
&& 
\!\!\!\!\!\!\!\!\!
=  
\ln(t_1/t_2) + 
\left[ a + b |\ln(t_1/t_2)| \right]^{\alpha} X_r(t_1,t_2)
\; , 
\label{eq:phi_scaling_form}
\end{eqnarray}
where, to Gaussian level $\alpha = 1/2$, and $X_r(t_1,t_2)$ is a random
Gaussian variable of unit variance and spatially correlated. Higher order
corrections could in principle modify this prediction significantly. It
turns out that simulational results~\cite{paper2} are consistent with
Eq.~(\ref{eq:phi_scaling_form}), although with $\alpha \neq 1/2$ and with a
non-Gaussian distribution for $X_r(t_1,t_2)$.

\section{Scaling of local correlations}
\label{scaling-local-corr}

After having introduced the local two-time 
quantities in Sec.~\ref{sec:definitions} and having discussed 
several scenarios for their behavior in Sec.~\ref{sigma},
in this section we study the dynamic behavior of the 
coarse-grained and noise-averaged local correlations
of the $3d$ {\sc ea} model using numerical simulations.

\subsection{Global correlation}

\begin{figure}[h]
\psfig{file=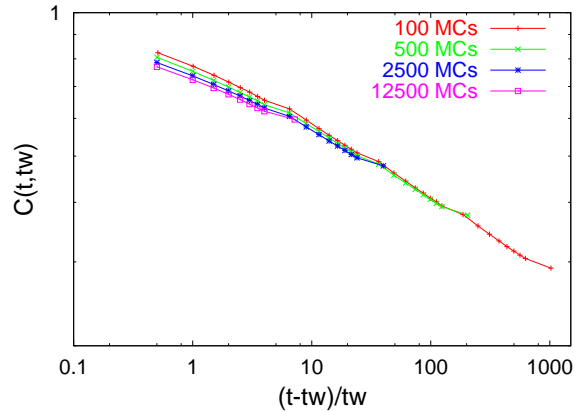,width=8cm}
\vspace{0.5cm}
\caption{Decay of the global correlation $C(t,t_w)$ in the 
$3d$ {\sc ea} model as
a function of $(t-t_w)/t_w$. The waiting-times are indicated 
in the key. $T=0.7$ and $L=32$.}  
\label{global-corr}
\end{figure}

We recall that, for the waiting and total times 
we use,  the aging decay of the global correlation [Eq.~(\ref{globalC})]
is rather well described with a simple $t/t_w$ scaling~\cite{Picco}.
Indeed, even without subtracting the first approach to the 
plateau at $q_{\sc ea}$ (that can occur with a very slow 
power law~\cite{Eric}) the scaling is quite acceptable. To illustrate this 
point, we show the global correlation against 
$(t-t_w)/t_w$ in Fig.~\ref{global-corr} for 
a system evolved with a single thermal history and using relatively
short waiting-times.

We stress that we do not claim that this scaling holds asymptotically.
It simply {\it approximately} describes the data for these short
waiting-times and up to the maximum waiting-time reachable in
simulations, which is of the order of $10^6$ Monte Carlo steps ({\sc
mc}s). A better description of the numerical data is obtained if one
uses $C(t,t_w) \approx f(h(t)/h(t_w))$ with
$h(t)=\exp(t^{1-\mu}/(1-\mu))$ and $\mu$ slightly larger than $1$ (see
Refs.~\cite{Picco,BerthierBouchaud} for a precise analysis where
longer waiting-times have been used and the approach to the plateau
has been taken into account).  In what follows, to keep the analysis of
the data as simple as possible, we adopt the simpler power law
expression for $h(t)$.

\subsection{Relation to random surfaces}
\label{random-surfaces}

The evaluation of the local correlations on the $d$-dimensional real-space
substrate allowed us to draw a parallel between the evolution of the local
coarse-grained correlations and the dynamics of a $d$-dimensional random
surface~\cite{paper2}, see Sec.~\ref{sec:randomsurfaceaction}. 
At any pair of times $(t,t_w)$ the random surface
fluctuates about the global value $C(t,t_w)$ and it is constrained to do so
between $-1$ and $1$ since $C_i \in [-1,1]$. Within our R$p$G invariant
theory, there is a one-to-one relationship between this random surface for
the coarse-grained correlator $C^{cg}(\vec r; t,t_w)$ and the random surfaces
$\phi(\vec r, t)$ (and $\phi(\vec r, t_w)$) discussed in
Sec.~\ref{sec:randomsurfaceaction}, as evident from the relation
Eq.~(\ref{eq:rs-q-and-phi}). The parallel between the evolution of the local
correlations and the dynamics of a $d$-dimensional random surface can also be
extended to the noise-averaged correlations (although we present no
analytical theory for this case).

\begin{figure}
\begin{center}
\vspace{0.75cm}
\psfig{file=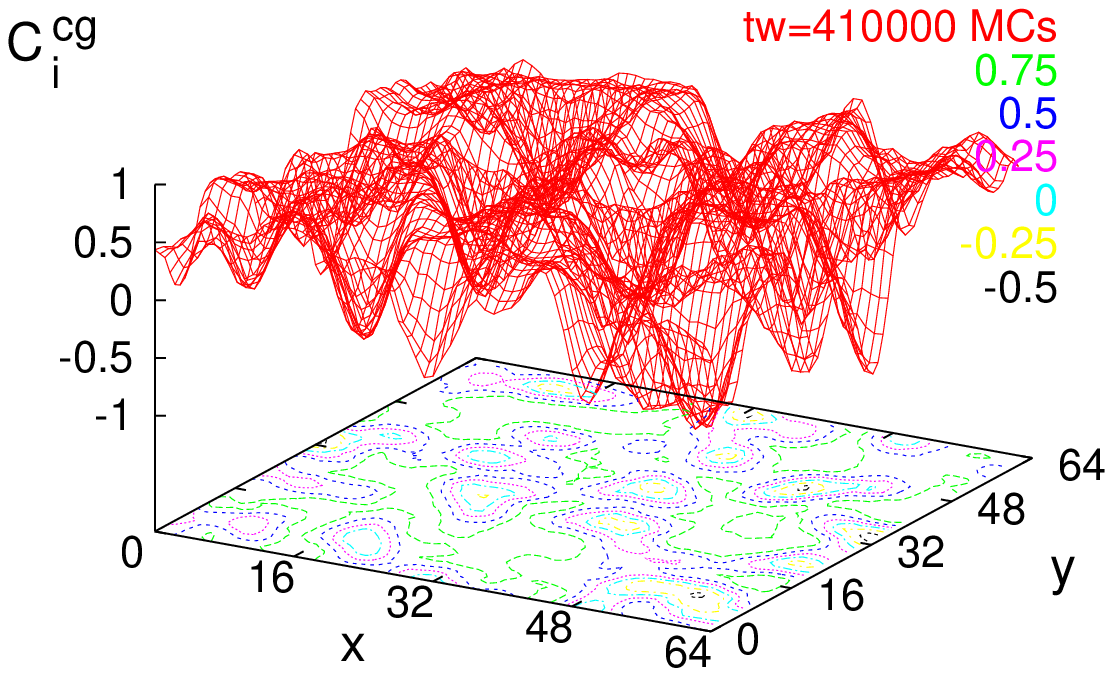,width=8cm}
\end{center}
\caption{Surface of local coarse-grained correlations
on a $2d$ cut of the $3d$ real space. $T=0.8$, $L=64$, $M=3$.
The local correlations are evaluated at $t_w=4.1\times 10^5$ and $t=2.8\times
10^6$, {\it i.e.} $t/t_w \sim 6.8$.}
\label{surface-coarse}
\begin{center}
\psfig{file=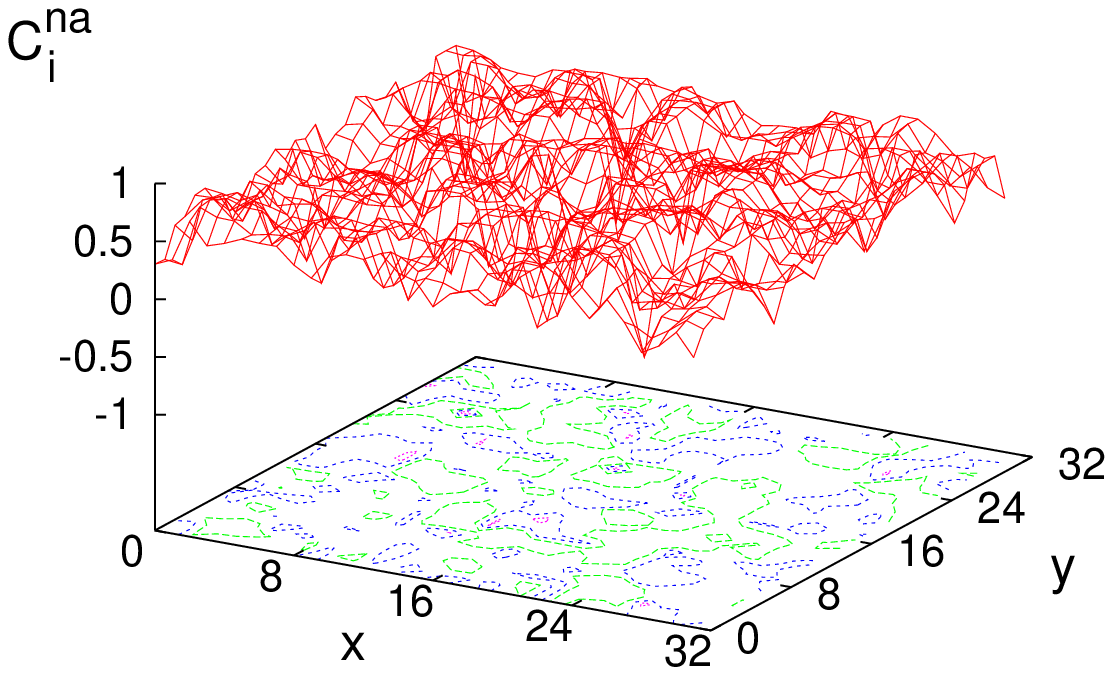,width=9cm}
\end{center}
\caption{Surface of local noise-averaged correlations
on a $2d$ cut of the $3d$ real space. $T=0.7$, $L=32$, $10^3$ noises.
The local correlations are evaluated at $t_w=1.6 \times 10^4$ and 
$t=4.8\times 10^4$, {\it i.e.} $t/t_w=3$. The fluctuations are reduced
with respect to the ones in Fig.~\ref{surface-coarse} since $T$ is lower and
$t/t_w$ is smaller here. Note that this surface fluctuates 
between $0$ and $1$ since the noise-averaging eliminates
negative values.}
\label{surface-noise}
\end{figure}

The statistical and dynamical properties of the surfaces in each case are not
necessarily the same.  In Figs.~\ref{surface-coarse} and \ref{surface-noise}
we show the values of the local coarse-grained and noise-averaged
correlations, respectively, on a $2d$ cut of the $3d$ real space. These
figures illustrate how the local correlations generate a surface with height
$C_i$ on each site $i$ of the $3d$ substrate. The statistical properties of
the $C_i$ (their distribution, geometric organization, etc.)  inform us about
the statistical properties of the random surface. Similarly, we can think
about the random surface generated by the coarse-grained and noise-averaged
local susceptibilities.  
In Sec.~\ref{geometric} we shall show the results of
the numerical analysis of the geometric properties of the
random surface of local correlations in the $2d$ and $3d$ {\sc ea} models.

\subsection{Coarse-graining volume}

\begin{figure}
\vspace{0.5cm}
\psfig{file=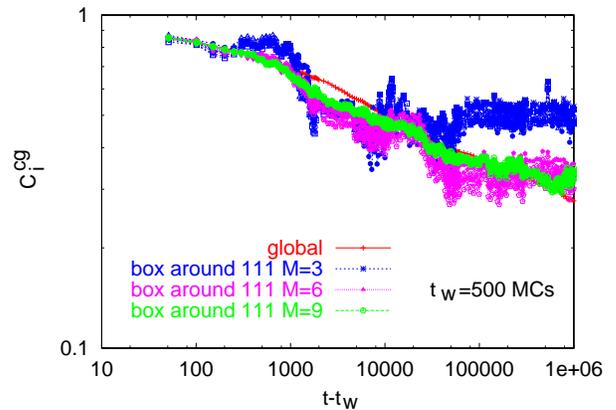,width=8cm}
\vspace{0.5cm}
\caption{The local correlation $C^{cg}_i(t,t_w)$ against $t-t_w$
for $t_w=500$ {\sc mc}s. $L=64$. The three groups of curves correspond to
coarse graining volumes $V=(2M+1)^3$ with $M=3,6,9$.
Different curves within each group correspond to the $7$ sites
in the cubic box centered on the site with coordinates $(1,1,1)$.
We use a variable coarse-graining time $\tau$ that starts at
$\tau=10$ {\sc mc}s and is multiplied by $5$ each time the total time
reaches $500\times 5^k$ with $k=1,\dots$
For comparison we also plot the global correlation $C$.}
\label{coarse1}
\end{figure}

Before analyzing the time dependence of the local correlations 
we briefly exhibit the dependence of the coarse-grained 
correlations on the coarse-graining volume.
In Fig.~\ref{coarse1} 
we plot $C^{cg}_i(t,t_w)$ against
the time-difference $t-t_w$ on seven
 sites around the site with
coordinates $(1,1,1)$, 
using three coarse-graining volumes $V=(2M+1)^3$ with $M=3,6,9$.
The waiting-time is $t_w=500$ {\sc mc}s.
 As expected, coarse-graining smoothens
the spatial variation of the local correlations and we see 
very little variation
between the local correlations on neighboring sites.
For $t_w=500$ {\sc mc}s, 
the curves for $M=3$ are rather noisy while 
those for $M=6$ and $M=9$ behave roughly in the same way.
For longer waiting-times  the curves for the
three coarse-graining volumes behave roughly in the same manner (not shown).
In what follows we shall typically use $M=1$, $M=3$ and $M=6$ that correspond 
to linear sizes $2M+1=3$, $2M+1=7$ and $2M+1=13$.
The first choice is the coarse-graining implicit in the analytic theory 
(see Eqs.~(\ref{localC1}) and (\ref{localR1}), 
except that here the central site is 
also included).
The two latter choices
are of the order of one fifth of the linear size of the system 
if $L=32$ and $L=64$, respectively. 

\subsection{Distribution functions}
\label{sec:distribution_functions}

What is the origin of the rather simple 
scaling of the global correlation for the times explored? 
Do all local correlations scale in the same way, and 
as $\approx t/t_w$, or is the global scaling the result of 
the combination of different behaviors on different
sites? 

\begin{figure}
\input{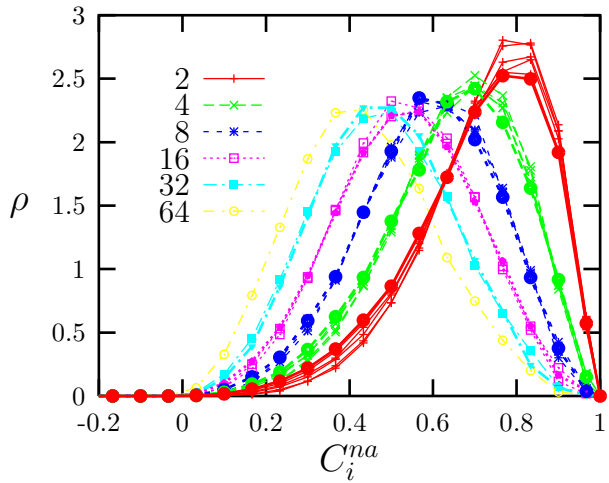}
\vspace{0.25cm}
\caption{The distribution of the noise-averaged local correlations
$C_i^{na}$ for several pairs of $t$ and $t_w$. The waiting-times are 
$t_w=10^3$ {\sc mc}s, $2\times 10^3$ {\sc mc}s, $4\times 10^3$ {\sc mc}s,
$ 8\times 10^3$ {\sc mc}s, $1.6 \times 10^4$ {\sc mc}s, 
and $3.2 \times 10^4$ {\sc mc}s, 
and the ratios $t/t_w$ are indicated in the key. 
We averaged over $10^3$ realizations of the
noise. $\tau_t=t/10$, $L=64$ and $T=0.7$.}
\label{distaverCi}
\end{figure}

In order to explore these questions we first study the 
time-evolution of the probability distribution function 
({\sc pdf}) of local correlations.
In Fig.~\ref{distaverCi} we show the distribution of $C_i^{na}$'s
for several pairs of $t$ and $t_w$ with ratios given in the key.
Only one disorder realization has been used and the curves 
are drawn with correlations that have been averaged over $10^3$
realizations of the noise. 
The full distributions scale approximately with $t/t_w$.
The peak moves towards smaller values of $C_i^{na}$ when the 
ratio increases ({\it cfr.} Fig.~\ref{distaverCr}) and the 
distribution gets slightly wider. 
The curves with wider lines correspond to the longest $t_w$.
For small $t/t_w$ a drift with increasing $t_w$ 
leading to a very mild decrease in the height of the peak
is visible in the figure. Note that when sufficiently
many noise realizations are averaged over, 
the distribution does not have any weight
on negative values of $C^{na}_i$. For even larger values of the ratio 
$t/t_w$ we expect to see a reverse trend in the sense that the 
distribution has to start getting squeezed around the value $C^{na}_i=0$.

\begin{figure}
\input{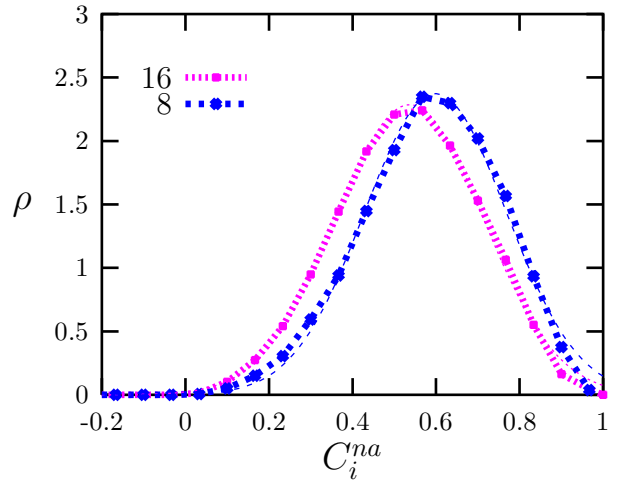}
\vspace{0.25cm}
\caption{The distribution of the noise-averaged local correlations
$C_i^{na}$ for two ratios $t/t_w$ given in the key, 
$t_w=3.2 \times 10^4$, and two Gaussian fits. 
Same parameters as in Fig.~\ref{distaverCi}. }
\label{Gaussian-fit}
\end{figure}

\begin{figure}
\psfig{file=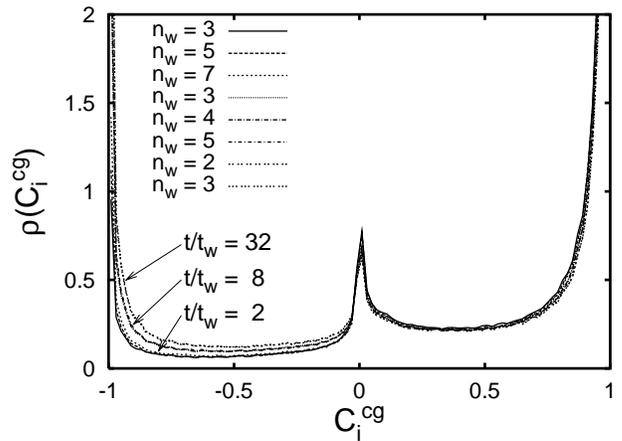,width=8.5cm}
\vspace{0.25cm}
\caption{The distribution of the coarse-grained local correlations
$C^{cg}_i$ for several pairs of $t$ and $t_w$.  $M=0$, [$2M+1=1
<\xi(t,t_w)$] $L=32$, $\tau=10^3$ {\sc mc}s and $T=0.8$.  The
parameters given in the key determine the waiting-times, $t_w=2^{n_w}
\times 10^4$ {\sc mc}s.}
\label{distaverCr}
\end{figure}

The central part of the distribution 
is described very well by a Gaussian distribution
for intermediate values of $t/t_w$ (before the 
{\sc pdf} starts to be squeezed on $C_i^{na}=0$).
Indeed, Fig.~\ref{Gaussian-fit} shows two Gaussian fits to the 
numerical {\sc pdf}s for the ratios $t/t_w=8,16$ and $t_w=3.2\times
10^4$. Note that since $C_i^{na} \in [0,1]$ the Gaussian fit cannot be 
perfect. In Sec.~\ref{random-surfaces} we mentioned the interpretation of 
the local correlations as generating a random surface on the $3d$ substrate.
The approximate Gaussian distribution of the $C_i^{na}$ implies 
that we can interpret the random surface of noise-averaged correlations 
as being approximately Gaussian. 

The distribution seems to become stable with respect to the number of
realizations of the noise, after a large enough number of such
realizations. More precisely, we do not see any noticeable variation
between the calculated distributions that are averaged over $500$ and
$1000$ thermal histories.  We expect this distribution to be
self-averaging, {\it i.e.}, independent of the particular realization
of the disorder for large enough systems. We have checked the
self-averaging property numerically, and also that, for large enough
values of $N$, the distribution becomes size independent.

\begin{figure}
\psfig{file=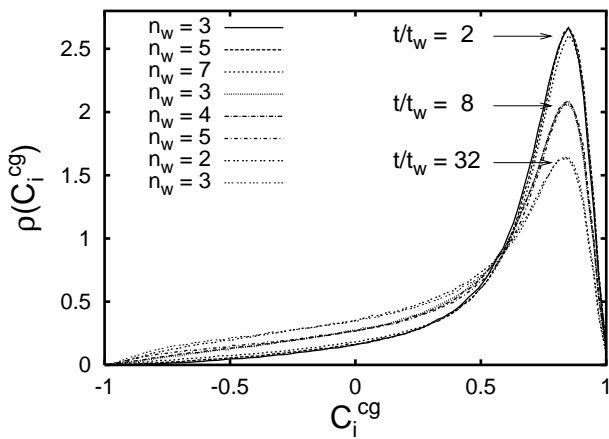,width=8.5cm}
\vspace{0.25cm}
\caption{The distribution of the coarse-grained local correlations
$C^{cg}_i$ for several pairs of $t$ and $t_w$. 
$M=1$ [$2M+1=3 \sim \xi(t,t_w)$], $L=32$, 
$\tau=10^3$ {\sc mc}s and $T=0.8$. The 
parameters in the key fix the waiting-times as in Fig.~\ref{distaverCr}.}
\label{distaverCr2}
\end{figure}

\begin{figure}
\psfig{file=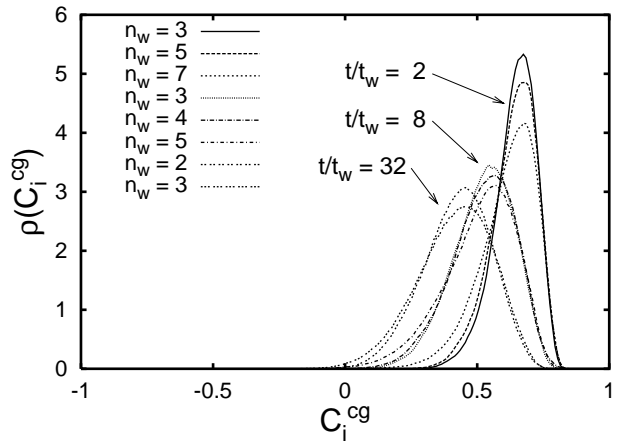,width=8.5cm}
\vspace{0.25cm}
\caption{The distribution of the coarse-grained local correlations
$C^{cg}_i$ for several pairs of $t$ and $t_w$. 
$M=6$ [$2M+1 \gg \xi(t,t_w)$], $L=32$, $\tau=10^3$ {\sc mc}s  and $T=0.8$.
The parameters in the key fix the waiting-times as in Fig.~\ref{distaverCr}.}
\label{distaverCr3}
\end{figure}

Figure~\ref{distaverCi} should be compared to 
Figs.~\ref{distaverCr}-\ref{distaverCr3}
in which the distribution of the coarse-grained 
local correlations, $C_i^{cg}$, is shown for several 
times and different values of the 
coarse-graining volume. 

In Fig.~\ref{distaverCr} the coarse-graining volume 
has a linear size $2M+1=1$. Thus, there is no spatial 
coarse-graining and the only reason why these curves are
not simple peaks at $-1$ and $1$ is that the coarse-graining
in time, done with $\tau_t=10^3$ {\sc mc}s 
for all times, slightly smooths the data.

In Fig.~\ref{distaverCr2} the coarse-graining volume 
has a linear size $2M+1=3$. This distance is slightly shorter
than the two-time dependent correlation length, $\xi(t,t_w)$, 
that we shall define and study 
in Sec.~\ref{correlation-length}. For this amount of coarse-graining
the {\sc pdf} has a nice scaling behavior.
The position of the peak is almost 
independent of the ratio $t/t_w$ in this case. 
Its height diminishes when the ratio $t/t_w$ 
increases and the tail at smaller values of $C_i^{cg}$ grows.
The scaling with $t/t_w$ is rather good in this case.

Finally, in Fig.~\ref{distaverCr3} the coarse-graining volume 
has a linear size $2M+1=13$ which is almost half the system size 
and much larger than the two-time dependent 
correlation  length $\xi(t,t_w)$ for all the waiting and 
total times studied. The width of the {\sc pdf}s has been considerably
reduced with respect to the previous case. In particular, the {\sc pdf}
does not have any weight on negative values, as opposed to 
what is shown in Fig.~\ref{distaverCr2}. It has also become quite
symmetric, centered at the global value which is also approximately the 
average value of this {\sc pdf}. Reasonably, the distributions 
drift towards smaller values of the correlations when 
the value of $t/t_w$ increases.
Moreover, the scaling with $t/t_w$ worsens with
too much coarse-graining, as is to be 
expected. Using such a large coarse-graining volume
one approaches the limit in which the distribution 
becomes a delta-function at the global value. The $t/t_w$ scaling 
is only an approximation to the true scaling, see Fig.~\ref{global-corr}. 
In order to improve the fit one should eliminate the contributions 
to the stationary decay but this is not easy to do at the level of
the full distribution.

It is clear that for a given disorder and thermal history
this distribution approaches a delta function on the 
time-dependent global value of the correlation 
when the coarse-graining volume reaches 
the size of the system. Even in this limit, 
for finite $N$ we still have sample-to-sample
and noise-to-noise fluctuations and hence a non-trivial 
distribution of these points.

To conclude, the noise-averaged {\sc pdf} and the coarse-grained {\sc
pdf} approximately scale with $t/t_w$ for  long waiting-times
when $2M+1 \sim \xi(t,t_w)$. However, even if for a given ratio $t/t_w$ the
mean of the distributions coincides (yielding the global value in both
cases), the form of the two {\sc pdf}s differs at equal $t/t_w$.  This
is most clearly seen by comparing, for instance, the curves for
$t/t_w=32$ in Figs. \ref{distaverCi} and \ref{distaverCr2}. The most
distinctive difference is given by the persistence of a peak in
$\rho(C^{cg})$ that is centered on a stable value (although its height
decreases with increasing $t/t_w$). Moreover, the {\sc pdf} of
noise-averaged correlations is approximately Gaussian for intermediate
$t/t_w$ values while the {\sc pdf} of coarse-grained two-time local
correlations is not when $2M+1 \sim \xi(t,t_w)$. For larger values of the 
linear size of the coarse-graining volume the {\sc pdf} looks more 
Gaussian, see Fig.~\ref{distaverCr3}, at least in its central part. 
We have observed the same features at other temperatures. 

We end this subsection by pointing out that the probability distributions for
noise-averaged and coarse-grained quantities can be interpreted as,
respectively, probabilities of averages and averages of probabilities. To see
this relation, consider a quantity that is noise averaged after being coarse
grained, and its probability density is $\rho(\langle C^{cg}_i \rangle)$. In
the limit when the coarse graining volume is a single spin ($M=0$), we
recover $\rho(C^{na}_i)$, using $C^{na}_i=\langle C_i \rangle$. Now, for a
large enough total system size, the probability distribution for
coarse-grained quantities $\rho(C^{cg}_i)$ becomes self-averaging,
$\rho(C^{cg}_i) \to \langle \rho(C^{cg}_i)\rangle$. Hence, the differences
between the noise-averaged and coarse-grained probabilities derive from the
fact that $\rho(\langle C^{cg}_i \rangle)\ne \langle \rho(C^{cg}_i)\rangle$.

\subsection{Scaling of noise-averaged local correlators}

We have just shown that the even if the probability distributions of 
noise-averaged and coarse-grained local correlations take different 
forms, they scale approximately 
as $t/t_w$. This does not mean, however, that each
site has a local correlation scaling with $t/t_w$.   
Equations~(\ref{localf}) and (\ref{eq:localh}) can now be put 
to the test by studying the decay in time of the individual local correlations 
$C_i^{na}$ and $C^{cg}_i$. 

\begin{figure}
\begin{center}
\psfig{file=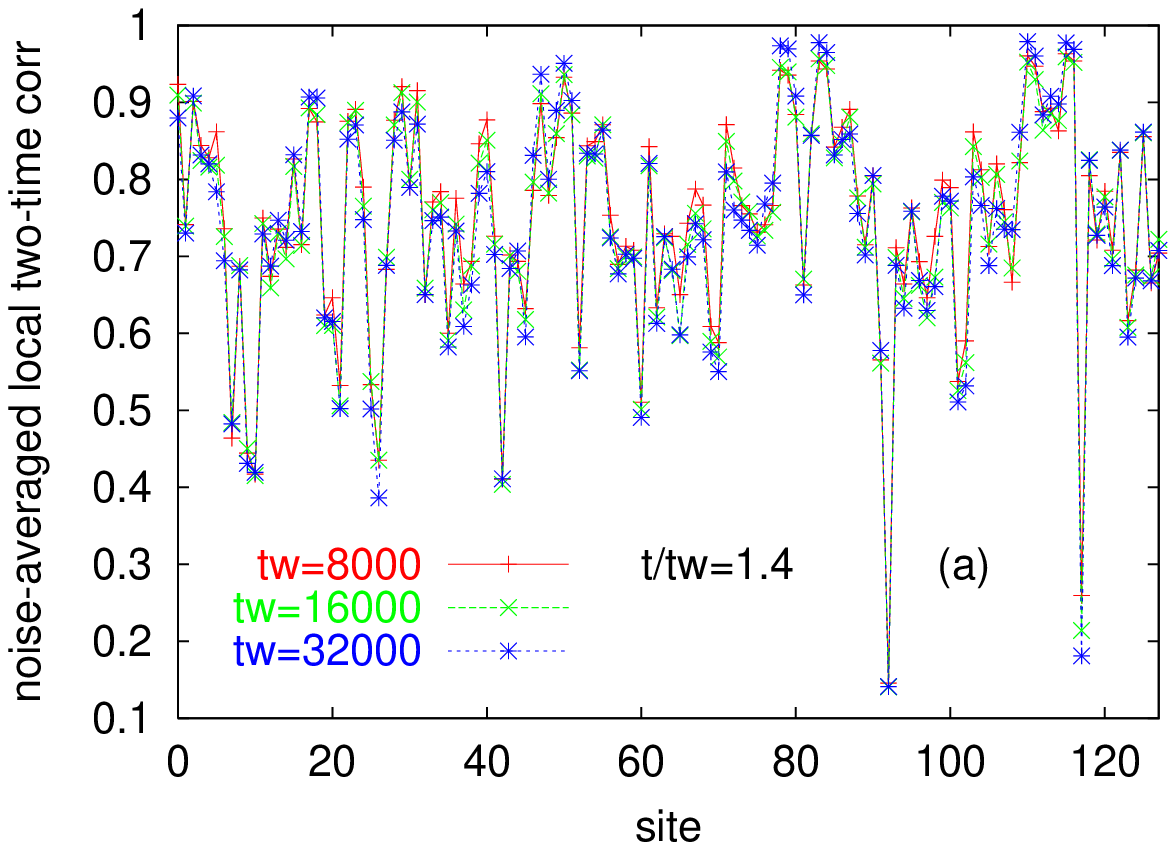,width=9cm}
\end{center}
\begin{center}
\psfig{file=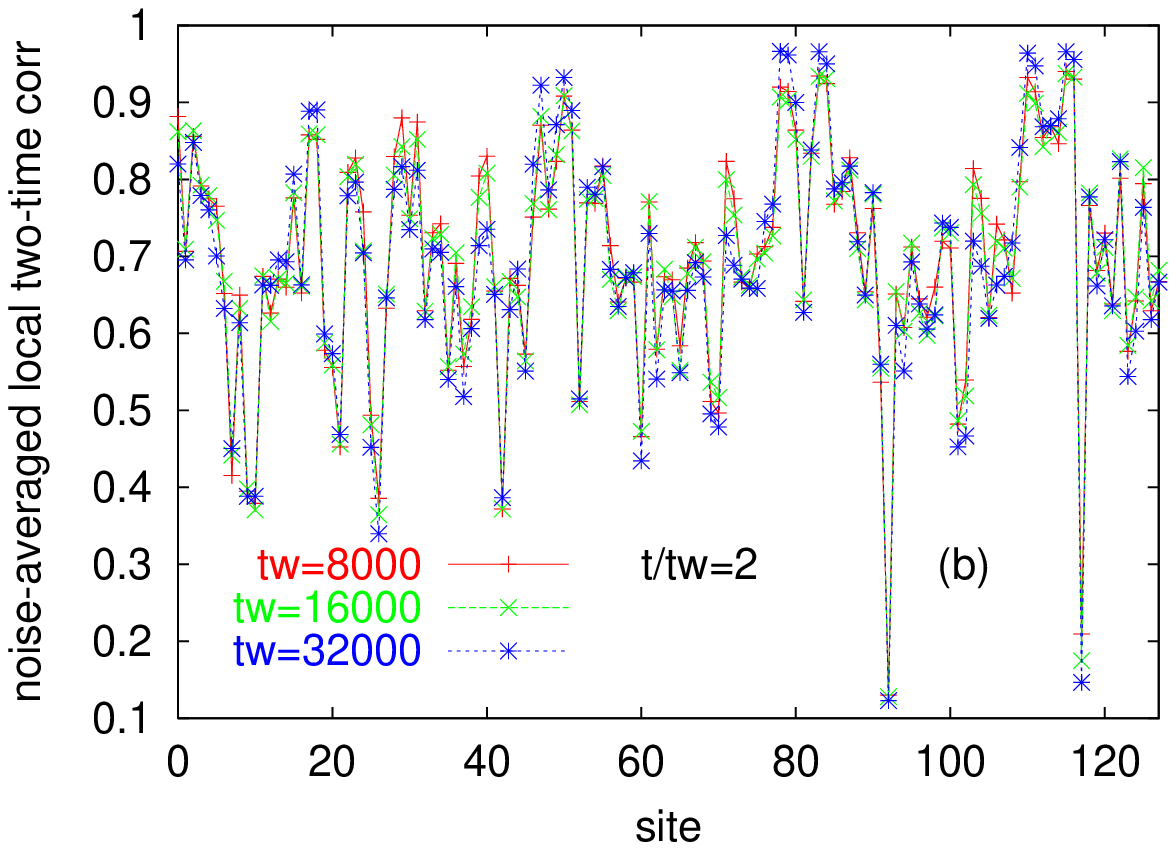,width=9cm}
\end{center}
\begin{center}
\psfig{file=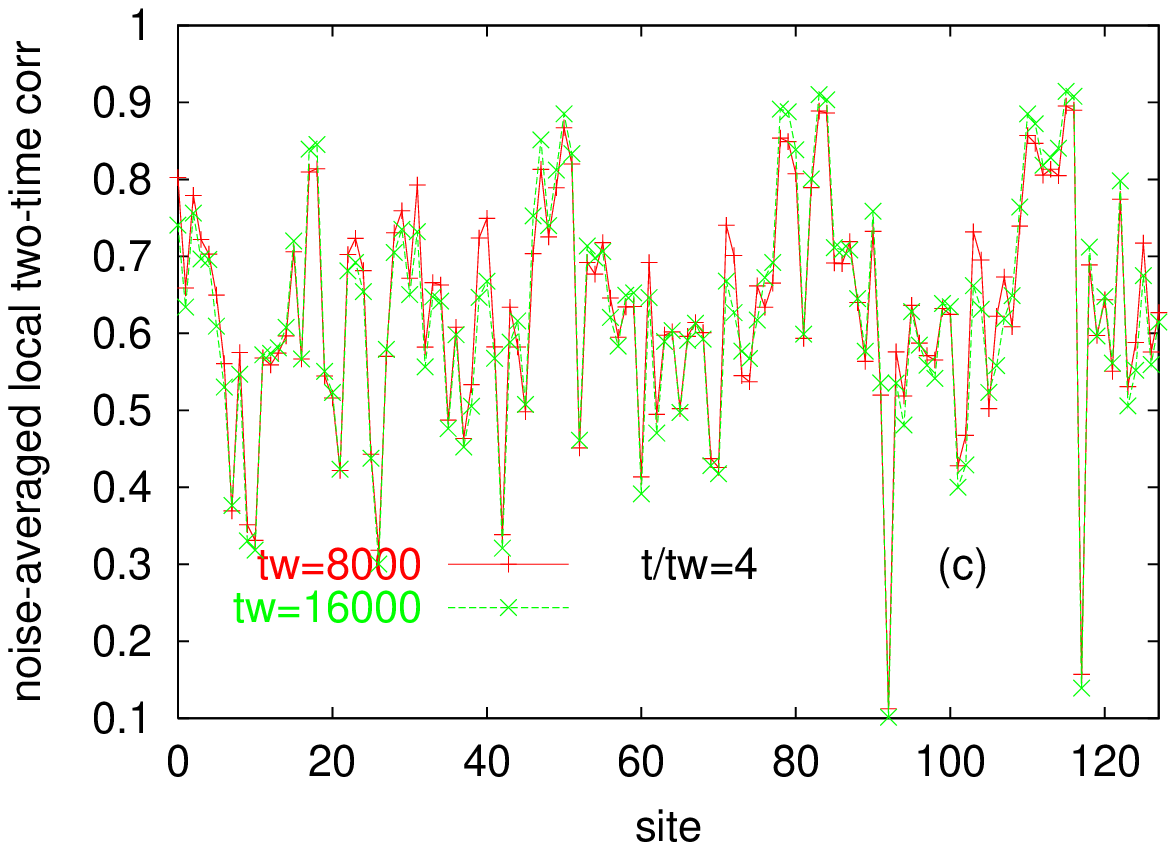,width=9cm}
\end{center}
\caption{The fingerprint of disorder. The noise-averaged local 
correlation on $128$ sites ordered along four adjacent rows
for three choices of the ratio $t/t_w$ indicated as a label 
in each panel. $L=32, T=0.7, \tau_t=t/10$.
The curves for different $t_w$'s fall on top 
of each other showing that $C_i^{na}$ scales as in Eq.~(\ref{localf}).
}
\label{test-localf}
\end{figure}

A simple way to test  Eq. (\ref{localf}) is to plot the 
values of the local correlations at different sites for different
pairs $(t,t_w)$ such that their ratio, $t/t_w$, is held fixed. If the 
hypothesis is correct, for a given site, its correlation must take a very 
similar value for all $t_w$'s. Figure~\ref{test-localf} shows this 
test for a $3d$ {\sc ea} model of linear size $L=32$ at 
$T=0.7$. 
The average involves $10^3$ noise realizations. In the 
three panels we used different values of the ratio $t/t_w$ as 
labelled.  The points represent
the values of the noise averaged local correlations at sites
in four rows in the cube. 
More precisely, the discrete points
on the axis labelled ``site'' correspond to $(x=0,y=0,z=0,\dots,L-1)$,  
$(x=0,y=1,z=0,\dots,L-1)$, $(x=0,y=2,z=0,\dots,L-1)$ 
and $(x=0,y=3,z=0,\dots,L-1)$ 
ordered in this 
way. This means that, for example, 
the values site $=0$, site $=32$ and site $=64$ are nearby 
sites on the lattice,  site $=0$ being  $(x=0,y=0,z=0)$,  
site $=32$ being  $(x=0,y=1,z=0)$ and site $=64$ being  $(x=0,y=2,z=0)$.
This explains the approximate periodicity  in the 
data: it  indicates that nearby sites have a rather 
similar behavior as most clearly indicated by staring at the points with 
rather small values of $C^{na}_i$. The lines are added as guides to the eye.
The figure shows  that 
the noise averaged correlations satisfy the hypothesis in Eq.~(\ref{localf})
for this range of times. The site-dependent external function $f_i$
in Eq.~(\ref{localf}) reflects the fingerprint of disorder.

\begin{figure}
\begin{center}
\psfig{file=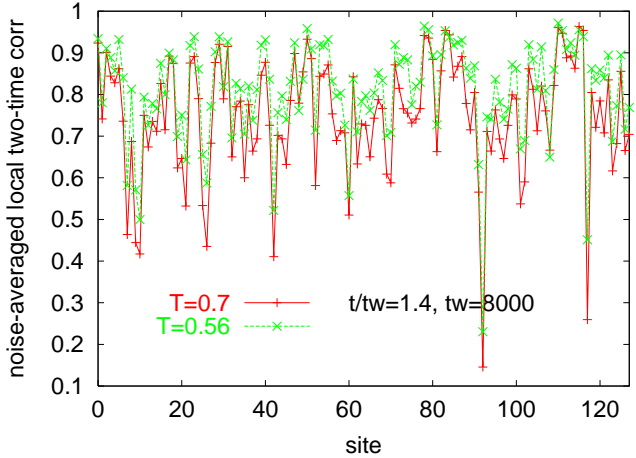,width=9cm}
\end{center}
\caption{Competition between temperature and quenched disorder. 
The noise-averaged local 
correlation on adjacent $128$ sites for two temperatures, $T=0.56$ and $T=0.7$.
The time ratio is $t/t_w=1.4$, $t_w=8\times 10^3$ {\sc mc}s and $\tau_t=t/10$. 
The two sets of data are surprisingly similar.}
\label{test-localT}
\end{figure}

Interestingly enough, even running at different temperatures (with the 
same seeds for the thermal noise) the individual evolution 
of the sites is still very similar (see Fig.~\ref{test-localT}). This issue 
deserves further investigation since it might be very useful 
in helping to explain the intriguing memory and rejuvenation effects 
seen in the dynamics of spin-glasses when temperature is modified. 
Nevertheless, we shall not expand on this  topic here. 

\subsection{Scaling of coarse-grained local correlators}
\label{section:scaling}

We can now analyze the coarse-grained local correlations 
in the same manner as the noise-averaged ones.

First, in Fig.~\ref{cg-test-localf} we show
$C_i^{cg}$ on $128$ sites chosen as in 
Fig.~\ref{test-localf} (see the text above) for three choices of 
$t/t_w=2,\, 4,$ and $64$ at $t_w=10^3$ {\sc mc}s.
The coarse-graining volume is small, $M=1$.
The curves for different $t_w$'s do not scale 
exactly since the noise induces fluctuations in 
addition to those that originate from the quenched disorder. 
$C_i^{cg}$ does not scale as in Eq.~(\ref{localf})
for any of the choices of the ratio $t/t_w$.

\begin{figure}
\begin{center}
\psfig{file=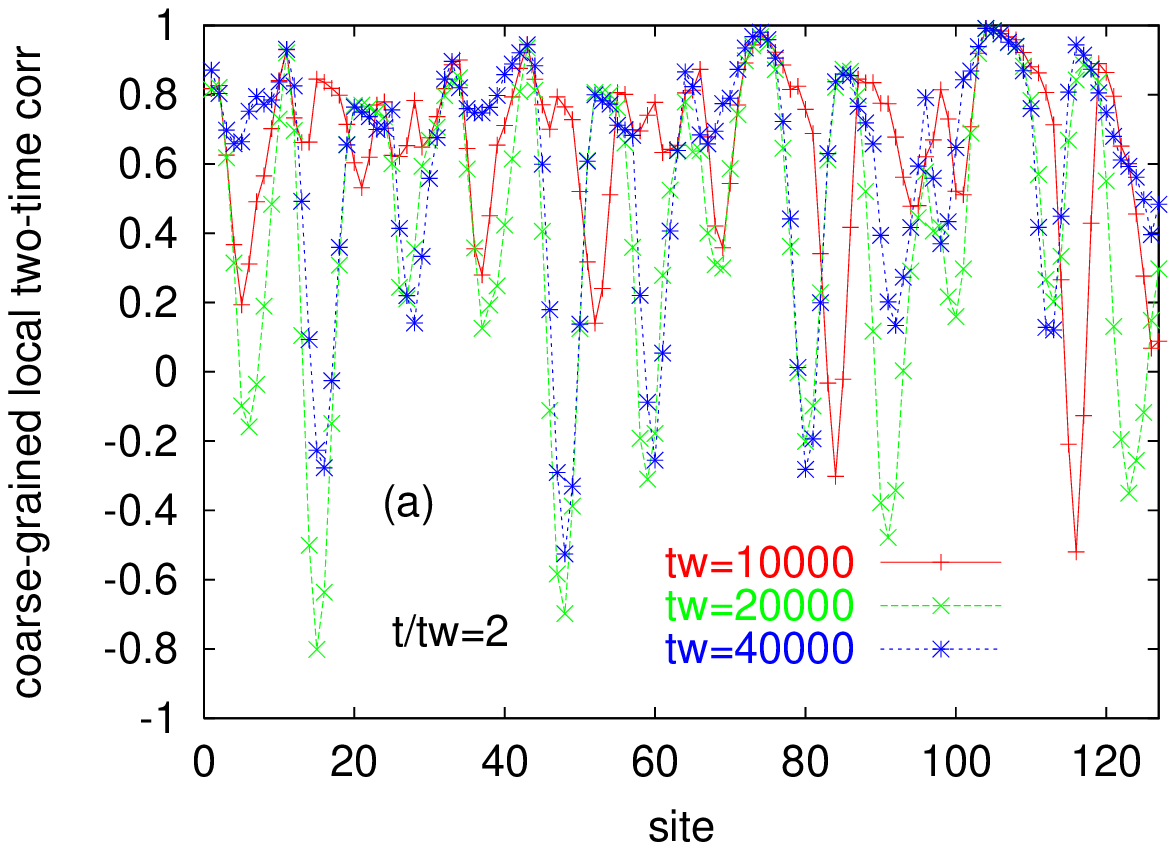,width=9cm}
\end{center}
\end{figure}
\begin{figure}
\begin{center}
\psfig{file=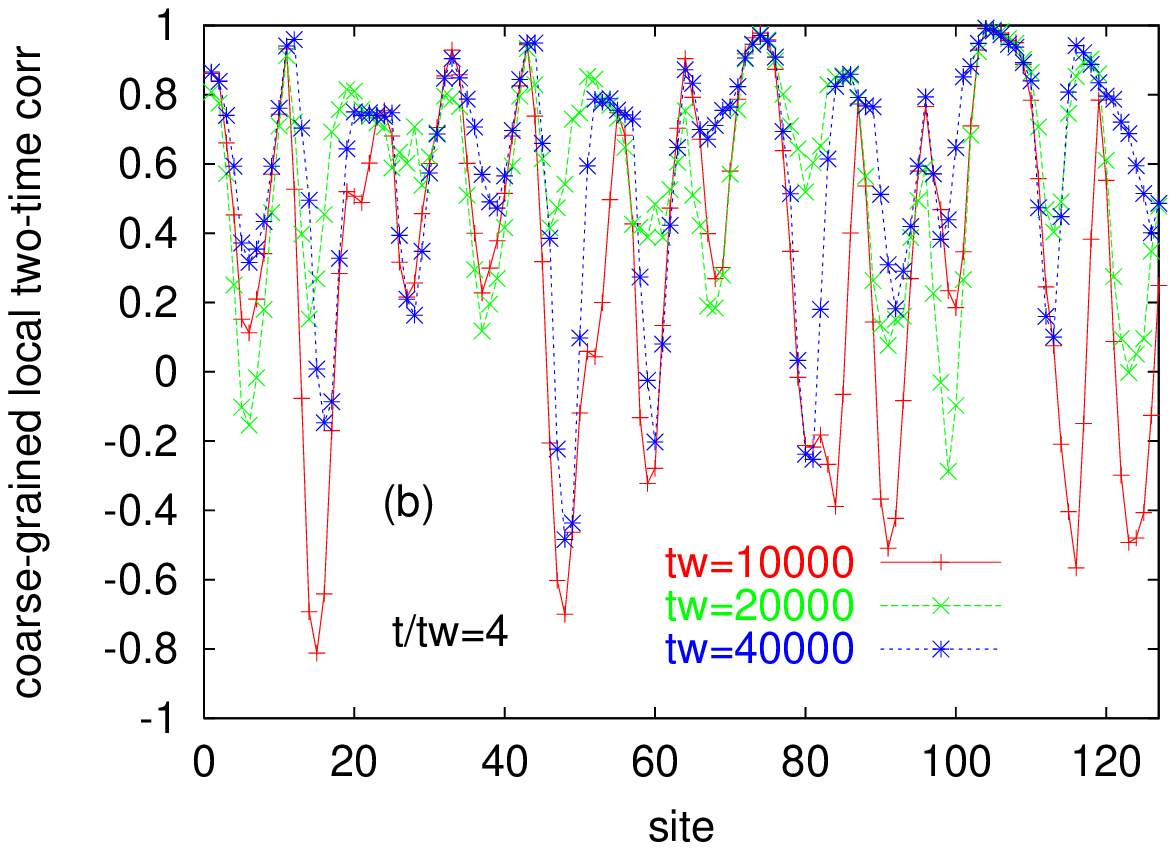,width=9cm}
\end{center}
\begin{center}
\psfig{file=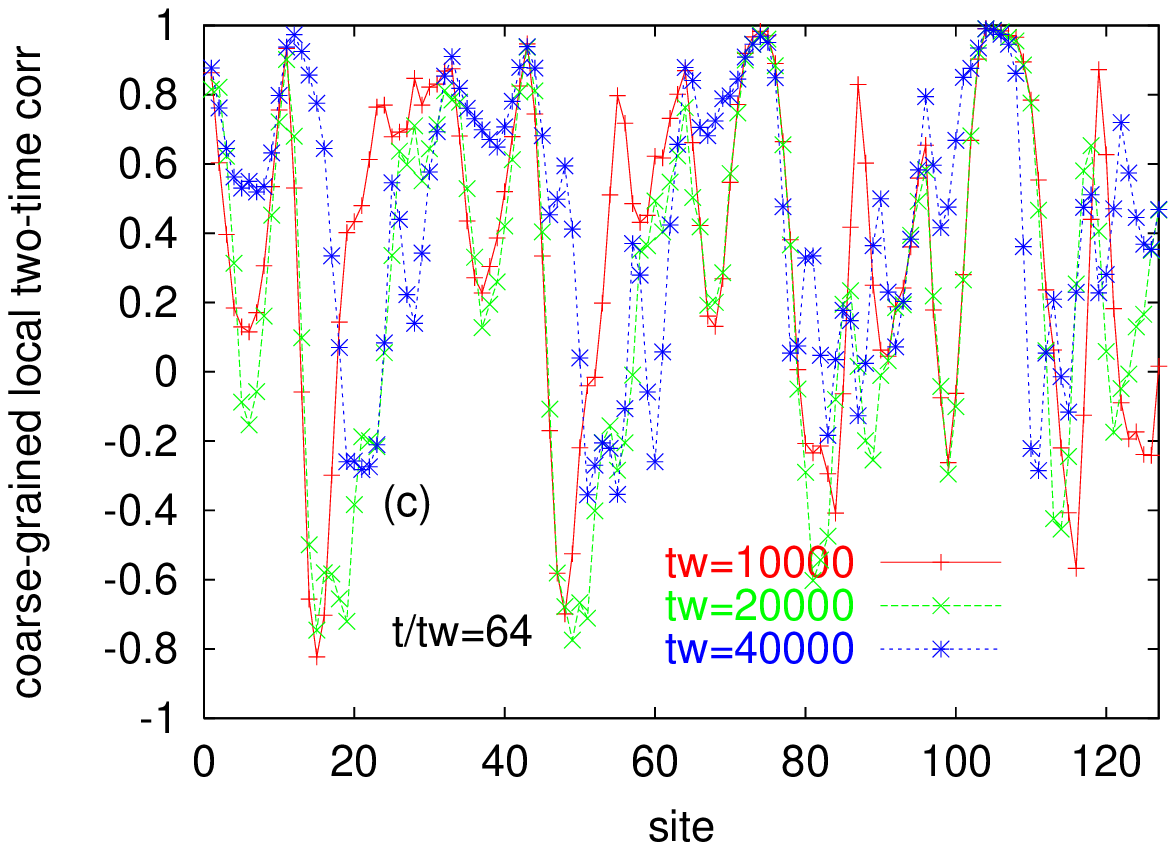,width=9cm}
\end{center}
\caption{Test of $t/t_w$ scaling in the coarse-grained
correlation. Values of this quantity on $128$ adjacent sites 
for three choices of the ratio $t/t_w$ indicated in each panel. 
$M=1$, $L=32, T=0.8$, and $\tau_t=t/10$.}
\label{cg-test-localf}
\end{figure}

Notice that the curves obtained for different pairs of time $(t,t_w)$, even
though the ratio $t/t_w$ is fixed, cross each other at many
points. Therefore, it is always possible to find sites $i$ and $j$ at
opposite sides of the crossing points for which $C_i^{cg}>C_j^{cg}$ for one
pair of $(t,t_w)$, but $C_i^{cg}<C_j^{cg}$ for another. In other words, the
relative age (as measured using the correlation value) between two sites $i$
and $j$ in the sample is not static, but fluctuates as a function of
time. These are exactly the {\it sorpassi} that we described in
Sec.~\ref{behavior-local}. Once again, notice that this is in sharp
contrast with the scaling Eq.~(\ref{localf}), where the relative age between
all sites in the sample keeps a constant, static, relative rank.

\begin{figure}
\begin{center}
\psfig{file=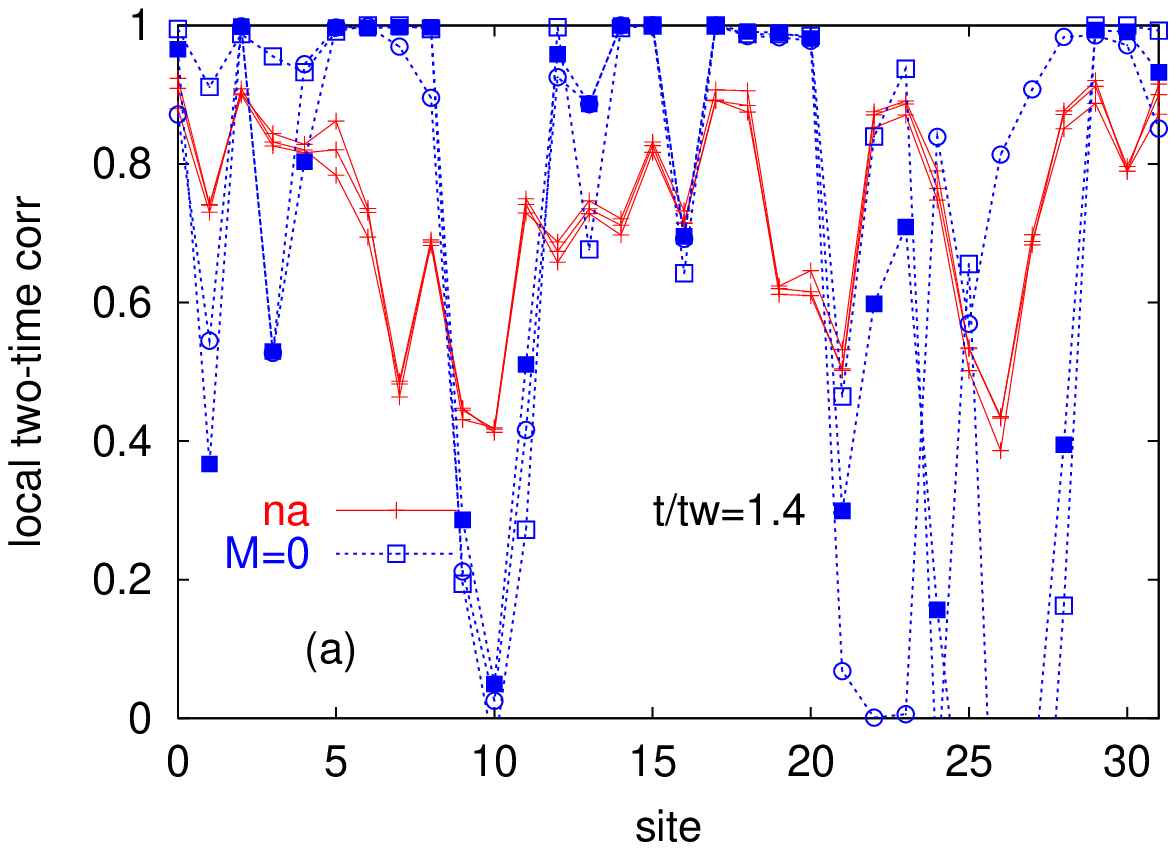,width=9cm}
\end{center}
\begin{center}
\psfig{file=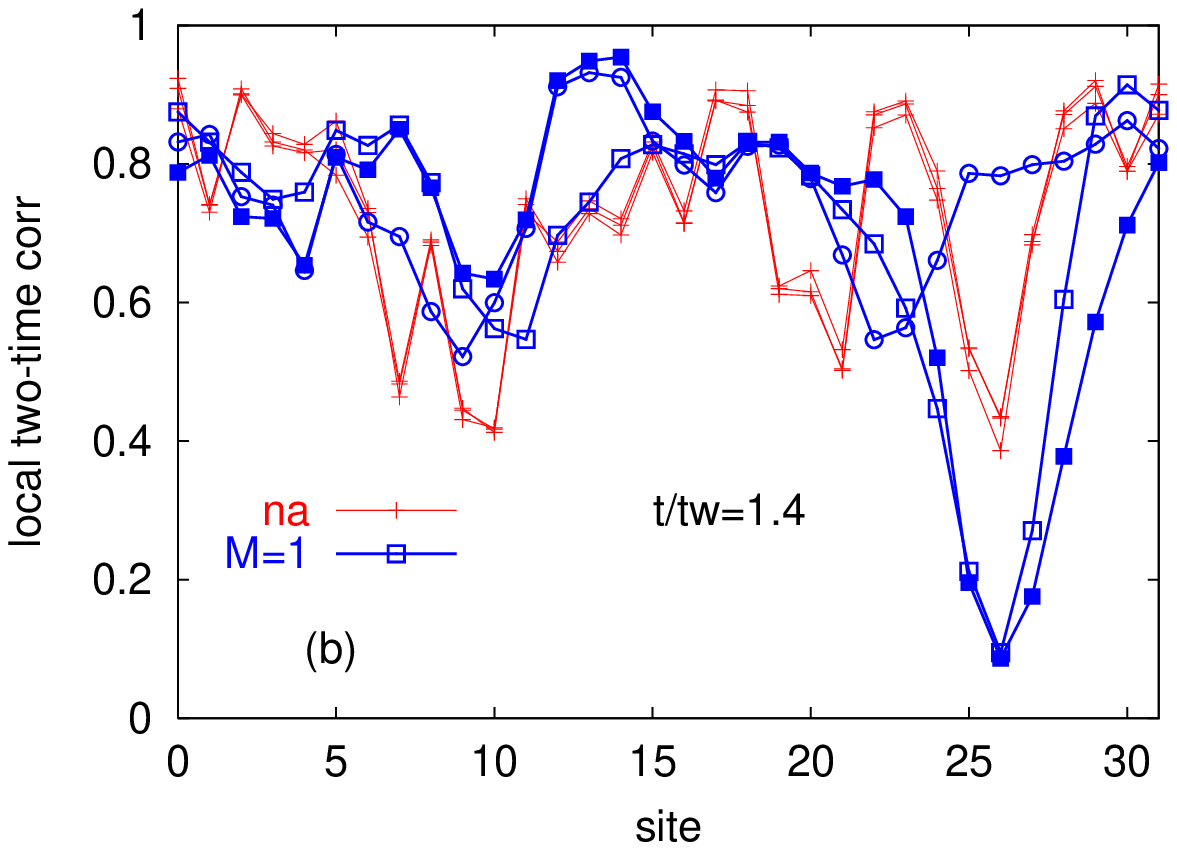,width=9cm}
\end{center}
\begin{center}
\psfig{file=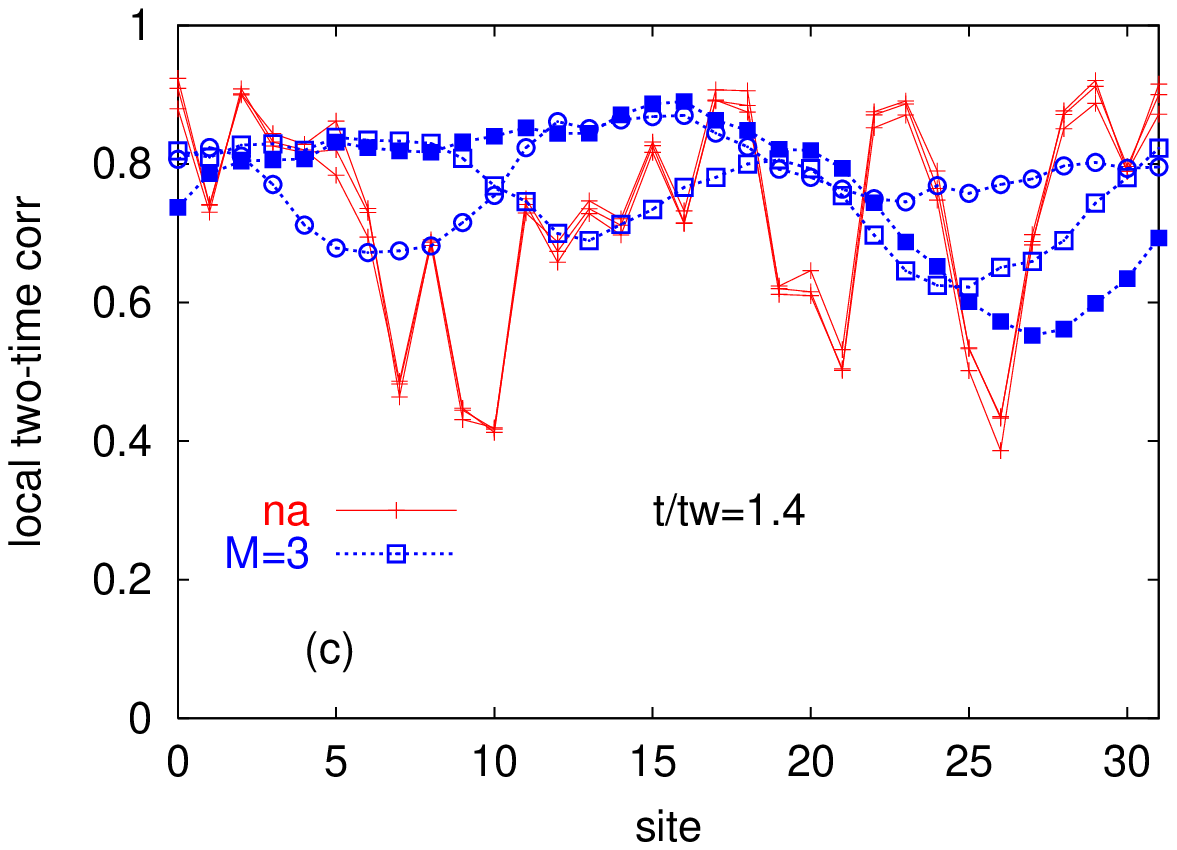,width=9cm}
\end{center}
\begin{center}
\psfig{file=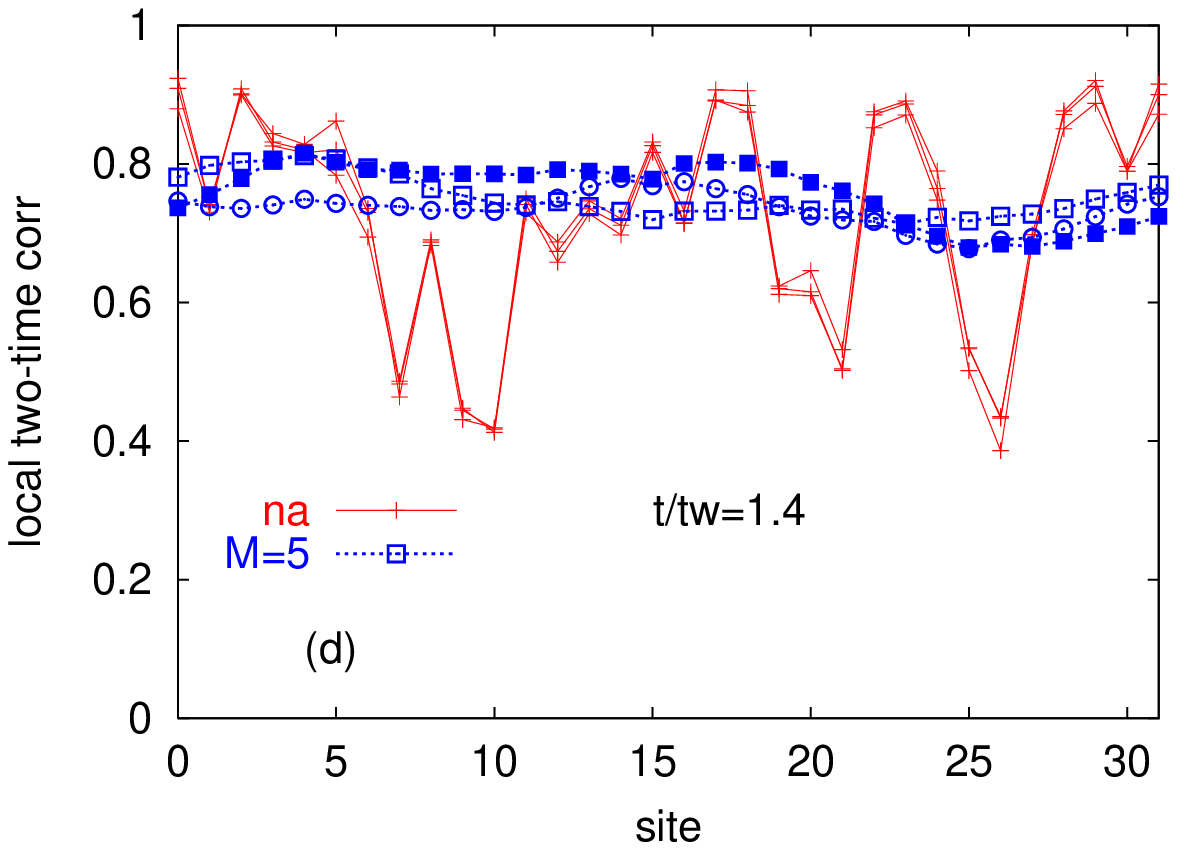,width=9cm}
\end{center}
\caption{Comparison between noise-averaged and coarse-grained 
local correlations. The three curves in each group correspond
to three waiting-times. $t_w=8\times 10^3$ {\sc mc}s (open squares),
$t_w=1.6\times 10^4$ {\sc mc}s (open circles) and 
$t_w=3.2\times 10^4$ {\sc mc}s (filled squares) for the coarse-grained data.
The coarse-graining volume is indicated 
in the labels. $L=32, T=0.7$. The coarse-graining times is $\tau=8\times 10^2$ 
{\sc mc}s when $t_w=8\times 10^3$ {\sc mc}s and is multiplied by $2$ 
when the time is doubled.}
\label{cfr}
\end{figure}

In Fig.~\ref{cfr}  we compare the coarse-grained and 
noise-averaged local correlations for two systems {\it with the same
quenched disorder}, evolved at the same temperature.
We plot the local correlations on one row of the $3d$ cube corresponding 
to $(x=0,y=0,z=0,\dots,L-1)$, with $L=32$. The temperature is $T=0.7$. 
In all panels we plot the noise-averaged data for three $t_w$'s,
$t_w=8\times 10^3$ {\sc mc}s, $1.6\times 10^4$  {\sc mc}s 
and $3.2\times 10^4$  {\sc mc}s,
with crosses and the coarse-grained data, for the 
same waiting-times, with open squares ($t_w=8\times 10^3$  {\sc mc}s),
open circles ($t_w=1.6\times 10^4$  {\sc mc}s) and dark squares 
($t_w=3.2\times 10^4$  {\sc mc}s). The lines joining the points 
are a guide to the eye.
The ratio between $t$ and $t_w$ is always $t/t_w=1.4$ but 
similar results are obtained for other choices.
In different panels in the figure we use different values of $M$, 
$M=0$ (a), $M=1$ (b), $M=3$ (c), and $M=5$ (d).

The first result from these figures is that, as already shown in 
Fig.~\ref{test-localf}, the noise-averaged data scale as in 
Eq.~(\ref{localf}). Instead, the coarse-grained data do 
not scale in this 
way for any of the values of $M$.

Let us now discuss in detail the effect of coarse-graining.  In panel a) we
compare the noise-averaged data to the results of a single run without
coarse-graining ($M=0$). We see that the $t/t_w$ scaling does not hold for
the curves with coarse-grained data, but the values of the local correlations
are very much influenced by the disorder. 
For instance, observe the low
values simultaneously taken by the correlation at site$=10$ 
for the noise-averaged
quantity and also in the single thermal history runs.

In panel b) we use a minimum volume
$V=(2M+1)^3=3^3$. We see that the ``surface'' created by the 
coarse-grained data has been smoothed with respect to the 
case $M=0$ shown in panel a) but there remains a memory 
of the underlying quenched disorder. The coarse-graining does
not help improve the $t/t_w$ scaling.

This trend is even clearer in panels c) and d) 
where we use $M=3$ and $M=5$, respectively. The
$t/t_w$ scaling is 
clearly broken as the surfaces become smoother and 
smoother when $M$ increases.
When $M=5$ the surface is almost totally flat
and the fingerprint of disorder has been washed out 
by coarse-graining. However, there are still soft local fluctuations
that break the $t/t_w$ scaling. 

The conclusion we draw from these plots, and from the theoretical
discussion in Sec.~\ref{sigma}, is that for a truly infinite system,
$N=L^3\to\infty$, we expect to find that when times diverge,
$t,t_w\to\infty$ with $C(t,t_w)$ fixed to a chosen value, say 
$C(t,t_w)={\tt C}$, thus selecting the
correlation scale, coarse-graining over a 
sufficiently large volume erases the fingerprint of disorder whilst still
allowing for fluctuations that change the reparametrization locally,
$h(t) \to h_i(t)=h(t)+\delta h_i(t)$. This argument implies that 
hypothesis (\ref{eq:localh}) should describe the data in this limit. 
In Sec.~\ref{correlation-length} we shall further discuss the
implications of coarse-graining.

The final test of this hypothesis would be to 
plot the value of the correlation on each site for
different pairs of times such that $h_i(t)/h_i(t_w)$ is 
held fixed. If the hypothesis in Eq.~(\ref{eq:localh}) holds, the 
surface of the resulting figure should be flat. 

An equivalent way to check hypothesis (\ref{eq:localh}) is to 
plot the decay of the local correlations on several sites, 
for several chosen values of $t_w$ and all subsequent $t$,
against different ratios $\lambda_i=h_i(t)/h_i(t_w)$. For each 
site $i$ the choice of an adequate scaling function $h_i(t)$ should lead 
to a collapse of each of the local correlations $C_i(t,t_w)$ 
corresponding to different $t_w$'s. Moreover, if the (in principle 
site-dependent) master curves, $f_i$, 
thus obtained are all identical, $f_i=f$ for all $i$, 
then the conjecture in Eq.~(\ref{eq:localh}) is satisfied. 

The careful implementation of these checks is rather tedious since one needs
an independent inspection of the dynamics of each site in the sample. We take
the flattening of the curves in Fig.~\ref{cfr} for increasing $M$ as strong
evidence for the scaling in Eq.~(\ref{eq:localh}) in the ``scaling limit''
defined in Sec.~\ref{scaling-limit}. 

\subsection{Effect of coarse-graining on already noise-averaged quantities}

\begin{figure}
\begin{center}
\psfig{file=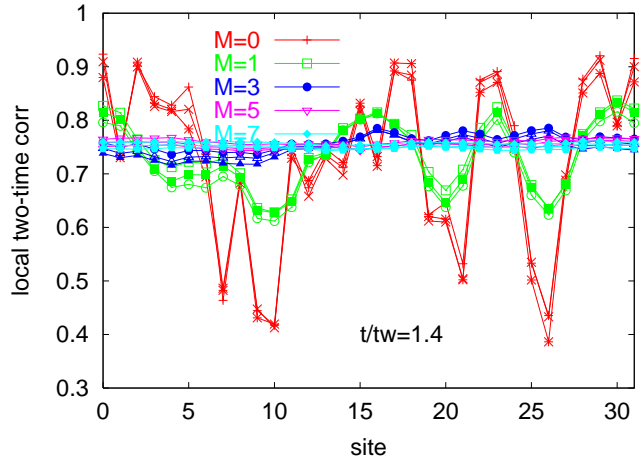,width=9cm}
\end{center}
\caption{Effect of coarse-graining on the local noise - averaged data.
The parameters are as in Fig.~\ref{cfr}}
\label{coarse-ontop}
\end{figure}

In Fig.~\ref{coarse-ontop} we coarse-grain the already noise-averaged
data for the local correlation. When $M=3$ the surface is already
quite flat (and $t/t_w$ dependent). This is to be compared with the 
the data in Fig.~\ref{cfr}c) for one run coarse-grained over the 
same volume. We see how the fingerprint of disorder progressively 
disappears with more coarse-graining.

\section{Two-time dependent correlation length}
\label{correlation-length}

In this section we define a correlation length from the
study of the spatial fluctuations of the two-time 
local correlators. We also discuss the interplay between 
times and length-scales in the analysis of our data.

\subsection{Definition}
\label{mass-goldstone}

The presence of a Goldstone mode in the dynamics gives rise to a specific
form for the spatial correlations of the fluctuating two-time fields
$Q_i^{\alpha\beta}(t,t_w)$. In the asymptotic limit, the R$p$G invariance
implies a true Goldstone, or zero mass,  mode. Therefore, the spatial
correlations in the fluctuations should show a power law decay $\sim 1/r$ in
$3d$. However, as mentioned before, we know that {\em for any finite time, the
R$p$G invariance is explicitly broken} by irrelevant terms that play the role
of symmetry breaking fields. As in the case of magnons in the presence of a
weak external magnetic field, the Goldstone mode in the dynamics acquires a
small mass, which vanishes in the asymptotic limit of small symmetry breaking
field, {\it i.e.} in the long time limit. Therefore, we expect that in a
simulation, the local two-time quantities will exhibit 
a finite correlation length for their fluctuations.
Fixing the 
relation between $t$ and $t_w$ to  have, say, a given global correlation,
one should find a correlation length $\xi(t,t_w)$
that increases monotonically for increasing $t_w$. 
Equivalently, the mass $m(t,t_w) = 1/\xi(t,t_w)$ should be a monotonically
decreasing function of the times~\cite{several-length-scales}.

Most of the results about the finite correlation length that we
present were obtained for a $3d$ {\sc ea} model with linear size
$L=32$ evolving at $T=0.8$. We considered $64$ realizations of the
disorder. The length of the simulation was $8.192 \times 10^7$ {\sc
mc}s.

The two-time quantities considered are the local two-time
correlations $C_{{\vec r}_i}(t,t_w) \equiv C_i(t,t_w) =  \overline s_i(t) 
\overline s_i(t_w) $, where the $\overline s_i(t)$ are
site magnetizations time averaged over $\tau=10^3$ {\sc mc}s.
The vector $\vec r_i$ is the position of site $i$ with respect 
to a chosen origin of coordinates.

We define a spatial correlator
\begin{equation}
\tilde A(r;t,t_w)
\equiv
\left[\frac{1}{N}\sum_i C_{{\vec r}_i}(t,t_w)\; 
C_{{\vec r}_i+\vec r}(t,t_w) \right] 
\end{equation}
that is averaged over disorder
realizations but is not averaged over the noise.
Since the problem is isotropic on average we 
expect the r.h.s. to depend only on the modulus of 
the distance between the positions of the spins considered, 
$r\equiv |{\vec r}|$.
Thus we write $r $ in the argument of $\tilde A$.
$N=L^3$ is the total number of spins in the sample.
As $r \to \infty$ we expect that this correlator 
will approach its disconnected part:
\begin{eqnarray}
\tilde A_\infty(t,t_w)&\equiv&  
\lim_{r\to\infty} \tilde
A(r;t,t_w)= \left[\left(\frac{1}{N} \sum_i C_{{\vec
r}_i}(t,t_w)\right)^2\right] 
\nonumber\\
&=& [C(t,t_w)^2]
\; ,
\end{eqnarray}
with $C(t,t_w)$ the global correlation.

In order to define a correlator that has a spatially asymptotic
value that is independent of the times $t$ and $t_w$ we
define the normalized correlator:
\begin{equation}
A(r;t,t_w)
\equiv 
\frac{\tilde A(r;t,t_w)}{\tilde A_\infty(t,t_w)}
\end{equation}
which tends to unity in the limit $r \to \infty$. 
Therefore, the connected part of $A$ is simply $A-1$. 
Note that the equal site correlator
$A(r=0;t,t_w)$ is very close to 
$1/\tilde A_\infty(t,t_w)$ since the zero-distance correlator,
$\tilde A(r=0,t,t_w)$, involves ${\overline s_i}^2(t)$ and ${\overline s_i}^2(t_w)$ that 
for short coarse-graining times $\tau$ and long times $t$ and $t_w$ 
are approximately equal to $1$. For any $t_w$ 
the denominator $\tilde A_\infty(t,t_w)=[C(t,t_w)^2]$ vanishes in the 
limit $t\gg t_w$. Consequently, as it is defined, the correlator 
$A(r;t,t_w)$ evaluated at zero spatial distance
diverges with increasing $t$ for any value of $t_w$. 
At equal times $A$ is 
approximately $1$ for all distances $r$; the reason
for this result is again the fact that  ${\overline s_i}^2(t) \approx 1$. 

One can also define a correlator that is forced to evolve,
for all pairs of times $t$ and $t_w$, 
between $1$ at zero distance and $0$ for infinitely separated
distances. Indeed, 
\begin{equation}
B(r;t,t_w) \equiv \frac{\tilde A(r;t,t_w)-\tilde A_\infty(t,t_w)}
{\tilde A(r=0;t,t_w)-\tilde A_\infty(t,t_w)}
\label{EQ:B_def}
\end{equation}
satisfies these requirements. 

\begin{figure}
\epsfxsize=8.5cm
\epsfbox{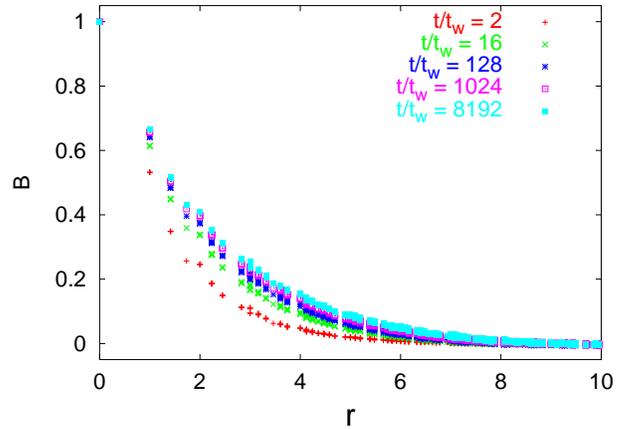}
\vspace{0.05cm}
\caption{Spatial correlations $B(r;t,t_w)$ against
distance $r$ for $t_w = 10^4$ {\sc mc}s and $t/t_w = 2$, $16$,
$128$, $1024$, and $8192$.}
\label{FIG:A_vs_x}
\end{figure}

\begin{figure}
\epsfxsize=8.5cm
\epsfbox{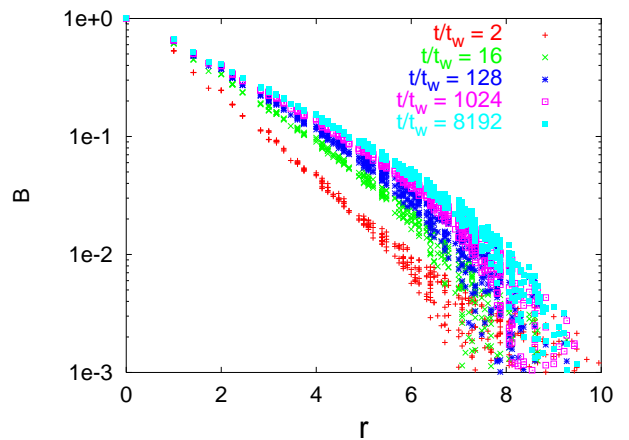}
\vspace{0.05cm}
\caption{Spatial correlations $B(r;t,t_w)$ against
distance $r$ in a semi-logartihmic scale 
for $t_w = 10^4$ {\sc mc}s and $t/t_w = 2$, 
$16$, $128$, $1024$, and $8192$. }
\label{FIG:A_vs_x_ln}
\end{figure}

In Fig.~\ref{FIG:A_vs_x}, we show the space dependence of 
$B(r;t,t_w)$, for various ratios $t/t_w$. In all cases we observe
that the correlation decays rapidly, consistent with an exponential
decay, for $r < 6$ lattice spacings.
To make this more apparent, we show a semilog plot for
the same  $t/t_w$ ratios in Fig.~\ref{FIG:A_vs_x_ln}. 

In the previous two figures, the horizontal axis represents the
distance $r$. One point is plotted for each vector ${\vec r}$,
therefore many points with similar $r$ but corresponding to
different directions for ${\vec r}$ appear close to each
other. The fact that in the figures the corresponding values of
$B$ are also close together indicates that the spatial
correlations are indeed spatially isotropic, as expected. 

\begin{figure}
\epsfxsize=8.5cm
\epsfbox{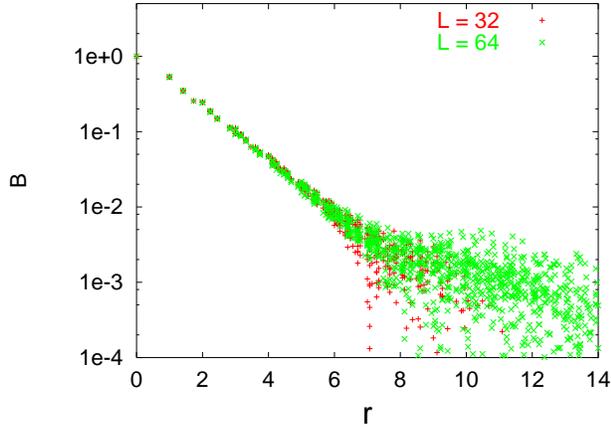}
\vspace{0.05cm}
\caption{Comparison between the spatial dependence of 
$B(r;t,t_w)$ for $L=32$ and
  $L=64$, with $t_w = 10^4$ {\sc mc}s and $t/t_w = 2$.}
\label{FIG:different_sizes}
\end{figure}

\begin{figure}
\epsfxsize=8.5cm
\epsfbox{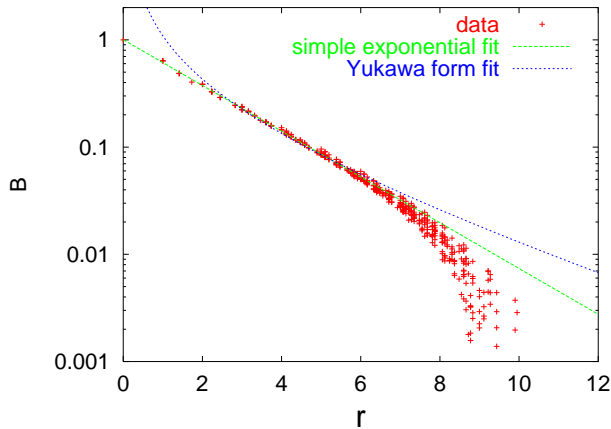}
\vspace{0.05cm}
\caption{Example of fit with an exponentially decaying function
given in Eq.~(\ref{exponential-fit})  and 
a Yukawa type of correlator in $3d$ given in Eq.~(\ref{yukawa-fit}) 
(see the text for details). 
The data correspond to $t_w = 3.2 \times 10^5$ {\sc mc}s, 
$t = 2.56 \times 10^6$ {\sc mc}s. The data
were fitted in the interval $r \in [1.5,6.0]$. } 
\label{FIG:fit_exp}
\end{figure}

In Fig.~\ref{FIG:different_sizes}, we show the decay of the
correlation for two system sizes $L=32$ and $L=64$. We observe that
the decay is roughly independent of the system size, up to distances
of the order of $r\sim 6$. One notices that the trend of the
points, after this limit, is to bend downwards for $L=32$ and to bend
upwards for $L=64$.  $r\sim 6$ is also the distance upto which the
exponential fit [see Eq.~(\ref{exponential-fit}) and Fig.~\ref{FIG:fit_exp}] 
is very good. (In this figure, $t_w=10^4$ {\sc mc}s
and $t/t_w = 2$.)

In Fig.~\ref{FIG:fit_exp}, we show a typical example of a fit of the
correlation to exponentially decaying functions. Two possible fits are
displayed, one with the simple exponential form 
\begin{equation}
B_{\rm exp}(r; t,t_w) = D(t,t_w) \; e^{-m(t,t_w) \, r} 
\; ,
\label{exponential-fit}
\end{equation}
the other with the Yukawa form 
\begin{equation}
B_{\rm Yuk}(r; t,t_w) = 
D(t,t_w) \; \frac{e^{-m(t,t_w) \, r}}{r} 
\; .
\label{yukawa-fit}
\end{equation}
The second form is the 
$3$-dimensional Fourier transform of the massive correlator $ 1/( k^2 +
m^2) $, and one would naively expect this form to
provide the best fit. 
However, on deeper examination, it is necessary to take into account the
form of the fluctuations, as discussed in
Sec~\ref{sec:randomsurfaceaction} (see Eq.~\ref{eq:rs-q-and-phi}), and in
particular the fact
that $C({\vec r},t,t_w)$ appears to depend {\em exponentially} on the two
fluctuating fields $\phi({\vec r},t_i)$ ($t_i = t, t_w$). Since the
fluctuations in $C({\vec r},t,t_w)$ are significant, even if the
propagator for each reparametrization fields $\phi({\vec r},t_i)$ had
a simple Yukawa form, the form of $B(r; t,t_w)$ is in
principle highly nontrivial, and in particular the value of the mass
$m(t,t_w)$ is not easy to predict. 

\begin{figure}
\epsfxsize=8.5cm
\epsfbox{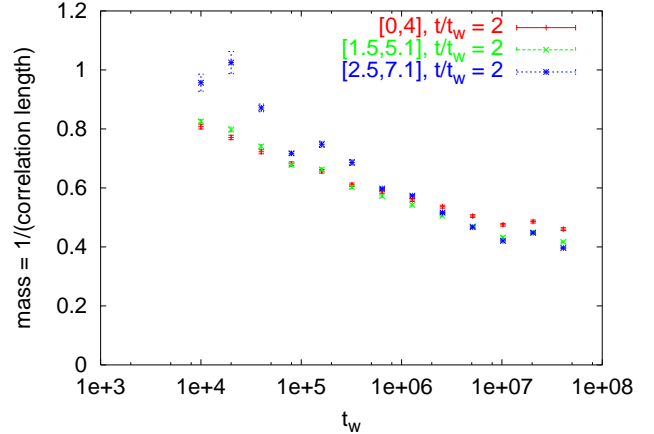}
\vspace{0.05cm}
\caption{Dependence of the fitted values of the mass on the interval
  chosen for the fit. Comparison between values of the mass obtained
  by fitting in the intervals $r\in [0.0,4.0]$, $r \in [1.5,5.1]$, and $r \in
  [2.5,7.1]$, displayed as a function of $t_w$. In all cases, $L=32$,
  and $t/t_w = 2$.}
\label{FIG:mass_interval}
\end{figure}

For the range of times and system sizes achieved in our simulations, 
it turns out that both the simple exponential and the Yukawa forms
provide good fits
for intermediate values of $r$, but the first one consistently
has a better $\chi^2$, and it consistently provides a better
extrapolation outside the fitting interval.
For larger values of $r$, finite size
effects seem to play an important role, producing an even steeper
decay of the correlations. 
In principle
there is no reason to expect the fit to be good for small values of
$r$. However, the simple exponential form actually does
provide a good fit down to $r = 0$. In what follows, all the
values of the mass (equivalently, the inverse correlation length) 
are obtained by using a simple exponential fit in the
interval $[0,4]$. Each fit has $52$ degrees of freedom, and typical
$\chi^2$ per degree of freedom are of the order of $10^{-3}$. 
From the definition in Eq.~(\ref{EQ:B_def}), we know that $B(r=0; t,t_w) =
1$. The fitting functions that we use do not need to have exactly this
value at the origin, but it turns out that, for the simple exponential
form, $D(t,t_w) \approx 1$ within an error of the order of 10 \%.

Figure~\ref{FIG:mass_interval} displays a comparison of results obtained
by using different fitting intervals. 
Changing the fitting interval does somewhat affect the results of the
fits, but the trends are consistent, and the interval chosen ($r \in
[0.0,4.0]$) seems to be the one that minimizes the noise in the mass
versus time curve.

\begin{figure}
\epsfxsize=8.5cm
\epsfbox{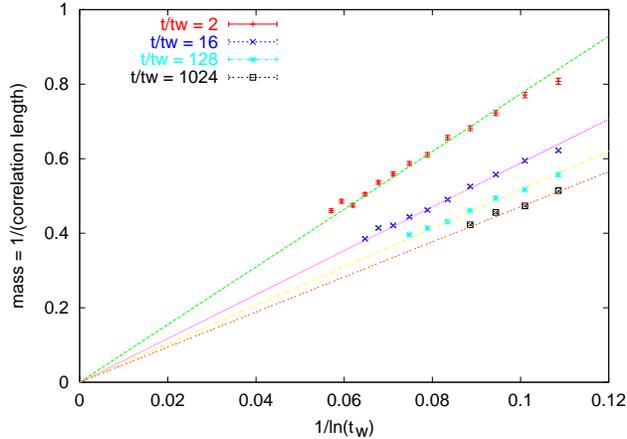}
\vspace{0.05cm}
\caption{Mass (inverse correlation length) as a function of the 
waiting-time. Each set
of points corresponds to a fixed ratio $t/t_w$. The horizontal axis
corresponds to $1/\ln{t_w}$.}
\label{FIG:mass_vs_time_ln}
\end{figure}

\begin{figure}
\epsfxsize=8.5cm
\epsfbox{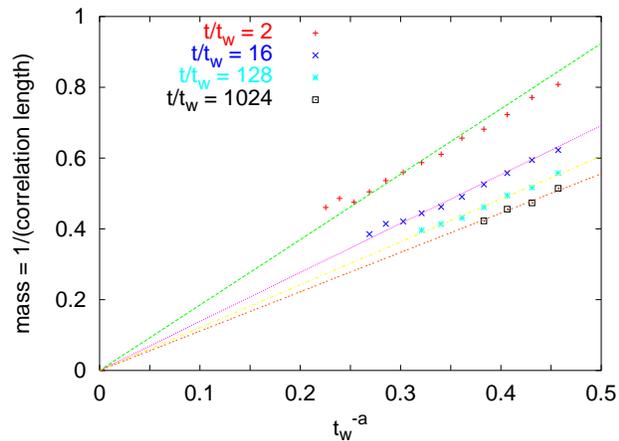}
\vspace{0.05cm}
\caption{Mass (inverse correlation length) as a function of the waiting-time.
Each set of points corresponds to a fixed ratio $t/t_w$. The
horizontal axis corresponds to ${t_w}^{-a}$ with $a= 0.085 $.}  
\label{FIG:mass_vs_time_pow}
\end{figure}

In Figs.~\ref{FIG:mass_vs_time_ln} and~\ref{FIG:mass_vs_time_pow},
we show the two-time dependence of the mass in our simulations. Each set
of points corresponds to a fixed ratio $t/t_w$. We attempt two fits of
the mass as a function of $t_w$. In Fig.~\ref{FIG:mass_vs_time_ln} we
show the fit $m(t,t_w)\equiv 1/\xi(t,t_w) \approx m_{0}(t/t_w)/\ln{t_w}$. 
In Fig.~\ref{FIG:mass_vs_time_pow}, 
we show the fit $m(t,t_w) \equiv 1/\xi(t,t_w) \approx m_{0}(t/t_w)
(t_w/t_0)^{-a}$, with $a = 0.085$. The simulation results are
consistent with both fitting forms. Due to 
the very slow decay of the mass with $t_w$, it is not possible to
distinguish between the two fits. 
Another important feature of the data is revealed by these figures:
the correlation length {\it increases} with increasing ratio $t/t_w$,
at fixed $t_w$. One must note, though, that the variation of the mass
with the ratio $t/t_w$ is also quite mild.

\begin{figure}
\epsfxsize=8.5cm
\epsfbox{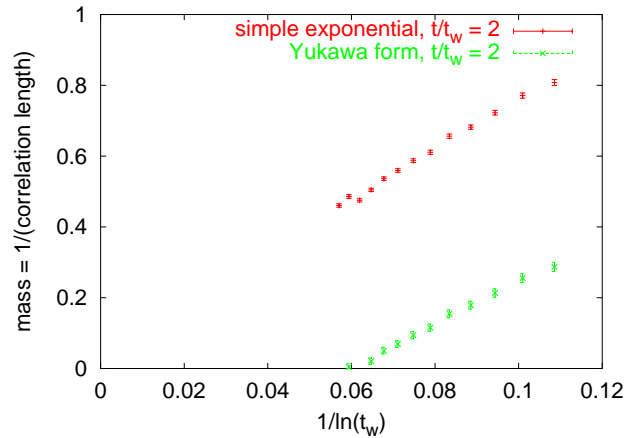}
\vspace{0.05cm}
\caption{Comparison between the values of the  masses obtained with 
the exponential fit given in Eq.~(\ref{exponential-fit}) 
  and the fit with the Yukawa form given in Eq.~(\ref{yukawa-fit}). In
  both cases $L=32$, $t/t_w=2$.}
\label{FIG:mass_yuk}
\end{figure}

An intriguing fact in our results is that the values of the
correlation length obtained are extremely short. Note that
Figs.~\ref{FIG:mass_vs_time_ln} and \ref{FIG:mass_vs_time_pow} show a
variation of the mass between $0.4$ and $0.8$ that corresponds to
lengths between $1.25$ and $2.5$ lattice spacings. As shown in
Fig.~\ref{FIG:mass_yuk} these values change quite a bit if one uses
instead a fitting procedure with the Yukawa form in
Eq.~(\ref{yukawa-fit}). As mentioned before, the simple exponential
form provides a better fit for the range of distances accessible in
our simulations. However, this conclusion might in principle change if
much larger sizes were simulated. (See,
{\it e.g.} Fig.~\ref{FIG:different_sizes}, where the data for $L=64$ seem to
deviate {\em above} the simple exponential fit for longer distances,
which would be consistent with a different functional form for $B$ and
possibly a smaller value of the mass.) The big difference in the mass
values for the two fits indicates that any quantitative results about
the precise value of the mass have to be taken with caution, and that
our conclusions should be restricted to indicate general trends in the
data.  Based on these considerations we shall postulate, somewhat
arbitrarily, that the correlation lengths for the times explored in
the rest of this paper are in between the two fits and of the order of
$3$ to $5$ lattice spacings.

Let us discuss a qualitative picture for describing the two-time
dependence of the $m(t,t_w)$ suggested by the random surface action of
Sec.~\ref{sec:randomsurfaceaction}. Although $m(t,t_w)$ depends on the
two times $t$ and $t_w$, an action such as that in Eq.~(\ref{eq:rf3a})
(with an added $\tau$-dependent mass term) suggests that different
time slices factorize in the effective theory for fluctuations.  This
picture is only approximate, in that for time ratios not far from one,
the subleading terms neglected in Eq.~(\ref{eq:rf3a}) will generate
corrections. Thus we expect that, for long times {\em and} large
$t/t_w$ ratios, $m(t,t_w)$ should be some kind of average of one-time
quantities associated with $t$ and $t_w$.  This picture suggests that
one should also analyze the correlation length data by plotting it
against an average between $t$ and $t_w$, the appropriate one being a
geometric average $t_{ave}=\sqrt{t t_w}$ [which corresponds to the
simple average $\tau_{ave}=(\tau+\tau_w)/2$ for the proper times,
where $\tau=\ln t$].

\begin{figure}
\epsfxsize=8.5cm
\epsfbox{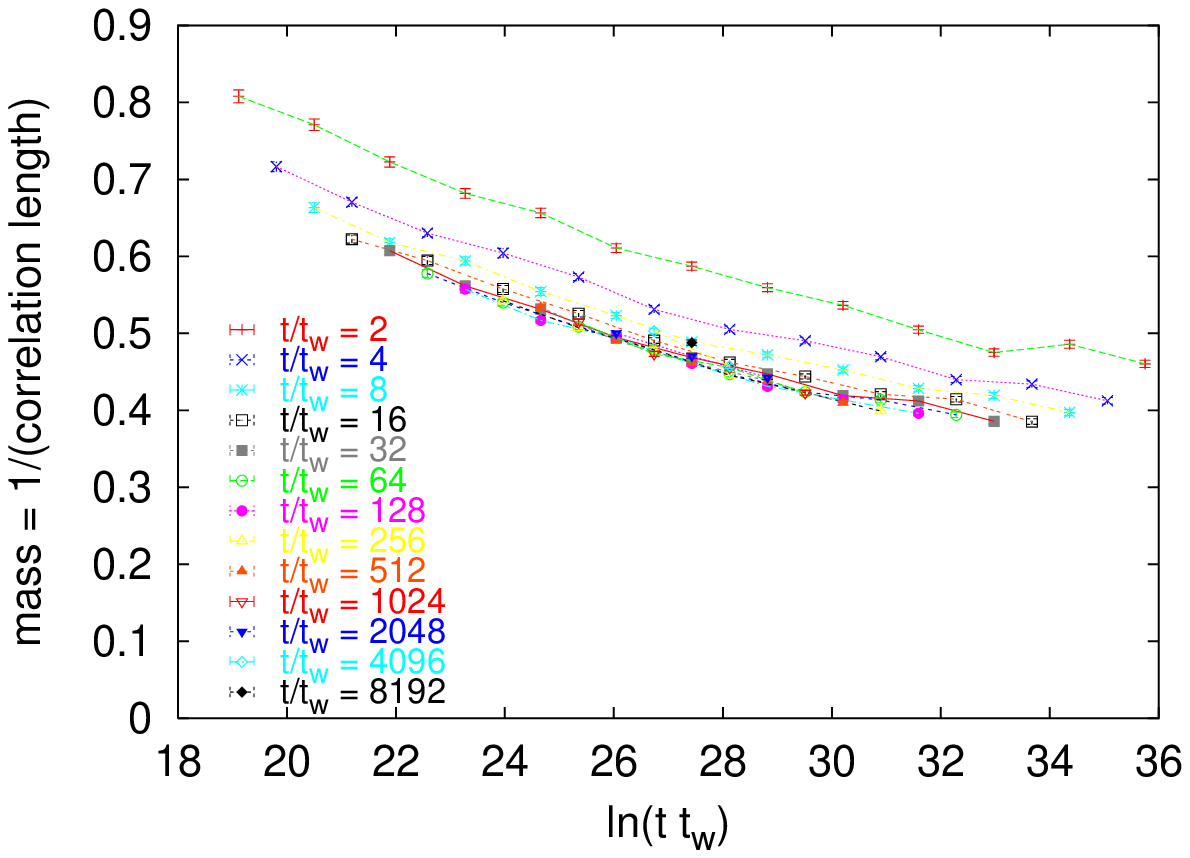}
\vspace{0.05cm}
\caption{Mass (inverse correlation length) as a function of the 
geometric average between $t$ and $t_w$. Each set
of points corresponds to a fixed ratio $t/t_w$. The horizontal axis
corresponds to $\ln(t t_w)$.}
\label{FIG:mass_vs_timeaverage}
\end{figure}

\begin{figure}
\epsfxsize=8.5cm
\epsfbox{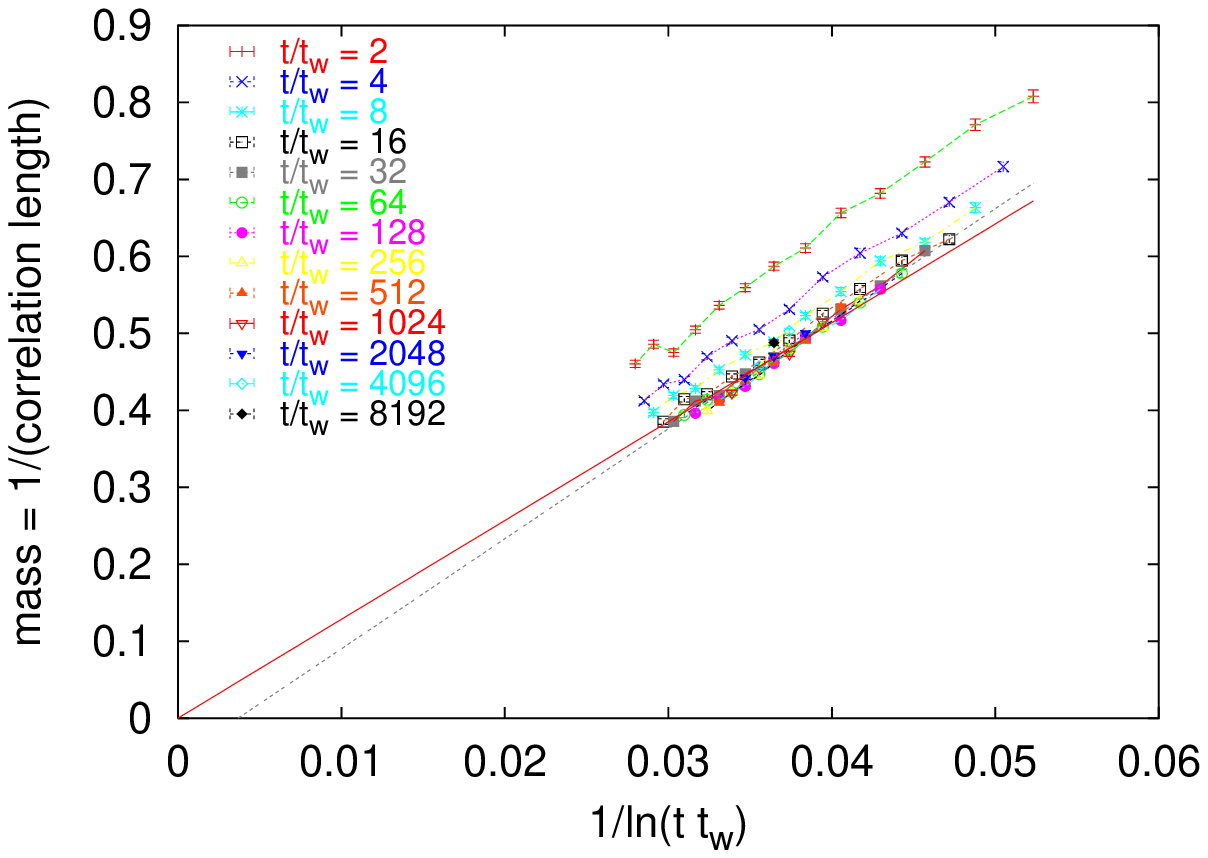}
\vspace{0.05cm}
\caption{Mass (inverse correlation length) as a function of the 
geometric average between $t$ and $t_w$. Each set
of points corresponds to a fixed ratio $t/t_w$. The horizontal axis
corresponds to $1/\ln(t t_w)$.}
\label{FIG:mass_vs_timeaverage_ln}
\end{figure}

With the above picture in mind, we now re-analyze the data in our simulations
by replotting the mass $m(t,t_w)$ as a function of $t t_w$ in 
Fig.~\ref{FIG:mass_vs_timeaverage}. 

\begin{figure}
\epsfxsize=8.5cm
\epsfbox{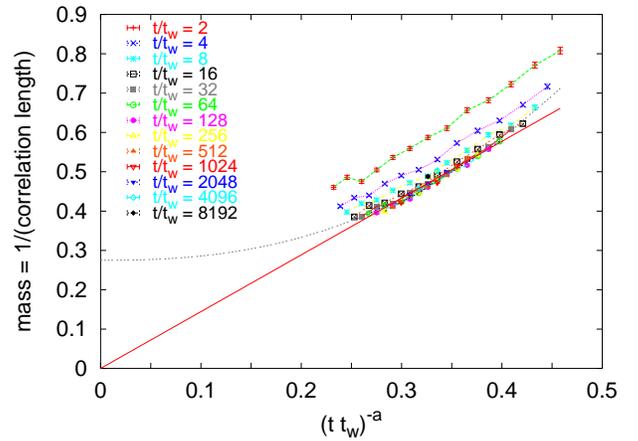}
\vspace{0.05cm}
\caption{Mass (inverse correlation length) as a function of the geometric
average between $t$ and $t_w$.  Each set of points corresponds to a fixed
ratio $t/t_w$. The horizontal axis corresponds to $(tt_w)^{-a}$ with $a=0.04$.}
\label{FIG:mass_vs_timeaverage_pow}
\end{figure}

As Fig.~\ref{FIG:mass_vs_timeaverage} shows, the curves for fixed
$t/t_w$ ratio eventually converge to an asymptotic curve for large
$t/t_w$. As discussed above, this is consistent with the subleading corrections
in Eq.~(\ref{eq:rf3a}) becoming negligible for large $t/t_w$
ratios. In Figs.~\ref{FIG:mass_vs_timeaverage_ln}
and~\ref{FIG:mass_vs_timeaverage_pow} we attempt to obtain more
detailed information about the time dependence of the mass. In
Fig.~\ref{FIG:mass_vs_timeaverage_ln} we show fits of $m(t,t_w)\equiv
1/\xi(t,t_w)$ to a $1/\ln(t t_w)$ dependence. If a nonzero value is
allowed for the mass at infinite time, the fitted value obtained is
negative, meaning that the mass actually goes to zero at very large,
but finite, times ($t t_w \sim 10^{276}$). In practice, this fit is as
good as a fit with the mass going to zero only at infinite time: in both
cases $\chi^2$ per degree of freedom is of order $10^{-4}$.

In 
Fig.~\ref{FIG:mass_vs_timeaverage_pow} we show the fit $m(t,t_w) \equiv
1/\xi(t,t_w)$ to a power law $(t t_w/t^2_0)^{-a}$. If a finite mass is
allowed at infinite times, an exponent $a \approx 0.10$ is obtained.
If, instead, the value of the mass is assumed to be zero at infinite
times, $a \approx 0.04 $ is obtained. The $\chi^2$ per degree of
freedom is again of the order of  $10^{-4}$ for both fits. 

In summary, the simulation results indicate that (i) $m(t,t_w)$
is asymptotically (for large $t$, $t_w$ and $t/t_w$) only a
function of $ t t_w $; (ii)  $m(t,t_w)$ is a very slow monotonous
decreasing function of $ t t_w $, consistent both with the $ m
\approx 1/\ln(t t_w)$ and with the $ m \approx (t t_w/t^2_0)^{-a}$
functional forms; (iii) the values of the correlation length $\xi=1/m$
are of the order of only a few lattice constants for the timescales
accessible in the simulations; (iv) the extrapolation of the data fits is
consistent with the expected $m \to 0$ behavior for long times, but a 
nonzero mass at infinite times cannot be excluded. 

Previous studies \cite{BerthierBouchaud2,Kisker} have found a correlation
lengthscale $\xi(t_w)$ growing with $t_w$ that is in the range of $3$ to $4$
lattice spacings for the waiting times and temperatures we have considered.
Those authors determined the correlation length from the spatial decay of
a four point correlation function \cite{Kisker}
\begin{eqnarray}
G(r,t_w) 
&=& 
\frac{1}{N}\sum_{i=1}^N \frac{1}{t_w}\sum_{\tau=t_w}^{2t_w - 1} 
\left[\langle
C^{ab}_{{\vec r}_i}(\tau) C^{ab}_{{\vec r}_i+\vec r}(\tau) 
\rangle
\right]
\; ,
\end{eqnarray}
with $C^{ab}_{{\vec r}_i}(\tau)\equiv 
s^a_{{\vec r}_i}(\tau) s^b_{{\vec r}_i}(\tau) $, 
involving thus two copies of the system $a$ and $b$, 
with identical exchanges, and an average over thermal histories and 
disorder. Note that this ``$2$-point correlator'' differs from ours in 
that (i) it is evaluated at equal times, 
(ii) there is a further average over the noise, and, more importantly,
(iii) the two-point functions that one
further correlates are evaluated between two copies of the system
evolving independently. The latter property implies that this length
can only be tested experimentally with indirect 
measurements~\cite{BerthierBouchaud,check-length}. 
Interesting enough, the length we 
define in this paper is, in principle, accessible directly with 
local experimental probes.
Even if representing
different quantities the values reached by the two correlation-lengths are
very short and very similar. 

\subsection{Interplay between the waiting-time,  linear coarse-graining
size and correlation-length}
\label{scaling-limit}

In this section we discuss the role played by the waiting and total
times, $t_w$ and $t$, the coarse-graining linear size, $2M+1$, and the
correlation length, $\xi(t,t_w)$, in our measurements. We assume that
the size of the system, $L$, is the largest scale in the problem that
has diverged at the outset of the discussion.

For a finite but very long waiting-time, ${t_w}_1$, and a time-window
$[{t_w}_1,t_1]$ on which one wishes to study the fluctuating dynamics,
the correlation length, $\xi(t_1,{t_w}_1)$ is finite but very long.
Choosing a cubic coarse-graining volume $V_1$ with linear size $2M_1+1
< \xi(t_1,{t_w}_1)$ one then accommodates
$[\xi(t_1,{t_w}_1)/(2M_1+1)]^d$ cells within each correlated volume
$[\xi(t_1,{t_w}_1)]^d$. According to the discussions in previous
sections, the local coarse-grained two-time functions, defined on each
cell, have different time-reparametrizations, $h_i(t)$, that vary
smoothly in real space until reaching the correlation length,
$\xi(t_1,{t_w}_1)$, when they completely decorrelate.  By taking a
large coarse-graining volume we ensure that the underlying effect of
disorder is erased. This kills the fluctuations in the external
function $f_i$, [see Eq.~(\ref{scaling1})] as well as erasing all
fluctuations in $q_{\sc ea}^i$.  The fluctuations in the local
time-reparametrizations give rise to a distribution of values of the
local coarse-grained correlations, $\rho(C_i^{cg}(t_1,{t_w}_1))$ that
we study numerically in Sec.~\ref{scaling-local-corr}.

For another finite, but longer waiting-time, ${t_w}_2 > {t_w}_1$, the
zero-mode discussed in Sec.~\ref{sigma} becomes flatter and one
expects the amount of fluctuations to increase.  If we also enlarge
the time-window $[{t_w}_2,t_2]$ to be analyzed in such a way that,
say, the global correlation, $C$, remains unchanged,
$C(t_1,{t_w}_1)=C(t_2,{t_w}_2)$, the new correlation length is longer
than the previous one, $\xi(t_2,{t_w}_2) >\xi(t_1,{t_w}_1)$ (see the
results in Sec.~\ref{mass-goldstone}). If we keep the same
coarse-graining linear size as before, {\it i.e.} $2M_1+1$, we expect
the level of fluctuations in the reparametrizations to increase and
the {\sc pdf} $\rho(C_i^{cg}(t_2,{t_w}_2))$ to be wider than
$\rho(C_i^{cg}(t_1,{t_w}_1))$.  If, simultaneously to increasing the
values of the times, we also increase the coarse-graining volume 
so as $(2M+1)$ scales with $\xi(t,t_w)$, we
should be able to maintain the amount of fluctuations. More precisely,
the {\sc pdf} computed with the new times, $(t_2,{t_w}_2)$, and the
new coarse-graining size $2M_2+1$, should be identical to the {\sc
pdf} computed with $(t_1,{t_w}_1)$ and $2M_1+1$.

This procedure can be taken further to postulate that a scaling limit
in which a stable distribution of fluctuations is reached.  This limit
is such that the waiting-time and the subsequent time go to infinity
together keeping the global correlation fixed to a prescribed
value ${\tt C}$ (note that this double limit can be more general than
the case in which $t_w$ and $t$ are proportional), and the
coarse-graining volume diverges together with
the correlation length between $t$ and $t_w$, {\it i.e.}
$\xi(t,t_w)/M$ is held fixed. If  in this 
limit the distribution of local two-time functions remains non-trivial, 
then dynamical heterogeneous regions of all sizes exist.

Even though for a system with an asymptotic reparametrization 
invariant action the correlation length is expected to diverge asymptotically,
in Sec.~\ref{mass-goldstone} we showed that $\xi(t,t_w)$ is still very short 
for the times accessible numerically in the spin-glass model that we 
study here. Thus, one cannot study the dynamics using 
a coarse-graining volume such that $a\ll 2M+1 \ll \xi(t,t_w)$
(with $a$ the lattice spacing). 

Another different, but also interesting, regime in which to study fluctuations
is the following. Let us take $t_w$ and $t$ very long but finite
leading to a finite correlation length $\xi(t,t_w)$. If one takes
the coarse-graining linear size to be of order $2M+1\sim
\xi(t,t_w)$ each block leading to a point in the construction of the
{\sc pdf}s is, roughly speaking, an independent model with finite size
$L=\xi(t,t_w)$.  The results in Sec.~\ref{finite-size-effects-3dea}
suggest that the fluctuations in the global two-time functions of
finite size systems have a very similar behavior to the local ones discussed
above.

Finally, if we use a coarse-graining 
volume with linear size $2M+1 \gg \xi(t,t_w)$, we do not expect to 
have any fluctuations and the {\sc pdf}s should be delta peaks
on single values.

\section{Scaling of local susceptibility}

In this section we study the dynamic behavior of the local 
integrated responses. First we select the magnitude of the 
applied perturbation by studying the behavior of the global 
susceptibility. Next we study the evolution of the 
probability distributions of coarse-grained and noise-averaged
quantities. 

\subsection{Choice of field strength}

\begin{figure}
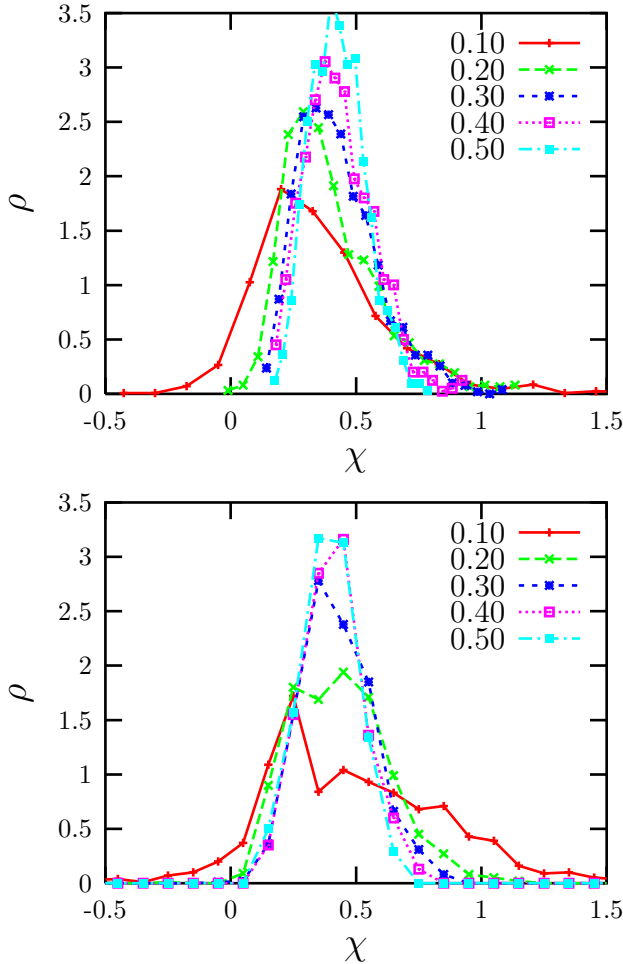

\begin{center}
\input{fields_field_aver.pslatex}
\input{fields_noise_aver.pslatex}
\end{center}
\vspace{0.15cm}
\caption{Distribution of {\it global} staggered 
susceptibilities for small $3$ {\sc ea} models. The field is applied at 
$t_w=10^4$ {\sc mc}s and held fixed subsequently. The integrated response is
measured at $t=4 \times 10^4$ {\sc mc}s. 
In the first panel we used one noise and 
$10^3$ field realizations; in the second panel 
we used one random field and $10^3$ noise realizations.
The curves correspond to different field strengths given in the key.
$L=8$, $T=0.7$ (noise averaged) and $T=0.8$ (field-averaged).}
\label{field-strength}
\end{figure}

We choose the strength of the perturbing field, $h$, 
in such a way that (i) there
are no negative global susceptibilities; (ii) there are no global 
susceptibilities with a magnitude larger than the value allowed
by {\sc fdt} $1/T$;
(iii) we select the range of field strengths such that linear response
holds, {\it i.e.} we find the maximum strength for which the 
distribution is stable. 
The first two conditions yield a lower bound on $h$, $h_{\sc min}$, the 
third determines its maximum possible value, $h_{\sc max}$.

In Fig.~\ref{field-strength} we show the outcome of these tests.
We plot the {\sc pdf} of the global staggered susceptibilities where we
used two different procedures to draw the histograms. In the first plot, 
we used $10^3$ realizations of a random field, $\eta_i=\eta \epsilon_i$ with 
$\epsilon_i=\pm 1$ with probability $\frac12$, and only one thermal noise
(note that perturbed and unperturbed copies are evolved with the same
thermal noise). We then draw the distribution of global values, 
\begin{equation}
\chi(t,t_w) \equiv \frac{1}{\eta} 
\sum_{i=1}^N [s_i^h(t) -s^0_i(t)] \; \epsilon^k_i 
\; ,
\label{field-aver}
\end{equation}
with one point per field realization, $k=1,\dots.\#_{fields}$. 
In the second case, we used only one random field realization and
we draw the distribution of global values, 
\begin{equation}
\chi(t,t_w) \equiv \frac{1}{\eta} 
\sum_{i=1}^N [{s_i^h}_k(t) -{s^0_i}_k(t)] \; \epsilon_i 
\; ,
\end{equation}
with one point per thermal noise realization, 
with $k=1,\dots,\#_{noises}$.
We see that for such a small system size the probability distributions 
$\rho(\chi)$ change quite a bit with the strength of the applied field
(even if the average and variance are quite stable, see Fig.~\ref{averaged-chi}
below). For larger system sizes, $L=16$, see Fig.~\ref{fieldsL16}, the 
probability distribution stabilizes for $\eta\geq \eta_{\sc min} \sim 0.2$.  

The plot in Fig.~\ref{averaged-chi} shows 
the average and the variance of the probability distribution
for systems of size $L=8$ and $L=16$ 
with $10^3$ noise realizations, and $L=8$ with $10^3$
field realizations. We confirm that $\eta_{\sc min}=0.2$ and 
we see that the behavior starts being non-linear at around $\eta_{\sc max}
\sim 0.7$. 

\begin{figure}
\begin{center}
\input{fieldsL16.pslatex}
\end{center}
\caption{Distribution of the global staggered susceptibilities for a 
$3d$ {\sc ea} model with 
linear size $L=16$ at $T=0.7$ using $10^3$ noise realizations
and one perturbing field with strength given in the key.
$t_w=10^4$ {\sc mc}s, $t=4\times 10^4$ {\sc mc}s and $\tau_t=t/10$ {\sc mc}s.}
\label{fieldsL16}
\begin{center}
\input{average.pslatex}
\end{center}
\vspace{0.15cm}
\caption{Average and variance of the distribution of 
global staggered  susceptibilities in the $3d$ {\sc ea} model
against the field strength. The parameters are as in 
Figs.~\ref{field-strength} and \ref{fieldsL16}.} 
\label{averaged-chi}
\end{figure}

We see that field strengths between $0.2$ and $0.7$ comply with our
criteria. In what follows we use $\eta=0.25$ and we focus 
on the noise-averaged and 
coarse-grained definitions of the local susceptibilities
given in Eqs.~(\ref{coarse-chi}) and (\ref{noise-averaged-chi})
eventually averaged over many field realizations.

\subsection{Time-dependent distribution functions}

\begin{figure}
\input{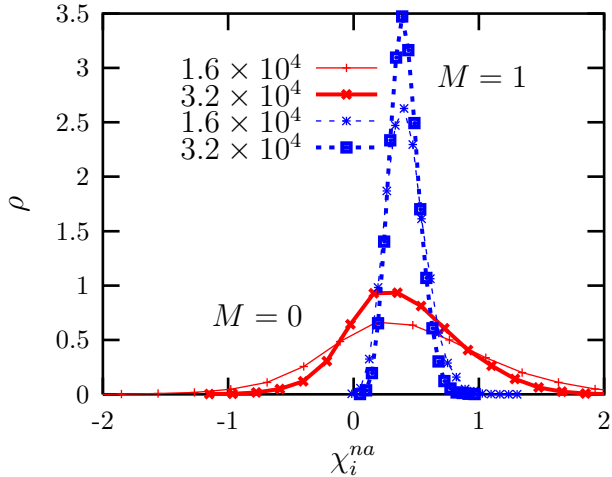}
\vspace{0.5cm}
\caption{Check of $t/t_w$ scaling of the distribution of 
noise-averaged local susceptibilities with no coarse-graining,
$M=0$ and with coarse-graining, $M=1$. The ratio is 
$t/t_w=2$. When $t_w=1.6\times 10^4$ {\sc mc}s we used $1200$ noise
realizations. When $t_w=3.2\times 10^4$ {\sc mc}s we used $200$ noise
realisations.}
\label{chi-ttw} 
\end{figure}

Here we study the evolution of the  
{\sc pdf} of local susceptibilities computed as in Eqs.~(\ref{coarse-chi})
and (\ref{noise-averaged-chi}). The figure shows that the $t/t_w$ scaling
is worse for the susceptibility than it is for the correlations.

\subsection{Coarse-grained data}

\begin{figure}
\begin{center}
\psfig{file=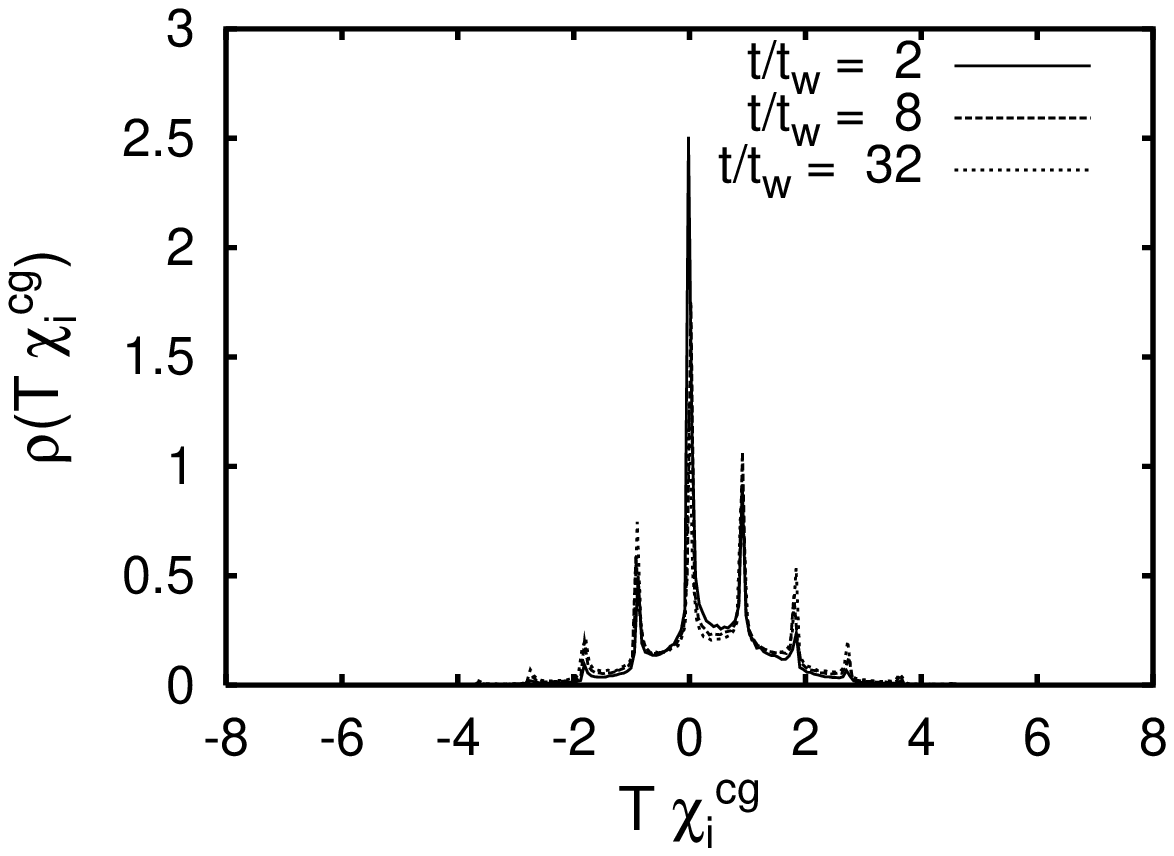,width=9cm}
\end{center}
\vspace{0.25cm}
\caption{Distribution of local coarse-grained staggered susceptibilities.
$L=32$, $T=0.8$ and $\eta=0.25$. 
The waiting times are $t_w=4\times 10^4$ {\sc mc}s, 
$1.6\times 10^4$ {\sc mc}s,  and $6.4 \times 10^4$ {\sc mc}s.
Three ratios are considered, $t/t_w=2,8,32$. The coarse-graining 
volume is $V=2M+1=1$.}
\label{distchi0}

\begin{center}
\psfig{file=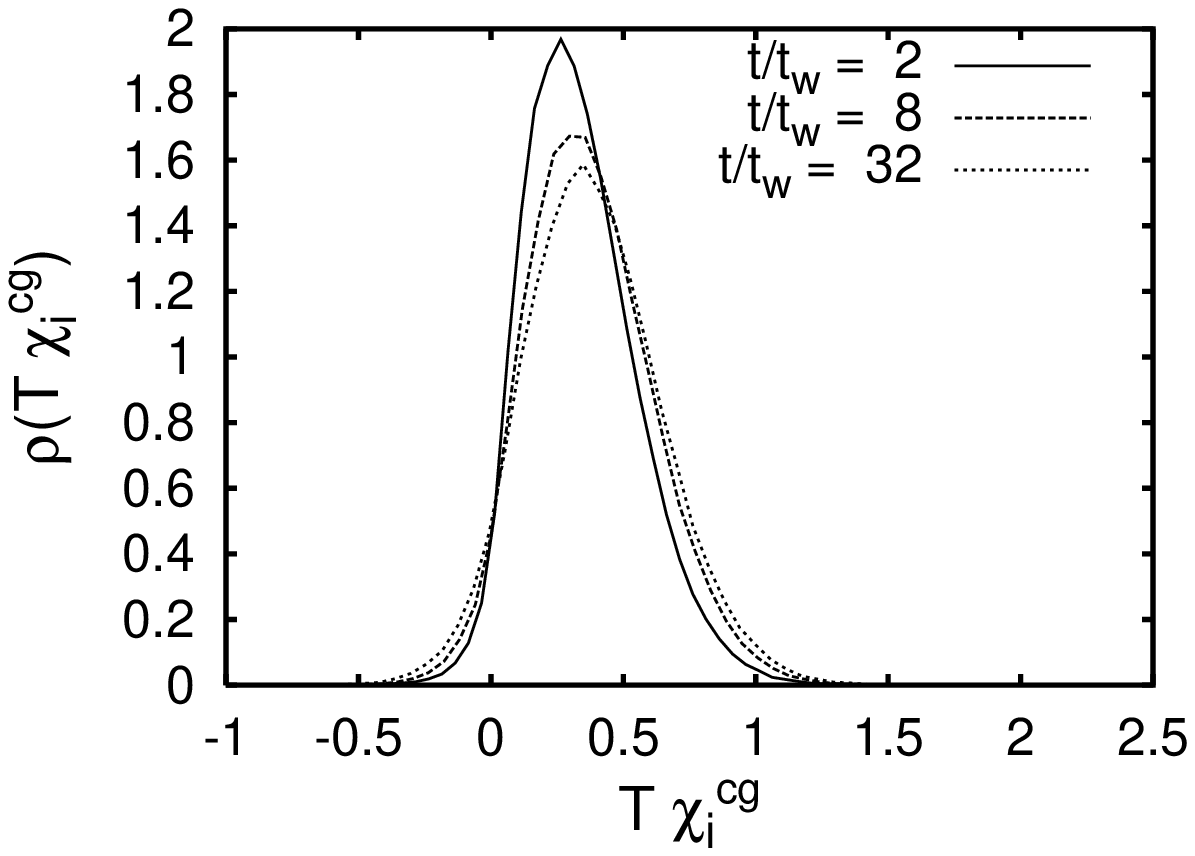,width=9cm}
\end{center}
\vspace{0.25cm}
\caption{Distribution of local coarse-grained staggered susceptibilities.
Same data as in Fig.~\ref{distchi0} with $M=1$.
}
\label{distchi1}
\end{figure}

\begin{figure}
\begin{center}
\psfig{file=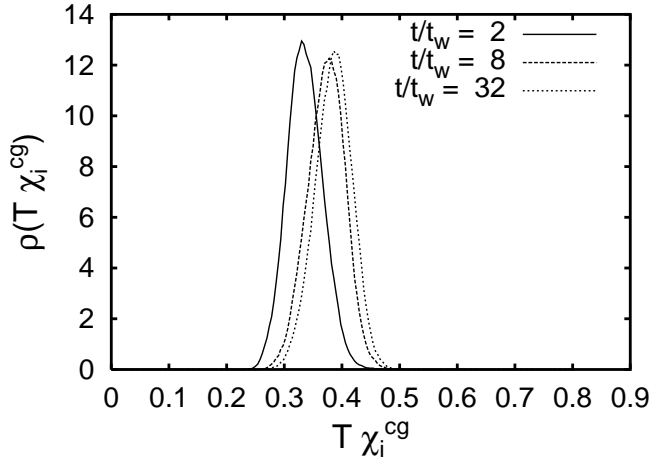,width=9cm}
\end{center}
\vspace{0.25cm}
\caption{Distribution of local coarse-grained staggered susceptibilities.
Same data as in Fig.~\ref{distchi0} with $M=6$.}
\label{distchi6}
\end{figure}

In Figs.~\ref{distchi0}-\ref{distchi6} we show the {\sc pdf} of 
local susceptibilities coarse-grained over 
different volumes. These figures are the counterpart of 
Figs.~\ref{distaverCr}-\ref{distaverCr3} 
where we showed the coarse-grained data for the
local correlations.

In Fig.~\ref{distchi0} $2M+1=1$, in 
Fig.~\ref{distchi1} $2M+1=3$, and in Fig.~\ref{distchi6} $2M+1=13$.
The strictly local susceptibilities shown in 
Fig.~\ref{distchi0}  are distributed almost in a discrete way, one sees
peaks at precise values of $\chi^{cg}_i$ that extend also
to the the negative  side of the axis. When $M=1$, 
the negative suseptibilities have not yet 
disappeared and the positive tail goes also beyond $T\chi^{cg}_i=1$,
see Fig.~\ref{distchi1}. Finally, when $M=6$ the {\sc pdf}s 
are much narrower with no support on negative values nor 
values that go beyond $T\chi_i^{cg}=1$. Note, however, that the
linear size of this coarse-graining volume, $2M+1=13$, 
is larger than the maximum correlation length reached with these 
times, say $\xi(t,t_w) \sim 5$ (see Sec.~\ref{correlation-length}).  

\section{The local fluctuation-dissipation relation}

Which of the possibilities listed in Sec.~\ref{local-resp-fdr}  
does the local {\sc fdr} follow? 
The aim of this section is to show numerical evidence for the 
following statements:

(i) The local coarse-grained two-time correlations and responses are
constrained to follow the global curve $\tilde \chi(C)$: the
dispersion along the $\tilde \chi(C)$ curve is much more important
than the dispersion perpendicular to this curve. For fixed $t_w$ and
increasing values of $t$ the dispersion along the $\tilde \chi(C)$
curve increases.  This is as predicted by the sigma-model argument of
Sec.~\ref{sigma}.  Equation~(\ref{localTidep}) holds for these
functions.

(ii) The local noise-averaged correlations do not necessarily follow
the global curve. Coarse-graining this data does not lead to
concentrating the distribution of $(C_i,T \chi_i)$ pairs around the
global curve. Equation~(\ref{localTidep}) does not hold. It is hard to
test numerically if Eq.~(\ref{localTiindep}) holds since there is no
constraint on the location of the pairs in the $C - T\chi$ plane.

(iii) The mesoscopic fluctuations (with no average over the noise)
behave as in (i). 

We provide numerical evidence for these statements by plotting the
joint probability distribution of pairs $(C_i,T\chi_i)$ evaluated at a
pair of times $(t_w,t)$, and its projection on the $2d$ plane
$C - T\chi$.

\subsection{Coarse-grained two-time functions}

\begin{figure}
\vspace{-2cm}
\begin{center}
\psfig{file=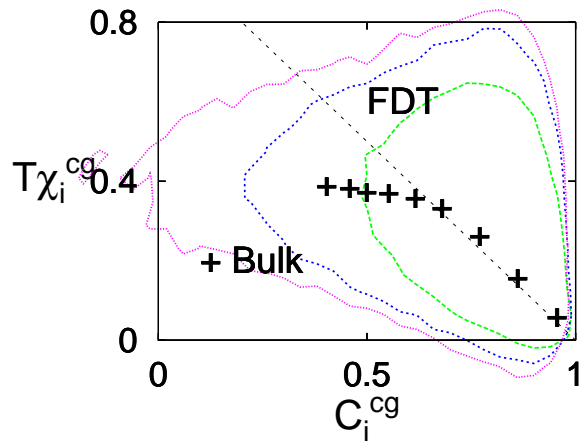,width=11cm}
\end{center}
\vspace{-1.cm}
\caption{Projection of the joint probability distribution of the
coarse-grained local susceptibilities and correlations. $L=32$,
$T=0.8$ and $M=1$ leading to $2M+1\sim \xi(t,t_w)$. 
The contour levels surround $66\%$ of the
weight of the distribution and each of them is drawn using a different
ratio between the two times. From inside to outside
$t/t_w=2,8,32$.}
\label{fig:coarse-grained-chiC}
\end{figure}

In Fig.~\ref{fig:coarse-grained-chiC}
we plot the projection of the   joint probability distribution of 
coarse-grained two-time functions $(C_i^{cg}, T\chi_i^{cg})$
on the $C - T\chi$ plane. 
The data correspond to a system of linear size 
$L=64$ at $T=0.8$. The coarse-graining linear size is $2M+1=3$
that is of the order of the correlation length $\xi(t,t_w)$.
We study simultaneously three ratios of times, $t/t_w=2,8,32$,
for $t_w=10^4$ {\sc mc}s in all cases. 
The contour levels are such that they include $66\%$ of the 
weight of the joint {\sc pdf} and they correspond to 
$t/t_w=2,8,32$ from inside to outside. Reasonably, as the times
get more separated the values of the local correlations vary more
and the region incircled by the contour level is wider.
This plot shows that these distributions are very wide, as also 
indicated by Figs.~\ref{distaverCr2}~and~\ref{distchi6} where we show 
$\rho(C_i)$ and $\rho(\chi_i)$, independently.
Even though the contour levels are also wide in the $T\chi$ direction,
they tilt in the horizontal direction of $C$, along the curve 
made by the crosses, that corresponds to $T\tilde\chi(C)$ for 
this $t_w$ and 10 values of $t$.  

\begin{figure}
\vspace{-2.5cm}
\begin{center}
\psfig{file=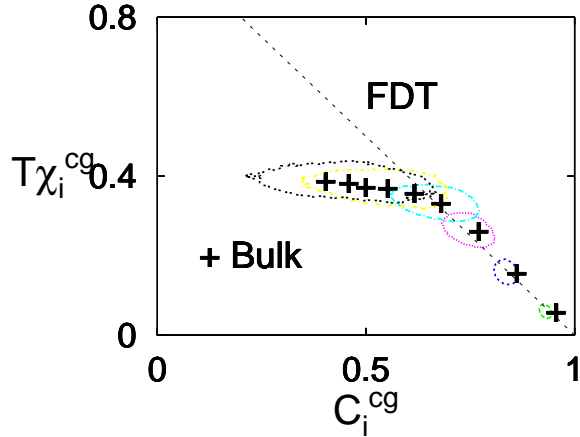,width=11cm}
\end{center}
\vspace{-1.cm}
\caption{Projection of the 
joint probability distribution of the coarse-grained 
local susceptibilities and correlations. $L=32$, $T=0.8$ and 
$M=6$. Note that  here $2M+1 \gg \xi(t,t_w)$.
The contour levels surround $66\%$ of the
weight of the distribution and each of them is drawn using a different
ratio between the two times with $t/t_w$ increasing from right to left.
}
\label{fig:furthercoarse-grained-chiC}
\end{figure}

In Fig.~\ref{fig:furthercoarse-grained-chiC} 
we see the effect of further coarse-graining 
the data corresponding to the same $t_w$ and $t$ as in 
Figs.~\ref{fig:coarse-grained-chiC}. 
In the projected plot we see how the  size of the
cloud around the $T \tilde \chi(C)$ plot is reduced. The longitudinal 
fluctuations that correspond to fluctuations in the function 
$f_i$ that characterizes the local correlations
are killed very quickly  by the coarse-graining
(see the theoretical background for this in Sec.~\ref{sigma} and 
the related effect in the values of the local correlation in 
Sec.~\ref{section:scaling}). The transverse fluctuations are also reduced
but in a weaker manner. These are related to the fluctuations 
in the local reparametrizations and, as we argued in Secs.~\ref{sigma}
and \ref{section:scaling}, they should survive in the limit of 
long times and large coarse-graining volumes. See 
Sec.~\ref{finite-size-effects}
for similar results for the joint {\sc pdf}s of the 
global quantities computed using sytems of small size.

\subsection{Noise-averaged two-time functions}
\label{noise-averaged-3dea}

\begin{figure}
\begin{center}
\vspace{-4cm}
\psfig{file=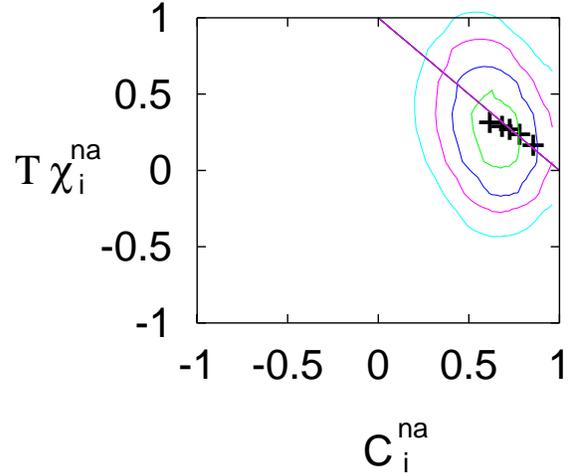,width=10cm}
\end{center}
\vspace{-1cm}
\caption{The projection of the joint {\sc pdf} on the $C - T\chi$
plane, noise averaged quantities, $L=32$, $t_w=1.6\times 10^4$ {\sc mc}s
and $t=4.8 \times 10^4$ {\sc mc}s. The {\sc fdt} prediction 
is represented with a straight line.}
\label{projM0}
\end{figure}

In Fig.~\ref{projM0} we show the joint probability distribution of the
noise-averaged local correlations and susceptibilities.  We see that
the distribution is concentrated about the {\sc fdt} line and does not
bend in the direction of the non-trivial part of the global $\tilde
\chi(C)$ as the coarse-grained data does, see
Fig.~\ref{fig:coarse-grained-chiC}.  This indicates that the disorder
induced fluctuations are not controlled by the sigma model argument of
Sec.~\ref{sigma}. Note that with no coarse-graining the
Edwards-Anderson parameter fluctuates from site-to-site and hence the
value of the correlation at which the sites enter the slow scale
varies. This result resembles 
Fig.~\ref{fig:sketch-chiC1step2} or Fig.~\ref{sketch-chiCthem}
only that the extent of the fluctuations in $C_i^{na}$ is narrower 
than in these sketches.

In Sec.~\ref{finite-size-effects} we shall
observe a very similar behavior of the noise-averaged 
mesoscopic fluctuations in finite size {\sc sk} models.

\subsection{Effect of coarse-graining on already noise-averaged
data}

\begin{figure}
\begin{center}
\psfig{file=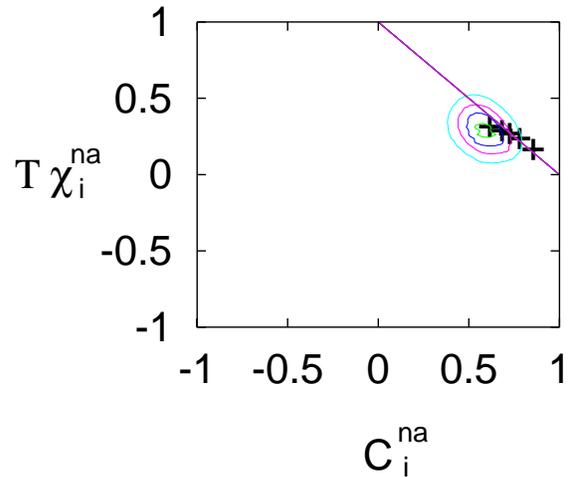,width=10cm}
\end{center}
\vspace{-1cm}
\caption{Same as in the previous figure, the  coarse-graining
volume has with linear size $M=1$.}
\label{projM1}
\end{figure}

\begin{figure}
\begin{center}
\psfig{file=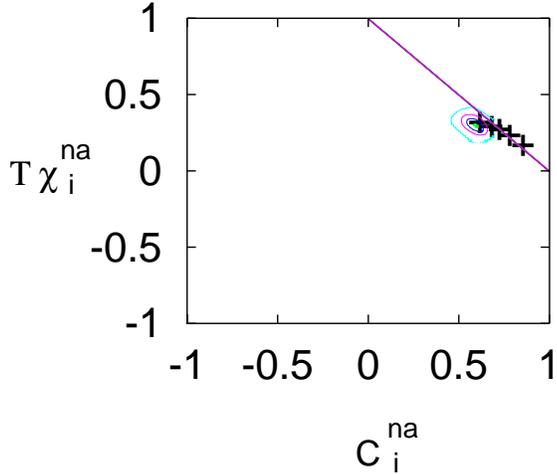,width=10cm}
\end{center}
\vspace{-0.5cm}
\caption{Same as in the previous two figure, the coarse-graining
has linear size $M=3$.}
\label{projM4}
\end{figure}

In Figs.~\ref{projM1} and \ref{projM4} we test the effect of 
coarse-graining on the joint probability distribution of 
noise-averaged local correlations and susceptibilities.
In Fig.~\ref{projM1} the linear size of the coarse-graining
box is $2M+1=3$ while in Fig.~\ref{projM4} it is 
$2M+1=7$. The system size is $L=32$ and $T=0.7$. We see 
that in all cases the distribution follows the {\sc fdt}
line. Indeed, averaging over the noise kills 
all thermal fluctuations and hence do not allow for 
the fluctuations in the time-reparametrization 
that cause the bending of the data 
along the $\tilde \chi(C)$ curve. This is consistent with the 
results in Fig.~\ref{coarse-ontop} where we checked the effect of 
coarse-graining the already noise-averaged local correlations.

\section{Finite size systems}
\label{finite-size-effects}

A. Barrat and Berthier~\cite{BarratBerthier}  showed that the 
modification of the {\sc fdt} in a $3d$ {\sc ea} model of finite
size that is evolving out of equilibrium (that is to say for 
$t$ and $t_w$ that are well below the finite equilibration time 
$t_{\sc eq}$),
has a very similar behavior to the one found in the thermodynamic limit. 
The system will eventually equilibrate but, as long as one keeps the 
times to be shorter than $t_{\sc eq}$, the global dynamics is very much the 
typical one of an out of equilibrium system. 

In this section we study the time-dependence of the 
distributions of the global correlation and 
susceptibility for finite-size samples of both finite-$d$ and infinite-$d$ 
spin-glasses. We show that the ``mesoscopic'' fluctuations, {\it i.e.} the 
fluctuations of the global quantities due to finite size effects, 
behave very similarly to the local coarse-grained ones in finite 
$d$. Moreover, we test the relation between the fluctuations in the 
susceptibility and the correlation and we find that they are also constrained
to follow the global parametric $\tilde \chi(C)$. 

\subsection{The $3d$ {\sc ea} model} 
\label{finite-size-effects-3dea}

In Figs.~\ref{dist-globalC-finiteN} and \ref{dist-globalchi-finiteN}
we show the evolution of the distributions of
the global correlation, $C(t,t_w)$,  and integrated response,
$\chi(t,t_w)$, with time $t$ and for fixed $t_w$, respectively. 
We construct the distribution functions using
one data point for the global correlation and linear 
integrated response of the full system obtained using one 
noise realization. 
The details of the system are given in the caption and key.  
We use $L=8$ so as to be able to access smaller values of the global
correlation than is possible in larger size systems. This small 
system size also allows us to have {\sc pdf}s that are rather wide.

\begin{figure}
\begin{center}
\input{dist-globalC-finiteN.pslatex}
\vspace{0.5cm}
\caption{Evolution of the 
{\sc pdf} of the global correlation for a system with linear size 
$L=8$ at $T=0.7$. The waiting-time is $t_w=10^4$ {\sc mc}s 
and the different values 
of the subsequent time $t$ used in each curve are given in the key.}
\label{dist-globalC-finiteN}
\vspace{0.5cm}
\input{dist-globalchi-finiteN.pslatex}
\vspace{0.5cm}
\caption{Evolution of the 
{\sc pdf} of the global susceptibility for the same system as above.
$\eta=0.3$.}
\label{dist-globalchi-finiteN}
\end{center}
\end{figure}

\begin{figure}
\begin{center}
\vspace{-2cm}
\psfig{file=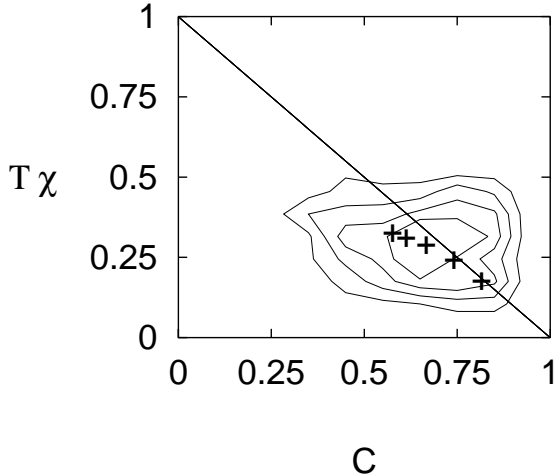,width=10cm}
\end{center}
\vspace{-0.75cm}
\caption{Projection of the joint probability distribution of the global 
correlations and integrated responses for a $3d$ {\sc ea} model with 
linear size $L=8$ at $T=0.7$. The strength of the 
applied field is $\eta=0.25$. The contour levels correspond to the 
joint {\sc pdf} for the waiting-time $t_w=10^4$ 
{\sc mc}s and $t/t_w=4$. One realization of the 
random exchanges and $10^3$ thermal histories are 
used to construct the {\sc pdf}. $\tau_t=t/10$ {\sc mc}s.
The straight line indicates the {\sc fdt} prediction and the 
crosses indicate the time evolution of the average $\chi(C)$.}
\label{dist-globalCchi-finiteN}
\end{figure}

In Figs.~\ref{dist-globalC-finiteN} and \ref{dist-globalchi-finiteN}
we display the evolution in time  
of the {\sc pdf}s for the global correlation and 
the global suseptibility of a system with $L=8$. We see that the 
distributions get wider as the separation between the times
increases. In Fig.~\ref{dist-globalCchi-finiteN}
we correlate the fluctuations in the global correlation and suseptibility.
We see that, as in the study of the local coarse-grained quantities
in a larger system, the contour levels of the joint {\sc pdf} are 
tilted in the direction of the global $\tilde \chi(C)$ that is indicated
with crosses.

\subsection{The {\sc sk} model}
\label{finite-size-effects-sk}

With the aim of giving additional support to our claim
that noise-averaged and coarse-grained dynamic fluctuations
behave differently we analysed the 
dynamic fluctuations in the fully connected {\sc sk} model on a finite 
size lattice.

The {\sc sk} model is defined in Eq.~(\ref{sk}), where the
couplings $J_{ij}$ connect all sites and are chosen randomly from a
bimodal distribution.  An infinite size system undergoes a
thermodynamic and dynamic phase transition at $T_c=1$.  When
$N\to\infty$ the dynamics in the low temperature phase was solved
analytically in the asymptotic limit of long-times (although finite with
respect to the size of the system)~\cite{Cuku2}. The solution has a
rather peculiar structure with the relaxation taking place in a
sequence of  hierarchically organized correlation scales.  These are
intimately related to the {\sc fdr}, $\tilde \chi(C)$, that takes a curved
form for values of the correlation that fall below the
Edwards-Anderson order parameter.

Several studies of the growth of the equilibration time with the 
size of the system, $N$, indicate that this increases approximately
as $t_{\sc eq} \propto e^{c N^{\alpha(T)}}$~\cite{sk-numerics1}
with $c$ a numerical constant and  the exponent $\alpha(T)$ 
increasing from $1/3$ at $T_c$ to $0.5$ at 
$T=0.4 \, T_c$ (see also Ref.~\cite{sk-teq}). Even though these studies used
a Gaussian distribution of couplings we take these results as an 
indication that even for small samples, {\it e.g.} $N=512$, we have a
very large time-window with nonequilibrium effects before equilibration
takes place. (We chose to use $N=512$ just to have the same number 
of spins as for the $3d$ {\sc ea} with $L=8$ used
in Sec.~\ref{finite-size-effects-3dea}.)

Numerical studies have found results in agreement with the 
analytic prediction of there being a sequence of global correlation
scales and a curve {\sc fdr} out of equilibrium~\cite{sk-analytics}.
Exactly how the scaling laws and 
the $\tilde\chi(C)$ prediction are modified due to finite size effects has
not been carefully investigated.

In this section we present results from a numerical simulation 
of the {\sc sk} model with bimodal interactions
using {\sc mc} dynamics with the heat bath algorithm at
$T=0.4$~\cite{Opper}. 
We pay special attention to the fluctuations induced 
by the finite size of the systems. 

\subsubsection{Finite size fluctuations of global quantities}

The natural counterpart to the coarse-grained local correlations
and responses in finite dimensional models is, for a fully 
connected model, the global quantity itself. The latter
fluctuates if the fully-connected system has a finite size.

\begin{figure}
\centerline{
\psfig{file=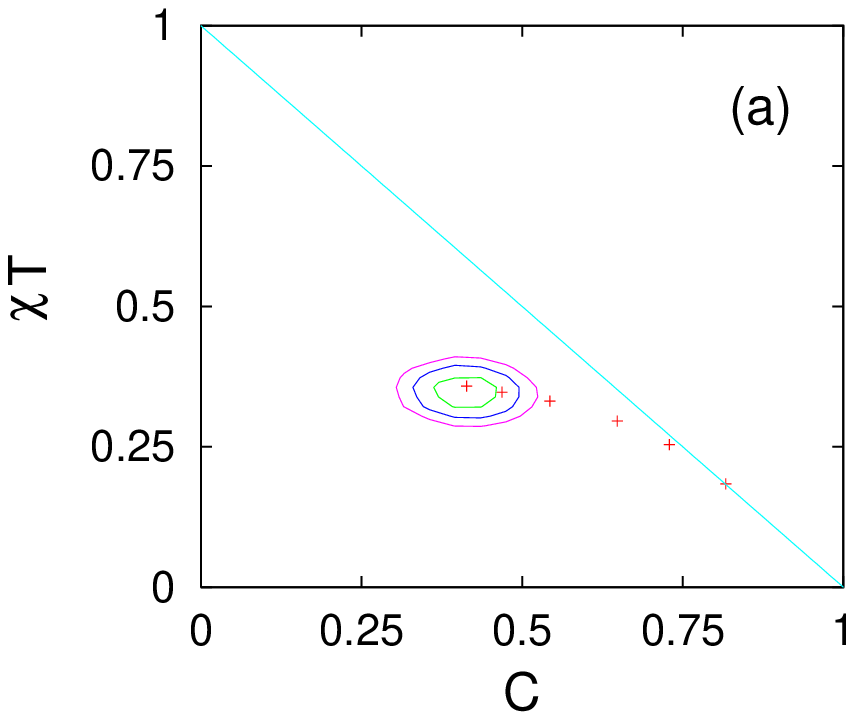,width=8cm}
}
\centerline{
\psfig{file=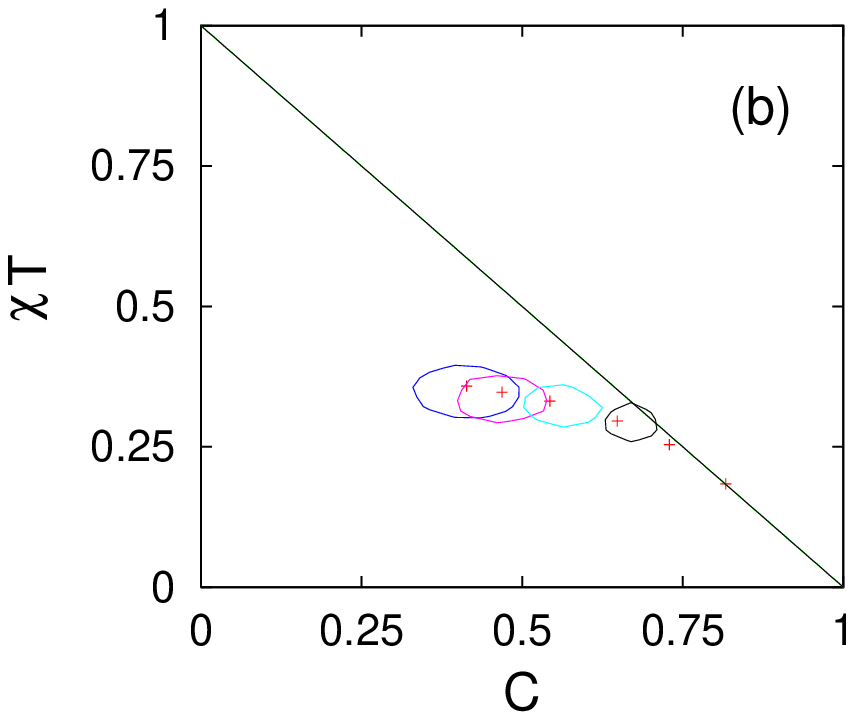,width=8cm}
}
\vspace{0.25cm}
\caption{Projection of the joint {\sc pdf} for the global
susceptibility and correlations of the {\sc sk} model with $N=512$ and
$\beta=2.5$. The strength of the applied field is $\eta=0.25$. The
coarse-graining over time is done using $\tau=2$ for $t_w=64$ {\sc
mc}s and $t=65,70$ {\sc mc}s, and $\tau_t=2$, $4$, $8$, and $16$ {\sc mc}s
for $t=128$, $256$, $512$,and $1024$ {\sc mc}s.  The crosses indicate values
averaged over the distribution, the straight line is the prediction
from the {\sc fdt}. In panel a) the contour levels are chosen at
heights corresponding to $95\%$, $90\%$, and 82\% of the maximum
in the {\sc pdf} for the global correlations evaluated at $t_w=64$
{\sc mc}s and $t=1024$ {\sc mc}s. In panel
b) the contour levels  are at $90\%$ of the maximum and they correspond
to the {\sc pdf}s calculated at $t_w=64$ {\sc mc}s and $t=128,256$,
$512,1024$ {\sc mc}s from right to left.}
\label{fig:sk-finite-size}
\end{figure}

In Fig. \ref{fig:sk-finite-size} we display the joint {\sc pdf} for
the global susceptibility and global correlation. We used $t_w=64$
{\sc mc}s and we evolved the systems until $t=1024$ {\sc mc}s.  We
constructed the distribution functions using $10^5$ pairs
$(C(t,t_w),\chi(t,t_w))$ calculated as follows. First we chose the
value of the total time $t$ with which to calculate the global
quantities, {\it e.g.} $t=128$ {\sc mc}s.  For a fixed realization of
the random exchanges we used $10^4$ noise histories and thus obtained
$10^4$ points. We repeated this procedure with $10$ different
realizations of disorder completing the set of $10^5$ data
points. With this data we obtained a probability distribution.  The
crosses represent the average over the $10^5$ points for $t=65$, 70,
128, 256, 512, and 1024 {\sc mc}s.  In Fig. \ref{fig:sk-finite-size}
we show the projection on the $C - T\chi$ plane of: three contour
levels at 95 \%, 90 \%, and 82 \% from the maximum of the joint {\sc
pdf} calculated for $t_w=64$ {\sc mc}s and $t=1024$ {\sc mc}s [panel
(a)]; one contour level at 90\% of the maximum of the joint {\sc pdf}
for four values of the total time, $t=128$, 256, 512, and 1024 {\sc
mc}s [panel (b)].  The straight 
represents the expected equilibrium curve.

We see that, similarly to
the local coarse-grained fluctuations for the $3d$ {\sc ea} model,
the distribution follows the global 
$T \tilde \chi(C)$ curve. The contour levels 
are inclined in the direction of the global curve.
Thus, despite having
very different time-scalings, these two models have very similar
parametric joint {\sc pdf} distributions.

\subsubsection{Fluctuations in the noise averaged local quantities}

\begin{figure}
\centerline{
\psfig{file=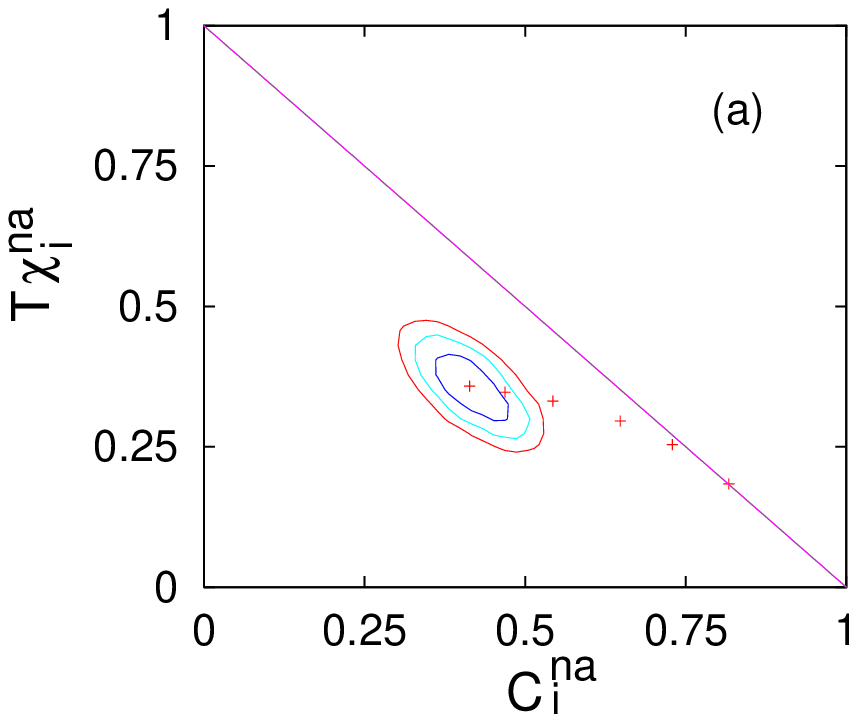,width=8cm}
}
\centerline{
\psfig{file=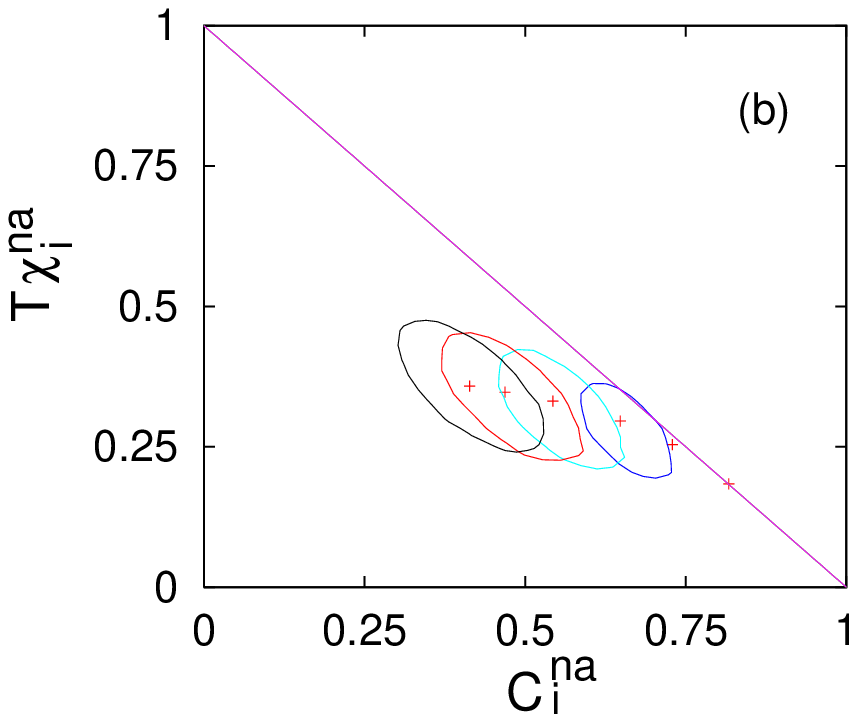,width=8cm}
}
\vspace{0.25cm}
\caption{Projection of the 
joint {\sc pdf} for the noise-averaged
``local'' susceptibilities and correlations
of the {\sc sk} model with $N=512$ and $\beta=2.5$. The strength of the 
applied field is $\eta=0.125$. The coarse-graining times $\tau$ are 
chosen as in Fig.~\ref{fig:sk-finite-size}. The crosses indicate 
values averaged over the distribution, the straight line
is the prediction from the {\sc fdt}. In panel a) the contour levels are chosen
at heights corresponding to $90\%, 85\%, 80\%$ and they correspond
to times $t_w=64$ {\sc mc}s and $t=1024$ {\sc mc}s. In panel b) the contour
levels  are at $80\%$ and they correspond to the joint 
{\sc pdf} at $t_w=64$ {\sc mc}s and $t=1024$ {\sc mc}s.}
\label{fig:sk-noise-averaged}
\end{figure}

In this section we follow a similar path to the one described in 
Sec.~\ref{noise-averaged-3dea} for the $3d$ {\sc ea} model. 
We simulated a {\sc sk} model with $N=512$ spins at $\beta = 2.5$. 
The waiting-time chosen was $t_w=64$ {\sc mc}s.
We averaged the spin-spin ``local'' 
self-correlation and integrated self-response
for chosen pairs of times $t,t_w$ with $t=128, 256, 512$, and $1024$ {\sc mc}s 
over $1600$ different noise realizations. With one realization of the 
random exchanges we thus obtained $N$ data points. To improve the 
statistics we repeated this procedure using $150$  different 
{\sc sk} models of the same size, {\it i.e.} with different realizations of
the coupling strengths. Thus, the {\sc pdf}s are constructed with 
$76800$ datapoints as is shown in Fig.~\ref{fig:sk-noise-averaged} 
(the crosses are the averaged values).
The qualitative form of the distribution is very different 
from the one in Fig.~\ref{fig:sk-finite-size}. The orientation 
of the contour levels does not follow the $T\tilde\chi(C)$ curve but, 
instead, it is approximately parallel to the {\sc fdt} straight line.
 
These results are similar to the ones displayed in
Sec.~\ref{noise-averaged-3dea} for the joint {\sc pdf} of the local
noise-avareged correlations and integrated responses in the $3d$ {\sc
ea} model.

\subsubsection{Effect of partial noise averaging}

\begin{figure}
\centerline{
\psfig{file=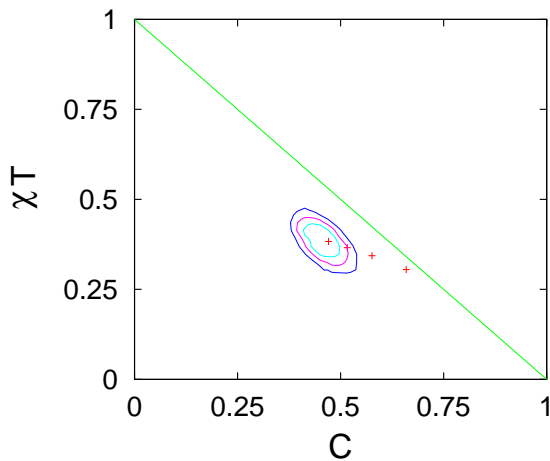,width=8cm}
}
\vspace{0.25cm}
\caption{Projection of the joint {\sc pdf} for the global
susceptibility and correlations of the {\sc sk} model with $N=128$,
$\beta=2.5$ and $\eta=0.1$. The global two-point functions for each
system have been averaged over $10^2$ noise realizations.  The
coarse-graining times $\tau$ are chosen as in
Fig.~\ref{fig:sk-finite-size}. The crosses indicate values averaged
over the distribution, the straight line is the prediction from the
{\sc fdt} and the contour levels are chosen at heights corresponding
to $80\%, 70\%$ and $60\%$ of the maximum in the {\sc pdf} evaluated
at $t_w=64$ {\sc mc}s and $t=1024$ {\sc mc}s.}
\label{fig:sk-partial-noise}
\end{figure}

Finally, we studied the effect of partial averaging over the noise the
global two-time functions in a still smaller system. By this we mean
that we averaged over $10^2$ realizations of the thermal history the
global correlation and integrated response of each of $6 \times 10^4$
{\sc sk} models with $N=128$ spins.  Thus, we constructed the joint
{\sc pdf} with $6 \times 10^4$ points. The result is displayed in
Fig.~\ref{fig:sk-partial-noise}. We see that averaging over the noise
we destroy the behavior in Fig.~\ref{fig:sk-finite-size}: the contour
levels are tilted toward the direction of the {\sc fdt} line. These
results approach the ones in Fig.~\ref{fig:sk-noise-averaged}.

\section{Geometric properties}
\label{geometric}

We have analyzed the local correlations and responses in terms of
their {\sc pdf}s and by looking directly at plots for their spatial
fluctuations along a plane or line of spins. An alternate way to
extract information about the spatial structure in spin glasses is
through an analysis of geometric properties, such as the fractal
properties of clusters of spins.

The clusters of spins that we have chosen to study are defined in a way that
we believe makes close contact with possible experiments with local probes.
The usual definition of a cluster in a spin glass is in terms of
spins belonging to different ground states -- since we have not calculated
the ground states in our simulations, such a definition is clearly not
available to us.  The definition we have chosen for clusters is as follows:
for a particular correlation ${\tt C}$, the cluster consists of all connected
spins with correlation $C_i$ in the interval $[{\tt C},{\tt C} + d{\tt C}]$
for a specified $d{\tt C}$.  This definition is used for both coarse-grained
and noise-averaged correlators. A similar approach could be easily
implemented to analyze experimental data on local noise in mesoscopic regions
in supercooled liquids and
glasses~\cite{heterogeneities,confocal,confocal2,dyn-heterogeneities}.

\begin{figure}
\psfig{file=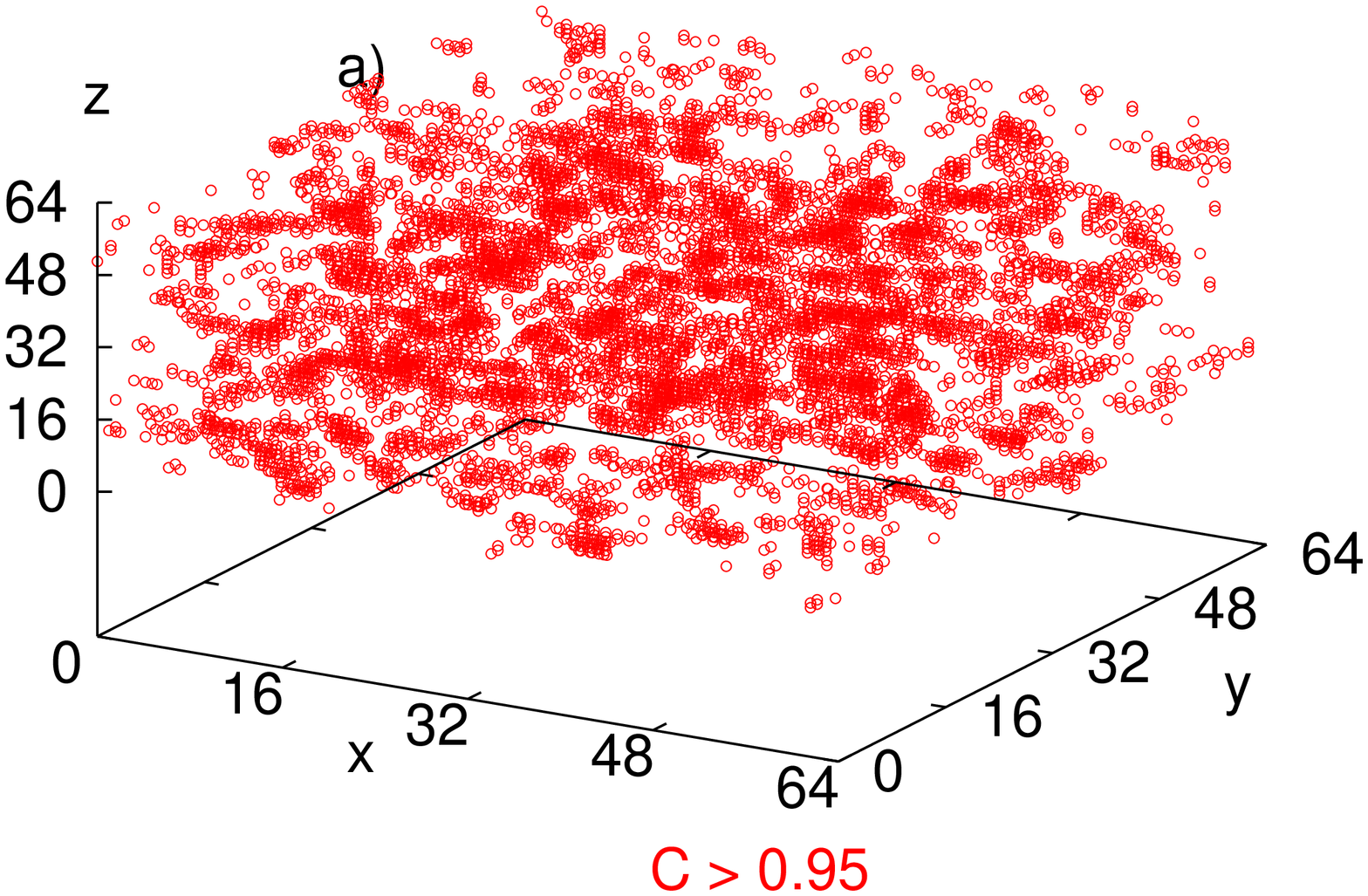,width=6cm,angle=-90}
\psfig{file=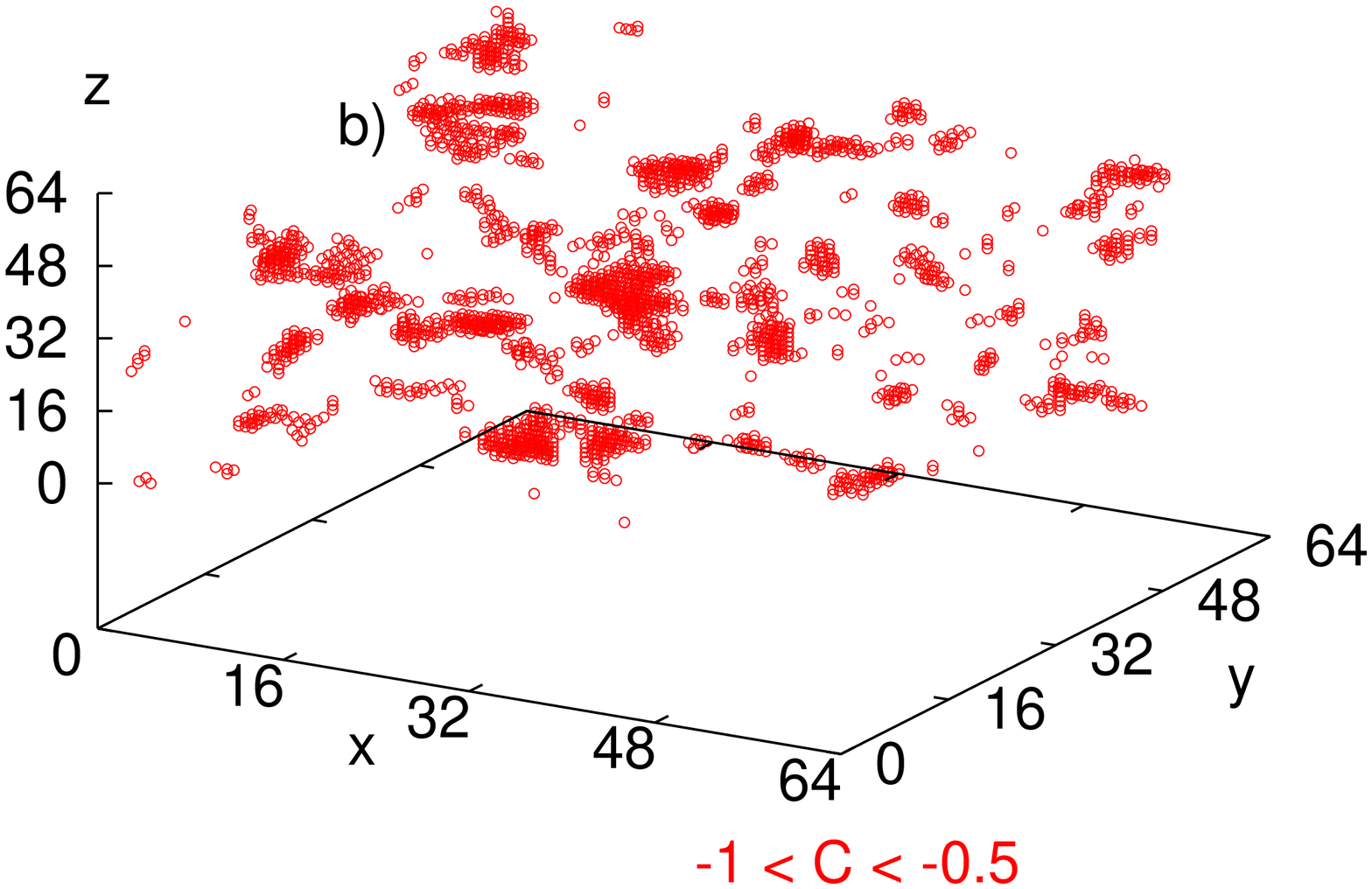,width=6cm,angle=-90}
\psfig{file=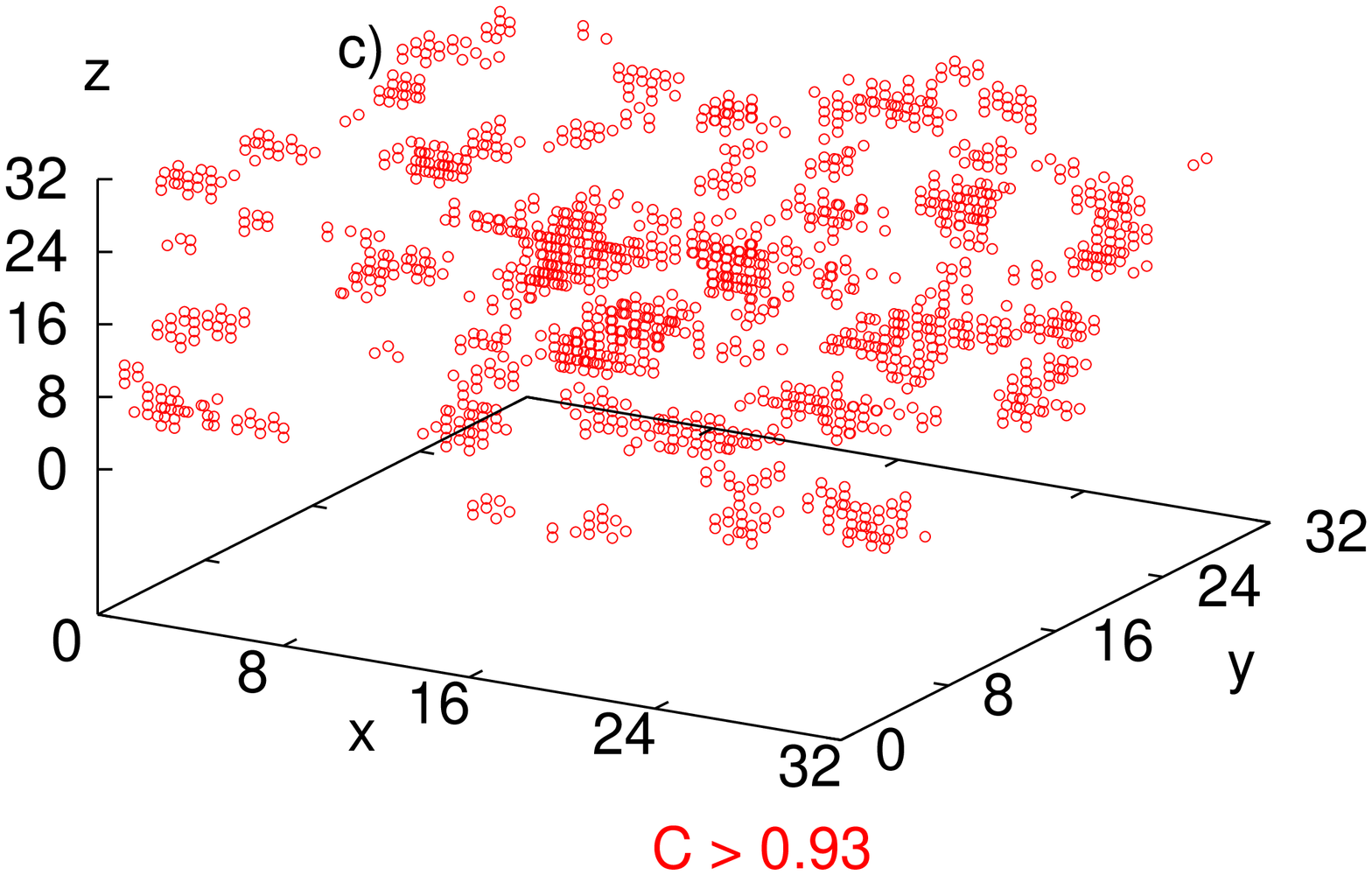,width=6cm,angle=-90}
\end{figure}
\begin{figure}
\psfig{file=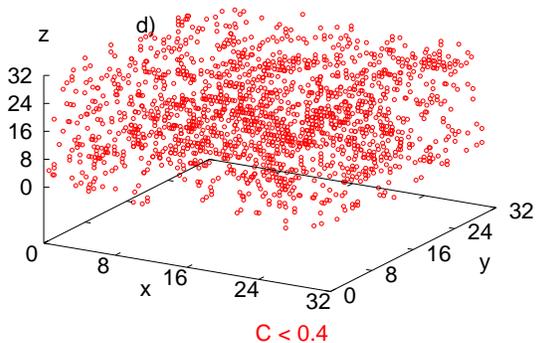,width=6cm,angle=-90}
\caption{a) Slices of $C^{cg}_i\in[0.95,1]$ (extended) and b) $-1 <
C^{cg}_i \leq -0.5$ (localized).  The coarse-graining length is 
$2M+1=3$. $L = 64$, $t/t_w = 2$, $t_w = 2 \times 10^4$ {\sc mc}s, $T = 0.72
\, T_c$.  Slices of c) $C^{na}_i\in[0.93,1]$ (localized) and d)
$C^{na}_i < 0.4$ (extended), and $L = 32$ with $t_w = 3.2\times 10^4$ 
{\sc mc}s,
$t/t_w = 2$, $\tau = 3200$ {\sc mc}s, $T = 0.7$. }
\label{fig1}
\end{figure}

The richness of the spatial structure is illustrated in
Fig.~\ref{fig1}a) and Fig.~\ref{fig1}b) by highlighting slices of the
correlation $C^{cg}_i$ taking values in a chosen interval.  We show
slices with ${\tt C}=0.95$ and $d{\tt C}=0.05$ in Fig.~\ref{fig1}a),
and slices of negatively correlated sites in Fig.~\ref{fig1}b).  The
waiting-time is $t_w=2\times 10^4$ {\sc mc}s and $t/t_w=2$. The
coarse-graining time is $\tau=10^3$ {\sc mc}s. Regions with negative
correlation are well localized in space while the sites with large
$C^{cg}_i$ are evenly distributed throughout the sample. We find an
essentially space-filling distribution of points for any choice of
positive $C^{cg}_i$ larger than about $0.1$ and a localized
distribution of points for $C^{cg}_i\in[-1,0]$. It should also be
noted that there is a distinct spatial anti-correlation between the
location of sites with positive and negative correlation.
We find that the number of points with $C\leq 0$ scales approximately
as $L^2$ and that the number of points with $C >0$ scales
approximately as $L^3$ for both $L = 32$ and $L = 64$.

When we look at the spatial structure for noise-averaged correlations, a
different picture emerges.  In this case, there appear to be space-filling
distributions of points for low correlations up to about $C_i^{na} \sim
q_{\sc ea}$ in Fig.~\ref{fig1}c) and localized regions of correlations for
$C_i^{na}$ close to 1 in Fig.~\ref{fig1}d). Indeed,
one might have expected that only sites that are
essentially forced by the disorder realization to take a particular
configuration and do not decay in time ({\it i.e.} those with large
values of $C_i^{na}$) would be likely to show a localized distribution
of points. In the noise-averaged case, there are essentially no sites
with negative correlations, as can be seen in Fig.~\ref{fig1}c)-d) (see
also Fig.~\ref{distaverCi} for the {\sc pdf} of noise-averaged local 
correlations).

\subsection{Fractal analysis}

We next studied the fractal dimension of clusters of connected spins with
$C^{cg}_i \in [{\tt C}, {\tt C}+d{\tt C}]$. The total number of spins in the
cluster is regarded as its ``mass'', $m$, and its radius of gyration, $R_g$,
is evaluated via the definition $R_g^2 = 1/(2m^2) \sum_{ij} |\vec r_i -\vec
r_j|^2$ where $\vec r_i$ is the position of spin $i$ \cite{Nakayama}. The 
fractal dimension $d_f$ of the clusters is defined from the scaling of 
the ``mass'' $m$ with the radius of gyration $R_g$, $m\propto R_g^{d_f}$.

\subsubsection{Coarse-grained correlations}

\begin{figure}
\begin{center}
\psfig{file=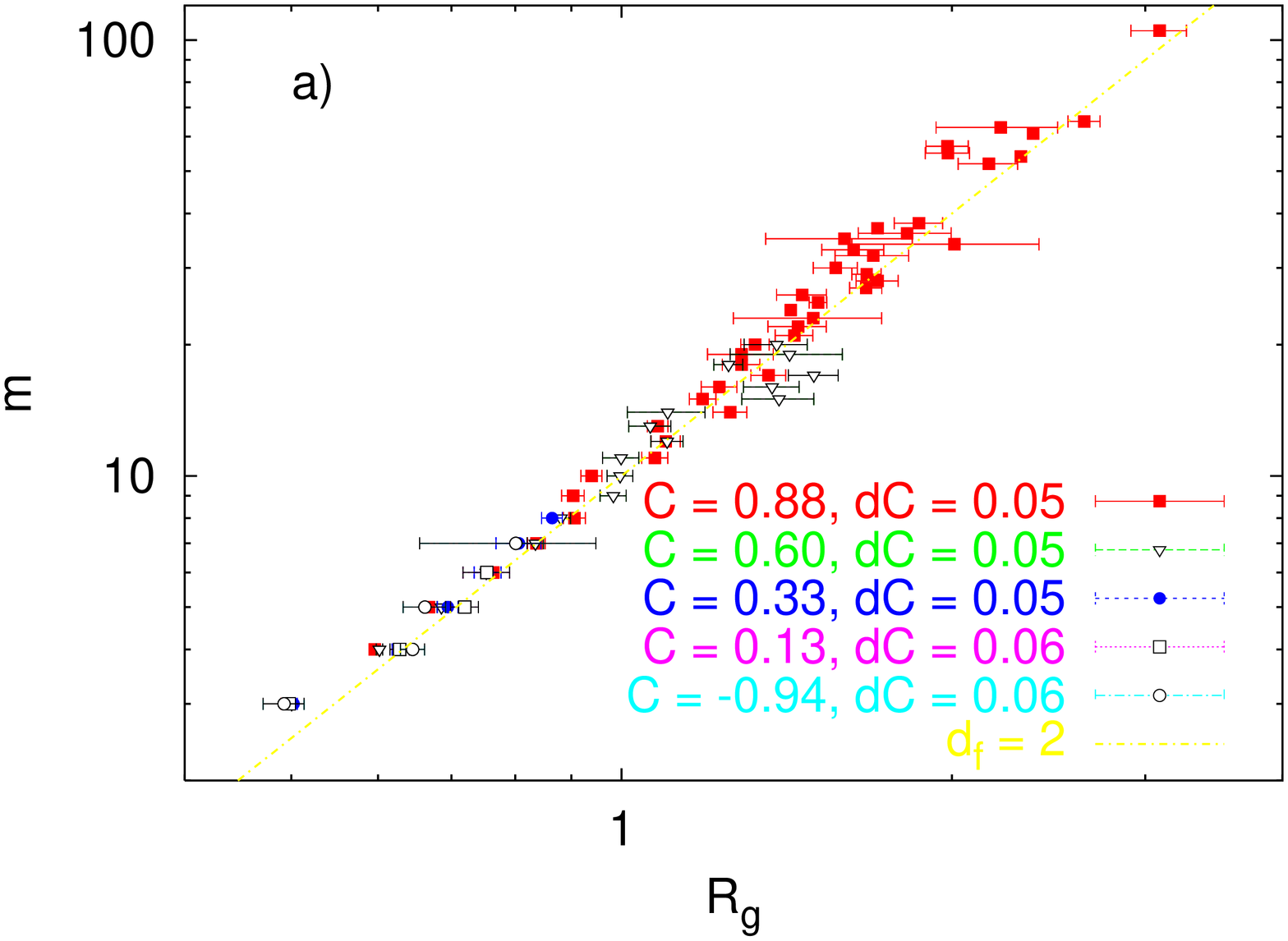,width=6.2cm,angle=270}
\psfig{file=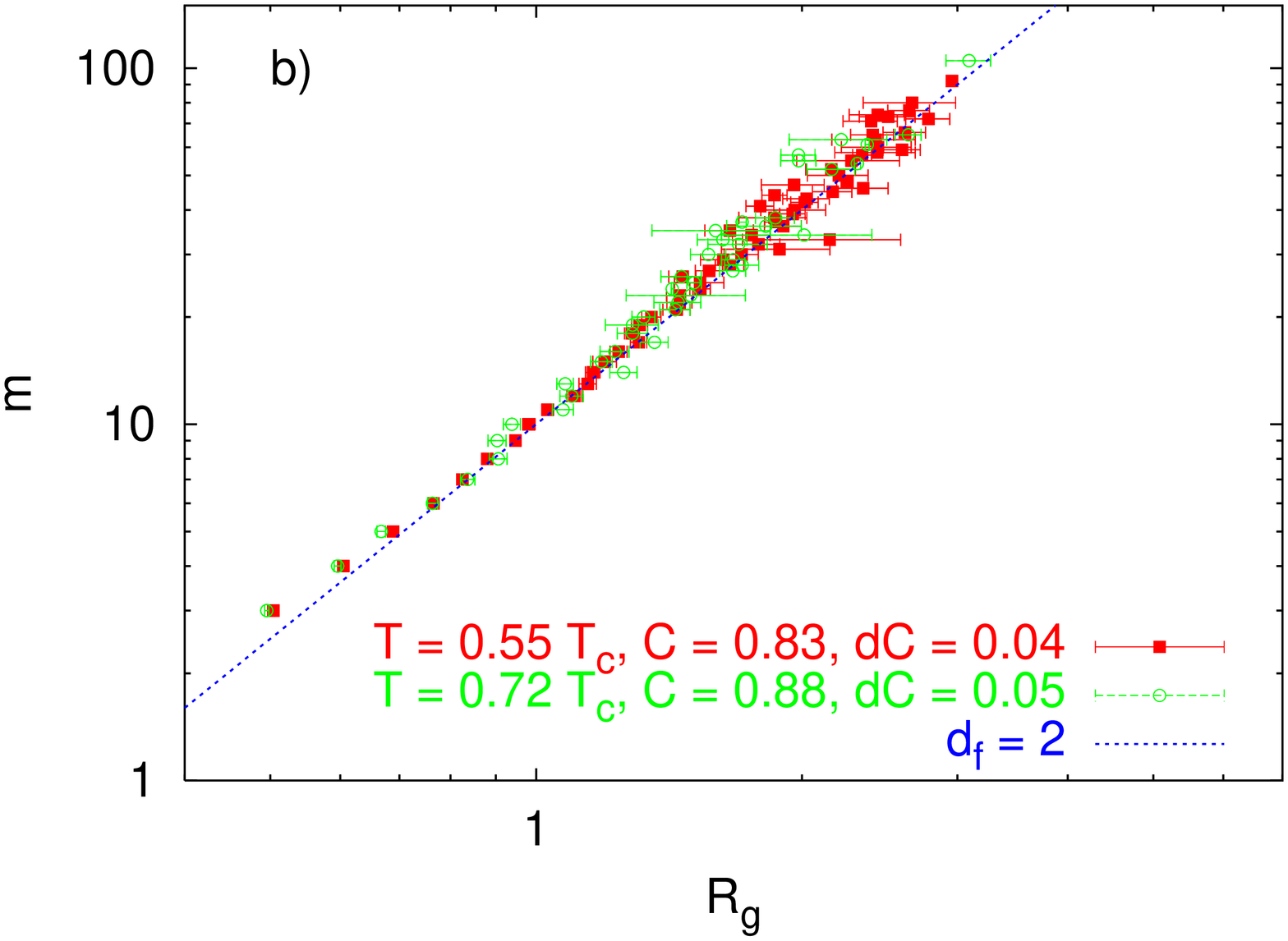,width=6cm,angle=-90}
\end{center}
\vspace{0.5cm}
\caption{Number of spins ($m$) against radius of gyration ($R_g$) for
the coarse-grained correlations in the 
$3d$ {\sc ea} model.  $M=1$.  Panel a) $T=0.72 \, T_c$ and three
pairs of ${\tt C}$ and ${\tt dC}$. $t_w=6.4 \times 10^5$ {\sc mc}s and
$t/t_w=2$.  Panel b) two values of $T$, $0.56 \, T_c$ ($t_w=3.2 \times
10^5$ {\sc mc}s and $t/t_w=2$), $T=0.72 \, T_c$ ($t_w = 6.4 \times 10^5$ {\sc mc}s
and $t/t_w=2$). The dotted lines correspond to $d_f=2$.}
\label{fig2}
\end{figure}

Figure~\ref{fig2} a) shows $m$ against $R_g$ for the $3d$ {\sc ea}
model and five pairs $({\tt C},d{\tt C})$ corresponding to ${\tt C} <
0$, $0 < {\tt C}<q_{\sc ea}$, ${\tt C}\sim q_{\sc ea}$ and ${\tt
C}>q_{\sc ea}$ (see the key).  The coarse-graining linear size is
$M=1$. There is no qualitative difference between the five sets of
data and they are consistent with $d_f=2.0\pm 0.1$.  It should be
noted, however, that we only fit over two decades in the number of
spins and it would be desirable to have a larger dynamic range to get
more precise results.  We obtained the same results for other values
of $t_w$ and $t$ and for different temperatures $T< T_c$, see panel b)
in the same figure. The results are relatively independent of the
value of $d{\tt C}$, the values of $d{\tt C}$ chosen are such that
they are big enough that they wash out some of the site-to-site
fluctuations due to thermal noise, but also small enough that they do
not lump very different values of the correlation in the same bin.
The results do not depend that strongly on the coarse-graining volume
until one gets to large coarse-graining volumes such as $M=13$, for
which $d_f\sim 2$ for small clusters with a cross-over to $d_f\sim d$
at larger cluster sizes.

In Sec.~\ref{correlation-length} we defined a two-time dependent
correlation length that is of the order of 3 to 4 lattice spacings for
the times and temperatures we have considered.  Notice that all the
clusters obtained using $M=1$ have $R_g<3$, which is of the order of
the correlation length $\xi$. This is consistent with having $d_f\sim
2< d=3$ for $R_g<\xi$.

Some recent work suggested that one class of low energy excitations in
3$d$ ${\sc ea}$ spin glasses have the properties of lattice animals
\cite{Lamarcq} and hence a fractal dimension of 2.  In that study, a
fractal dimension $d_f\sim 2$ was found for the lowest energy
excitations above the ground state made of a connected cluster with
chosen number of spins and a given site. The precise connection
between these objects and the ones studied here is, however, not
clear.  However, it is intriguing that the fractal dimension observed
here is identical to the one they observed.

\begin{figure}
\centerline{
\psfig{file=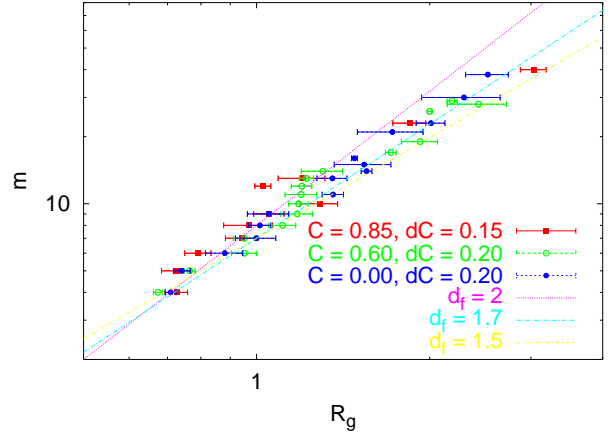,width=6cm,angle=270}}
\vspace{0.5cm}
\caption{Number of spins ($m$) against radius of gyration ($R_g$) for
the coarse-grained correlations in 
the $2d$ {\sc ea} model.  $t_w=5 \times 10^3$ {\sc mc}s and $t/t_w=16$. $T =
0.8$. $M=1$.}
\label{fig3}
\end{figure}

In $d=2$ the glass transition occurs at $T_c=0$. However, the dynamics
at low $T$ and large finite times strongly resembles those seen in
$3d$.  Several ``glassy'' features are observed, such as aging
phenomena and a non-trivial relation between global correlation and
response~\cite{BarratBerthier}, that eventually disappear at very long
times.  In Fig.~\ref{fig3} we analyze the fractal dimension of a two
dimensional system with $L=128$, $M=3$, $t_w=5 \times 10^3$ {\sc mc}s and
$t/t_w=16$. The dotted lines indicate where the points would be
expected to lie for $d_f = 2$, $d_f = 1.7$ and $d_f = 1.5$.  Lattice
animals in two dimensions have $d_f \simeq 1.5$ below the percolation
threshold, $d_f \simeq 1.85$ at the percolation threshold, and $d_f
\simeq 2$  for percolating clusters \cite{Ball}.  The
results here are not inconsistent with lattice animals in 2$d$, but
the large error bars make it hard to definitively establish a
connection.  (The errorbars are much more important than in $d=3$
since we work with many fewer spins: $128^2=16384$ as opposed to
$64^3=262144$.) The ratio $t/t_w$ used in the plot is quite large,
similar results are obtained for other (smaller or larger) ratios.

\subsubsection{Noise averaged correlations}

\begin{figure}
\centerline{
\psfig{file=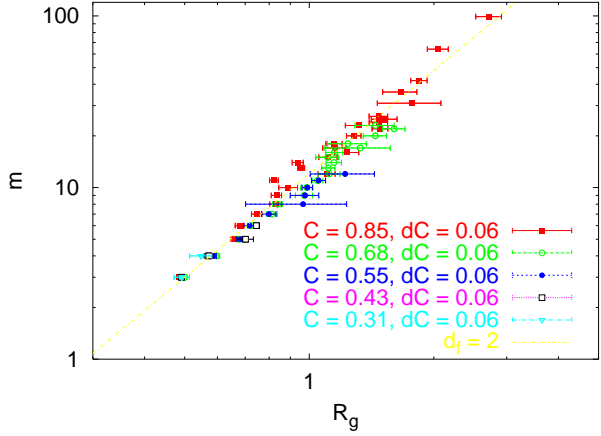,width=6cm,angle=270}
}
\vspace{0.5cm}
\caption{Number of spins ($m$) against radius of gyration ($R_g$) for
the noise-averaged correlations in the 
$3d$ {\sc ea} model. $L = 32$, $t_w = 8 \times 10^3$ {\sc mc}s, 
$t =1.28 \times 10^4 $ {\sc mc}s,
$ T = 0.7$ and the averaged has been done using $868$ noise realizations
for the same random exchanges.}
\label{fig:cnafig1}
\end{figure}

In addition to looking at the coarse-grained correlations, we also performed
a fractal analysis on the noise-averaged correlations.  Whilst there are a
number of similarities between the data, there were also some significant
differences.  At relatively short $t_w$ and $t$, essentially all values of
${\tt C}$ appear to have $d_f \simeq 2$, as can be seen in
Fig.~\ref{fig:cnafig1}.

\begin{figure}
\centerline{
\psfig{file=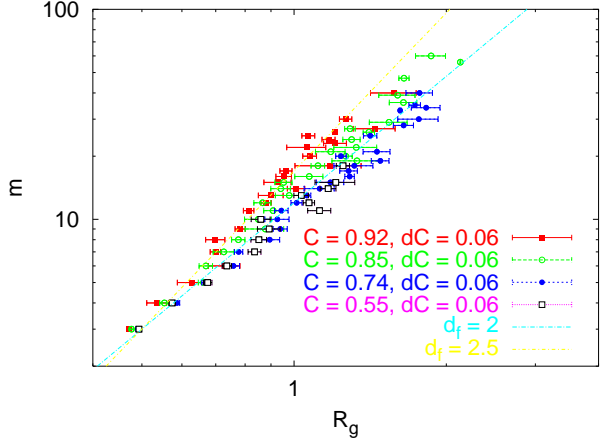,width=6cm,angle=270}
}
\vspace{0.5cm}
\caption{Number of spins ($m$) against radius of gyration ($R_g$) for
the $3d$ {\sc ea} model. $L = 32$, $t_w = 3.2 \times 10^4$, $t =48000 $, $ T
= 0.7$ 822 samples.  }
\label{fig:twodf}
\end{figure}

However, at longer $t$, or $t_w$ there are two types of behavior, depending
on the value of the correlation.  For correlations ${\tt C}$ less than the
peak in the distribution of $C_i^{na}$, the fractal dimension is close to 2,
as found in all other cases.  However, for large values of the correlation
(close to 1), it appears that the $d_f \sim 2.5$ for small clusters, with a
crossover to $d_f \sim 2$ at larger cluster sizes. This is illustrated in
Fig. \ref{fig:twodf}.  This large value of $d_f$ suggests that the clusters
contributing for these values of the correlation may be those that do not
have the effect of domain walls passing through them many times, as they are
the same population of spins that are localized in the 3 dimensional plot.

Note that the same phenomenology shown in Fig. \ref{fig:twodf} was also seen
at lower temperatures, $T = 0.56 \,T_c$.  There is no qualitative difference
between the data at that temperature and the data shown here.  

\subsubsection{Average cluster sizes}

\begin{figure}[htb]
\centerline{
\psfig{file=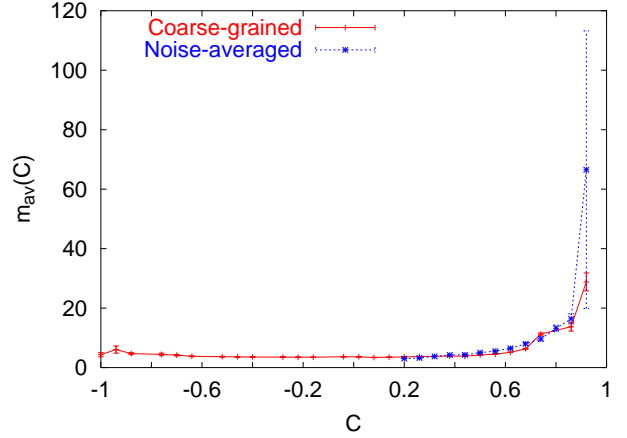,width=6cm,angle=270}
}
\vspace{0.25cm}
\caption{Average cluster size 
$m_{\sc av}$ against the value of the correlation on the cluster 
${\tt C}$.  ${\tt dC} = 0.06$. The system size is $L = 64$,
$T = 0.72 \, T_c$, $t/t_w = 2$ and $t_w = 6.4 \times 10^5$ {\sc mc}s
for the coarse-grained data. 
The noise-averaged data is for $L=32$, $T=0.7$, 
$t_w = 3.2 \times 10^4$ {\sc mc}s and $t/t_w=2$.}
\label{fig:avm}
\end{figure}

Another way to analyze the clusters defined above is to look at their average
size for a given value of the correlation. As illustrated in
Fig. \ref{fig:avm}, it is clear that the largest clusters are found in the
vicinity of $C_i > q_{\sc ea}$ for both the noise-averaged and coarse-grained
cases.   This average cluster size for a given
correlation $C$ is defined as (note that $m_k(C)$ is the ``mass'' of the
$k^{th}$ cluster at correlation $C$)
\begin{equation}
m_{\sc av}(C) \equiv
 \frac{1}{n(C)}
\sum_{k=1}^{n(C)} m_k.
\end{equation}

\begin{figure}
\psfig{file=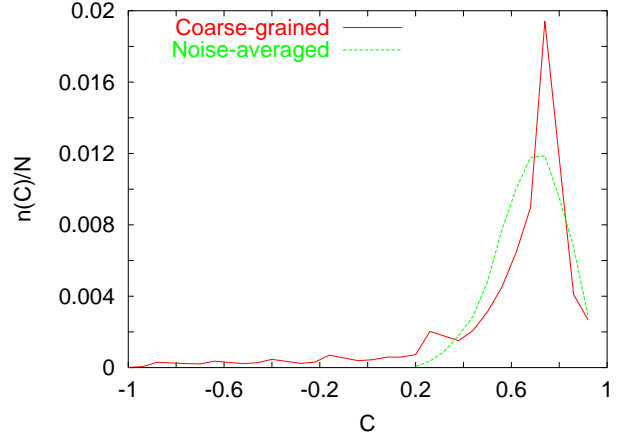,width=6cm,angle=270}
\vspace{0.25cm}
\caption{Number of clusters normalized by the size of the system
$N$ as a function of the correlation. Parameters are the same as 
for Fig.~\ref{fig:avm}.
}
\label{fig:numberclusters}
\end{figure}

We also plot the number of clusters at a given correlation, $n(C)$, for
both the noise-averaged and coarse-grained cases in
Fig. \ref{fig:numberclusters}.  Both have relatively similar forms to
the {\sc pdf} of the correlations, however, in the coarse-grained case
there are very few clusters at negative $C$, consistent with the
existence of a few localized clusters evident in Fig.~\ref{fig1}.  The
peak in cluster sizes also does not coincide with the peak in $n(C)$,
suggesting a characteristic state of a large number of moderately
small clusters with only a few large clusters even when ${\tt C} \sim
q_{\sc ea}$.

\subsection{Multifractal analysis}

Another approach to investigating the spatial structure of
correlations in aging dynamics is to perform a multifractal scaling
analysis. In a disordered system, different moments of a probability
distribution may scale in different ways as the length scale
changes. A multifractal analysis seeks to determine the scaling
behavior of each moment of the distribution. We do not find any
multifractal behavior here, but it would be interesting to look for
it in other glassy models or in larger simulations.  The method used
here, as in the method of fractal analysis, is very simple to implement
and could also be used to examine experimental data, such as that
collected in confocal microscopy experiments.

We follow a similar procedure to that outlined by Janssen \cite{Janssen}, and
define a box probability $p_i$ such that

\begin{equation}
\label{eq:botox}
p_i = \frac{C_i \theta(C_i)}{\sum_j C_j \theta(C_j)}
\end{equation}
where $C_i$ is taken to be the average correlation in the box. Let the box
size be $L_b$ and the system size $L$, which leads to a dimensionless
parameter $\lambda = L_b/L$. The theta function in Eq.~(\ref{eq:botox}) is to
enforce $p_i > 0$ and thus $p_i$ has an interpretation as a probability. In
practice, the theta function is not very important, since the average
correlations in the data considered here are positive in all but a very small
proportion of the boxes, even at the smallest box size $L_b = 2$ in 3 dimensions.  The
distribution of box probabilities $P(L_b)$ is such that when all boxes are
included, the total probability is unity.  For a given $\lambda$ there are
$N(\lambda) \sim \lambda^{-d}$ boxes, where $d$ is the dimension of space.
Moments of the distribution are defined via the relation

\begin{equation}
\left<P^q(L_b)\right>_L = \frac{1}{N(\lambda)} \sum_i p_i^q
\; ,
\end{equation}
and the scaling of these moments is investigated. Multifractal scaling should
be valid for

\begin{equation}
\label{eq:mfcondition}
l \ll L_b < L \ll \xi 
\; ,
\end{equation}
where $l$
is a microscopic length and $\xi$ is the correlation length, and we expect
the moments to scale as
\begin{equation}
\left<P^q(L_b)\right>_L \sim \lambda^{d + \tau(q)}
\; ,
\end{equation}
where $\tau(q)$ defines a generalized dimension $d(q)$ via $\tau(q) =
d(q)(q-1)$ for $q > 1$.

\subsubsection{3$d$ {\sc ea} model}

\begin{figure}
\centerline{\psfig{file=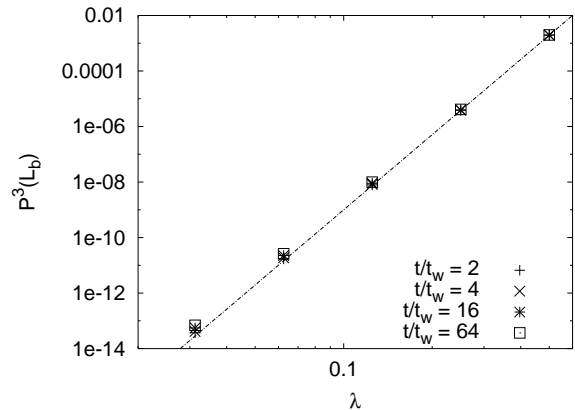,height=8cm,angle=270}}
\vspace{0.25cm}
\caption{Multifractal scaling of $P^3$ in the $3d$ {\sc ea} model.
The straight line is a fit to $P^3(L_b) \propto \lambda^9$, {\it i.e.}
$d(3) = 3$.}
\label{fig:3Dp3}
\end{figure}

In the 3$d$ {\sc ea} model it is found that $\tau(2) = 3$ and $\tau(3)
 = 6$, corresponding to $d(2) = 3$ and $d(3) = 3$.  The same behavior,
 {\it i.e.} $d(q) = 3$ is observed for $q =4$, 5, and 6.  As is evident in
 Fig.~\ref{fig:3Dp3}, which is for data with $L=64$, at $T=0.72 \,
 T_c$ with $t/t_w$ = 2, 4, 16, and 64, there appears to be very little
 dependence on the ratio of the two times.

\subsubsection{$2d$ {\sc ea} model}

\begin{figure}
\centerline{\psfig{file=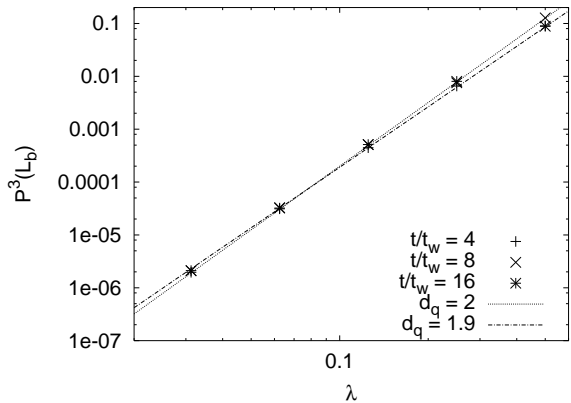,height=8cm,angle=270}}
\vspace{0.25cm}
\caption{Multifractal scaling of $P^2$ in the $2d$ 
{\sc ea} model. $T = 0.8$, $t_w = 5\times 10^3$ {\sc mc}s,
$L = 128$, $t/t_w$ as shown in the key.}
\label{fig:2Dp2}
\end{figure}

In performing a multifractal analysis of the 2$d$ {\sc ea} model, the
same approach to the analysis was used as outlined for the 3$d$ {\sc ea}
model.  The results were similar to the 3$d$ case, in that $d(q) \sim
d$ for the times available.  In the data shown in Fig.~\ref{fig:2Dp2},
it is clear that for $q = 3$, $1.9 < d(3) < 2$, with perhaps a
slightly lower value of $d(q)$ at long waiting times.  The data is
from a run with $L = 128$, $T = 0.8$ and $t_w = 5000$.  The time
ratios considered are $t/t_w$ = 4, 8 and 16.  (Note that alternative
definitions of the box probability in 2$d$, such as using $|C_i|$
rather than a theta function can lead to $d(q) \neq d$.  In that case
it appears to be due to the mixing of non-equilibrium correlations
with equilibrium correlations.  There will still be such mixing using 
the definition above, but all negative correlations are excluded, and
these may cover a non-trivial number of boxes for small box sizes.)

\subsubsection{Summary}

The picture obtained from the above analysis is that the multifractal
structure of aging dynamics in 2 and 3 dimensional {\sc ea} spin
glasses is very similar. In both cases, all moments of the probability
distribution of correlations appear to scale with the same generalized
dimension, which is equal to the dimension of space. The reason why this is
so here is that the correlation length $\xi$ is small, of the order 3 to 5
lattice spacings, and therefore we are not truly in the scaling regime
defined in
Eq.~(\ref{eq:mfcondition}). Nonetheless, this method of examining spatial
heterogeneities via multifractal analysis may be a useful tool for
investigating the behavior observed in different glassy models and in
experimental systems.

\section{Conclusions}

In this paper we showed further evidence 
that the coarse-grained 
two-time correlators are the ones that reflect the existence of an
asymptotic zero mode in the underlying theory.  We defined a 
two-time correlation length that controls the spatial fluctuations 
in the coarse-grained local two-time correlators and we 
showed numerical evidence for the asymptotic divergence 
of this length in the glassy phase of the $3d$ {\sc ea} model, 
as was to be expected 
from the existence of the asymptotic zero mode. It should be noted, 
though, that for the times reached numerically the correlation 
length is still very short. We argued that in the limit in which 
the coarse-graining linear size $M$ is taken to diverge together with the 
correlation length the individual and joint 
distributions of coarse-grained local correlations and integrated
responses should reach a stable form. This means that heterogeneities
of all sizes exist in the system. This is another consequence 
of the asymptotic zero mode.

In disordered systems 
one can also define noise-averaged, as opposed to
coarse-grained, fluctuating two-time functions. 
We showed numerically that the fluctuations in these quantities
are not controlled by the zero mode and that they behave 
rather differently than their coarse-grained cousins.  
In slightly more technical terms, we claim that:
\newline 
(i) the fluctuations in the local time-reparametrization
$h_i(t)$ [see Eq.~(\ref{eq:localh})] are coupled to the thermal noise
and hence manifest in the fluctuations of the coarse-grained 
local correlations and responses.
\newline
(ii) the fluctuations in the external functions
$f_i$ [see Eq.~(\ref{limit1})], are coupled to the quenched disorder
and hence manifest in the fluctuations of the noise-averaged 
local correlations and responses.

Consequently,
in a system with disorder the 
noise-averaged local quantities show fluctuations in $f_i$ but 
average out  the ones in $h_i$. In a system without disorder
these quantities do not fluctuate. In contrast, for any coarse-graining
volume if the times are long-enough the coarse-grained local 
quantities keep the fluctuations in $h_i$ while the ones in 
$f_i$ are erased since an effective average over the random 
exchanges is performed.

We related the study of the fluctuations in the local 
correlations (and susceptibilities) to the study of 
the evolution of random surfaces. The local two-time functions
correspond to the ``local heights''  of a fluctuating random 
surface on the $d$ dimensional substrate.
We presented a phenomenological effective action 
for the fluctuations in the local quantities. This allowed us to 
predict several dynamic properties of their distributions. 
 On the other hand, the 
geometric analysis of clusters of spins that we introduced in 
Sec.~\ref{geometric} also has a counterpart in the theory of random 
surfaces; it corresponds to the analysis of contour levels of the 
surfaces~\cite{contour}. 
A complete study of the statistical and dynamic properties of these
surfaces might be useful to determine the 
lower critical dimension of different glassy models. 

The analytic calculations that we use as a guideline in this 
paper where performed using the finite $d$ spin glass 
Hamiltonian~\cite{paper1}. The numerical data that we present 
also correspond to this glassy problem. However, we believe
that our results are more general and should apply also, with 
a few modifications, to other glassy problems. In the next 
section we discuss several possible spin-offs of our results as
well as a number of models in which the ideas here discussed 
can be put to test.

\section{Perspectives}

Several questions remain open even within the study of the 
finite $d$ {\sc ea} model. In particular, we have not checked numerically 
that a scaling limit is reached by taking the limit of long-times 
and large coarse-graining volumes while keeping $(2M+1)/\xi(t,t_w)$ fixed.
This check remains out of the reach of present 
computer simulations.
Another independent issue that deserves further study is the intriguing 
fact that the noise-averaged local correlations are quite insensitive 
to temperature and its possible relation to the 
observation of memory and 
rejuvenation in spin-glass experiments. Moreover, the study of the geometric 
properties of clusters can be improved and made more complete in 
several directions, {\it e.g.} by examining lower temperatures in the $2d$ case
and higher temperatures in the $3d$ case in order to identify 
similarities and differences, etc.

We have stated that our approach and its results, derived in this 
manuscript for spin models with disorder,  can be adapted to 
describe the structural glass problem. Let us summarize what 
we expect in this case. 

Dynamic heterogeneities in the super-cooled liquid phase have been
identified numerically~\cite{heterogeneities-num,Heuer} and
experimentally~\cite{Ediger,heterogeneities,confocal}. In the case
of particle (or molecular) interacting systems the natural and
simplest two-time local correlation that makes contact with our
approach is the density-density correlator defined on boxes with
volume $V_B=(2M+1)^d$. Having partitioned the total volume $V$ in
$V/V_B$ such boxes, one then has $V/V_B$ local correlators and local
responses (if a perturbation coupled to the density is applied) with
which one can construct the {\sc pdf}s.

In the super-cooled liquid phase we expect that the local correlations
and integrated responses defined in boxes of finite size will be
typically stationary (after a sufficiently long waiting-time that goes
beyond the equilibration time) but with different finite structural
relaxation times. This is consistent with the experimental observation
that dynamic heterogeneities in supercooled liquids seem to have a
lifetime of the order of the relaxation time. At high temperatures the
size of the heterogeneities is finite and hence one should suppress
the fluctuations by using sufficiently large coarse-graining
volumes. The correlation length $\xi(t,t_w)$, that is also stationary,
should remain finite, even in the limit of long-times. From a
theoretical point of view, this picture is, in a sense, similar to the
one that describes the paramagnetic phase in the $O(N)$ model, just
above the ordering transition temperature.

When lowering the temperature the size and life-time of the
heterogeneities increases~\cite{heterogeneities-num}.  A
mean-field-like, or mode-coupling-like approach predicts that their
typical size will diverge at the mode coupling transition
temperature~\cite{Silvio}.  We expect then that the correlation length
$\xi(t,t_w)$ will saturate at a higher value when $T$ decreases
approaching $T_c$.  In real systems the divergence at $T_c$ is
smoothed and hence $\xi(t,t_w)$ should not strictly diverge.

At still lower temperatures the bulk quantities age and we expect then 
to observe heterogeneous aging dynamics of the kind described in this 
paper, with a two-time dependent correlation length for the 
local fluctuations. The heterogeneities will age too, in  a
``dynamic'' way. By this we mean that if a region 
 looks older than another one 
when observed on a given time-window, it can reverse its status and 
look younger than the same other region when observed on a different
time-window. 

The numerical studies of the global two-time correlations and 
integrated susceptibilities of Lennard-Jones 
mixtures~\cite{KobBarrat,JLBarrat} have shown a remarkable 
accord with the predictions 
from the analytic solution to mean-field-like glassy models~\cite{yo,chicos}. 
We then expect that the {\sc pdf}s of  local correlations and local 
integrated responses, for the same time-scales used for the bulk calculations, 
will show the main features described in this 
paper. In particular, we expect the joint {\sc pdf}s 
to reproduce the sketch shown in 
Fig.~\ref{fig:sketch-chiC1step2}.

In the glassy phase, 
we expect the correlation length of structural systems
to grow for increasing times
roughly in the manner here described. However, this growth might be 
modified for long enough times when the 
dynamics crosses over to a different, activated-like, nonequilibrium 
regime that we cannot characterize theoretically~\cite{yo,clarify1}. 
We cannot predict what happens to the correlation length on these 
extremely long time-scales.

The ideas discussed in this paper should not only apply to 
systems that relax in a nonequilibrium manner as glasses
but also to systems that are kept out of equilibrium with a 
(weak) external forcing.
As shown in Ref.~\cite{Cuku3} the 
time-reparametrization of the bulk quantities 
that is selected dynamically is very easy 
to modify with external perturbations. Indeed, a small force 
that does not derive from a potential and is applied on every
spin in the model renders an aging $p$ spin model 
stationary~\cite{Cukulepe} while the model maintains a separation 
of time-scales in which the fast scale follows the temperature of the bath, 
$T$, while the slow scale is controlled by an 
an effective temperature, $T_{\sc eff}>T$. In this case, 
the aging system selects a time-reparametrization 
$h(t)=t$~\cite{Cuku,KimLatz} while in the perturbed model
$h(t)=e^{-t/\tau_\alpha}$. Similarly, the aging of a Lennard-Jones 
mixture is stopped by a homogeneous shear~\cite{JLBarrat}. A
different way to modify the time-reparametrization that characterizes
the decay of the correlations is by using complex thermal baths~\cite{Cuku3}.
Recently, there has been much interest in the appearance of shear
localization, in the form of shear bands, in the rheology of complex
fluids~\cite{shear-bands,shear-bands-num}. Along the lines here
described it would be very interesting to analyze the fluctuations in 
the local reparametrizations in the fluidized shear band and the 
``jammed'' glassy band.

The appearance of an asymptotic 
invariance under time-reparatrizations in the mean-field 
dynamic equations was related to the reparametrization invariance 
of the replica treatment of the statics of the same models~\cite{Sompo,Brde}. 
The latter remains rather abstract. Recently, Br\'ezin and de 
Dominicis~\cite{Brde} studied the consequences of twisting the 
reparametrizations
in the replica approach. Interestingly enough, this can be simply done 
in a dynamic treatment either by applying shear forces, as discussed in the 
previous paragraph or by applying
different heat-baths to different parts of the system. More precisely,
using a model with open boundary conditions 
one could apply a thermal bath with a characteristic time-scale 
on one end and a different thermal bath with a different characteristic
time-scale on the opposite end and see how a time-reparametrization 
``flow'' establishes in the model. 

We would like to end this article by mentioning a number of 
other tests and 
interesting applications of the ideas here described
to other models with a slow relaxation.

(i) In this paper we studied the distributions of the two-time, 
local in space, spin-spin correlations and their associated 
responses. In a finite dimensional system one can construct many 
other two-time correlations that are still local in space.
The question then arises as to if all the distributions of all 
possible correlators have the same qualitative features. 

(ii) An important property of the interpretation of 
the {\sc fdr} in terms of effective temperatures is that 
in systems that reach an asymptotic regime with slow dynamics 
and small entropy production one expects that
all observables evolving in the same time-scale partially 
equilibrate and hence have the same value of the effective 
temperature~\cite{Cukupe}. Related to this question one can 
try to determine if the joint probability 
distributions of the local correlations and susceptibilities
of other pairs of observables are the same and have the same time
evolution.

(iii) We expect to see a similar behavior of the local 
{\sc fdr} in simple systems undergoing domain growth~\cite{Bray}. 
More precisely, using
the {\it coarse-grained} local correlations between the fluctuations in the 
magnetizations and their associated coarse-grained local susceptibilities, 
we expect to find a joint {\sc pdf} that is mostly concentrated along a
{\it flat}, global $\tilde \chi(C)$ curve when $d > 1$.
This statement can be checked rather easily with 
Monte Carlo simulations of the finite dimensional ferromagnetic 
Ising model. 

(iv) Related to (iii), the ferromagnetic Ising chain is a 
particularly interesting case of study. It has been proven 
that at zero temperature (or when the coupling strength diverges) the 
global {\sc fdr} takes a curved form $\tilde \chi(C)$~\cite{chain}. For this 
model it might be possible to derive an analytic expression for the 
joint {\sc pdf}. Similarly, one can attempt an analytic calculation 
at criticality in coarsening models as done in Ref.~\cite{critical-dyn}.

(v) Kinetically constrained lattice models~\cite{Felix}
capture many of the characteristic features of glasses. 
An analytical study of dynamic heterogeneities in one such spin model
has recently appeared~\cite{Juanpe}. Versions in which one works with
particles on a lattice are also 
rather simple to simulate. In  these models one can 
partition the full lattice in boxes of size  $(2M+1)^d$ and 
define the local two-time density-density correlators 
within them. A local susceptibility can also be easily defined 
following Ref.~\cite{Selitto}. A check of the form of the joint probability 
distribution and its evolution in time is an interesting problem.

(vi) One would like to study realistic models of 
glass formers with molecular dynamics and test the scaling laws and 
qualitative features in these cases. 

(vii) 
Herisson and Ocio~\cite{Herisson} studied recently 
the bulk two-time correlation between magnetic fluctuations 
and the bulk two-time integrated response to an external magnetic field 
(magnetic susceptibility) of an insulator spin-glass. 
Their aim was to test the
modifications of the global {\sc fdt} in this nonequilibrium system.
In order to have smooth data, they averaged these quantities 
over many repetitions of the experiment done after 
heating the sample above the transition. In our terms, the
bulk correlations were averaged over different noise-realizations.
Another experimental system in which the two-time evolution of 
a  bulk two-time correlation has been recorded, 
in this case the voltage noise self-correlation,
is laponite~\cite{dyn-heterogeneities}. Large fluctuations appear.
It will be very 
interesting to analyze the {\sc pdf}s of the two-time noise-noise 
correlation and th
e two-times integrated response along the lines 
here described.

Last, but not least, the approach based on reparametrization
invariance suggests that it may be possible to search for universality
in glassiness. For example, a Ginzburg-Landau theory for phase
transitions captures universal properties that are independent of the
details of the material. It is symmetry that defines the universality
classes. For example, one requires rotational
invariance of the Ginzburg-Landau action when describing
ferromagnets. Reparametrization invariance may be the
underlying symmetry that must be satisfied by the Ginzburg-Landau
action of all glasses.
What would determine if a system is
glassy or not? We are tempted to say the answer is if the symmetry is
generated or not at long times. Knowing how to describe the universal
behavior may tell us all the common properties of all glasses, but
surely it will not allow us to make non-universal predictions, such as
what is the glass transition temperature for a certain material, or
whether the material displays glassy behavior at all. This quest for
universality is a very interesting theoretical scenario that needs to
be confronted.

\vspace{0.5cm}
\noindent{\bf Acknowledgements} This work was supported in part by the
NSF grants DMR-98-76208 and INT-01-28922 and the Alfred P. Sloan
Foundation (C.~C.), CNRS-NSF grant, an ACI-France, the Guggenheim
Foundation, and ICTP-Trieste (L~.F.~C.), CONICET and the Universidad
Nacional de Mar del Plata, Argentina (J.~L.~I.).  C.~C. thanks the
LPTHE - Jussieu and M.~P.~K. thanks BU for their hospitality. The simulations
were done with the BU Scientific Supercomputing and Visualization
facilities.

We especially thank J. Kurchan for very enlightning
discussions throughout the development of this work.
We also thank L.~Balents, A.~Barrat, S.~Franz, D.~Huse, A.~Montanari,
G.~Parisi, L.~Radzihovsky, D.~Reichmann, and X.-G.~Wen for  
suggestions and  discussions.

\appendix


\begin{thebibliography}{99}
\bibitem{clarify} The existence of a well-defined plateau
for the $3d${\sc ea} model is not established beyond any doubt.
We shall use it as a hypothesis admitting the possibility that 
future generation simulations might show that $q_{\sc ea}\to 0$ 
even in $d=3$.
\bibitem{clarify0} The dynamic equations are derived in the limit $N\to\infty$.
In this case, the dynamic correlations and responses are self-averaging and 
independent of the nosie realization. For this reason we do not worry about 
having to average the two-time quantities in Eqs.~(\ref{eq:Crpgtransf})
and (\ref{eq:chirpgtransf}).
\bibitem{several-length-scales} We shall not consider the possibility
of there being several (hierarchical) length-scales. Moreover, to keep the 
discussion simple as elsewhere in this article, we study the asymptotic 
properties of the mass keeping $t/t_w$ fixed [using a ratio $h(t)/h(t_w)$ 
fixed might be more precise].
\end{thebibliography}
\end{document}